\documentclass[12pt]{article}
\usepackage{lmodern}
\usepackage[english]{babel}
\usepackage[T1]{fontenc}
\usepackage[latin9]{inputenc}
\usepackage{geometry}
\usepackage{amsmath}
\usepackage{amssymb}
\usepackage{graphicx}
\usepackage{color}
\usepackage{esint}
\usepackage{authblk}
\usepackage{etoolbox}
\usepackage{subfigure}
\usepackage{helvet}
 
\makeatletter
\@addtoreset{figure}{section} 
\providecommand{\tabularnewline}{\\}
\patchcmd{\@maketitle}{\sfdefault \@title}{\selectfont\@title}{\sfdefault}{}
\makeatother

\def\beq{\begin{equation}}
\def\eeq{\end{equation}}
\usepackage{amsthm}

\newtheorem*{lemma*}{Lemma}
\newtheorem*{theorem*}{Theorem}

\newcommand{\R}{\mathbb{R}}
\newcommand{\N}{\mathbb{N}}

\renewcommand{\d}{\mathrm{d}}

\newcommand{\ket}[1]{|#1\rangle}

\newcommand{\scal}[1]{\langle #1\rangle}

\newcommand{\threej}[6]{\begin{pmatrix}  #1&#3&#5\\#2&#4&#6\end{pmatrix}}
\newcommand{\sixj}[6]{\left\{ \begin{array}{ccc}  #1&#3&#5\\#2&#4&#6\end{array}\right\}}

\date{\normalsize \fontfamily{lmss}\selectfont{\today} }
\author[1]{\fontfamily{lmss}\selectfont{Klaus Liegener} \thanks{klaus.liegener@gravity.fau.de}}
\affil[1]{ \fontfamily{lmss}\selectfont{Institute for Quantum Gravity, Friedrich-Alexander University Erlangen-N\"urnberg, Germany}}
\author[1]{ \fontfamily{lmss}\selectfont{Thomas Thiemann} \thanks{thomas.thiemann@gravity.fau.de}}
\title{\fontfamily{lmss}\selectfont{ Towards the fundamental spectrum of the Quantum Yang-Mills Theory}}
\begin{document}
\maketitle
\begin{abstract}
\fontfamily{lmss}\selectfont{ In this work we focus on the quantum Einstein-Yang-Mills sector quantised by the methods of Loop Quantum Gravity (LQG). We point out the improved UV behaviour of the coupled system as compared to pure quantum Yang-Mills theory on a fixed, classical background spacetime as was considered in a seminal work by Kogut and Susskind. Furthermore, we develop a calculational scheme by which the fundamental spectrum of the quantum Yang-Mills Hamiltonian can be computed in principle and by which one can make contact to the Wilsonian renormalization group, possibly purely within the Hamiltonian framework. Finally, we comment on the relation of the fundamental spectrum to that of pure Yang-Mills theory on a (flat) classical spacetime.}
\end{abstract}
\section{Introduction}
\label{c1}
\numberwithin{equation}{section}
The Hamiltonian approach to pure quantum Yang-Mills theory on Minkowski space was much developed by Kogut and Susskind \cite{KS75}. These authors regularised the classical expression for the Yang-Mills Hamiltonian on a regular spatial lattice of cubic topology embedded in $\R^3$, which comes with a lattice length parameter $\epsilon$ as measured by the spatial Euclidean background metric induced by the Minkowski metric on spatial hypersurfaces of Minkowski space. The quantum Hamiltonian was written in terms of non-abelian fluxes through the faces of the cubic cell complex dual to the lattice for the electric degrees of freedom and in terms of non-abelian holonomies along the plaquette loops of the lattice. Furthermore, those authors assumed a representation of holonomies and fluxes on a Hilbert space of square integrable functions of the magnetic loop functions just introduced, where the natural Haar measure on the compact gauge group is used in order to define the Hilbert space measure.\\
While well defined at finite $\epsilon$, the necessary continuum limit $\epsilon \rightarrow 0$ is problematic in this approach: Namely, the regularized Hamiltonian involves an inverse power of $\epsilon$ and thus blows up at fixed Yang-Mills coupling. This leads to the conclusion that the Yang-Mills coupling entering the Hamiltonian is to be considered a bare coupling that must be renormalized suitably in the continuum limit. Since the renormalization is, arguably, easier to study in the path integral formulation, the Hamiltonian approach to quantum Yang-Mills theory was basically dropped and research focused on the functional integral approach, whose underlying mathematical framework is the constructive Euclidean program \cite{OS72,GJ87,Bal89a,Bal89b,Fr78,Riv00}. Starting from the Euclidean action, not the Hamiltonian, hence involves an additional integral and thus in 4 spacetime dimensions does not involve $\epsilon$ explicitly. The well established and very active research field of Lattice Quantum chromodynamics (LQCD) is the practical implementation of that program and has produced many spectacular results, see e.g. \cite{GL10,Cr83}, yet the existence of pure quantum Yang-Mills theory has not been proven. In fact, the Clay Mathematical Institute \footnote{ http://www.claymath.org/millenium-problems/yang-mills-and-mass-gap} has devoted one of its millennium prizes to this research topic.\\
To circumvent these problems this paper does not deal with the Euclidean formulation at all. Futhermore, we will leave the realm of QFT on curved spacetime \cite{BFV03,Buch00, FRS07, JR06} completely and pass to quantum gravity, because we wish to examine here the old idea that quantum gravity itself resolves the UV divergences of QFT. We do this in the Hamiltonian approach to quantum gravity, one incarnation of which is Loop Quantum Gravity (LQG) \cite{ Thi07, Rov04, GS13}. This approach is ideally suited to Yang-Mills theory, because the gravitational field, in its canonical formulation, can be viewed as a Yang-Mills theory for the gauge group SU(2) with a very complicated interaction. Thus the quantisation methods developed for Yang-Mills fields, in fact pioneered by Kogut and Susskind, can also be applied to the gravitational degrees of freedom, as has been done in \cite{Thi98}.\\
Indeed, a rigorous Hilbert space representation can be found for the so called holonomy flux algebra, in fact for any compact gauge group and any spacetime dimension, which consists of holonomies along one dimensional paths and non-Abelian fluxes through two dimensional surfaces (in 3+1 spacetime dimensions). This is in fact very similar to the Kogut-Susskind program, but the difference is that in LQG there is no fixed lattice and dual cell complex, there is also no lattice regulator $\epsilon$ at all. Rather, one considers {\it all paths and all surfaces} in one big Hilbert space, that is to say, one considers all graphs and dual cell complexes. LQG is therefore a continuum theory without a lattice cut-off.  We will see that in the corresponding quantum operator the factor $1/\epsilon$ of the Kogut-Susskind Hamiltonian is replaced by the $1/\ell_P$ where $\ell_P$ is the Planck length. At that level therefore, there is no problem in taking the continuum limit. However, renormalisation group ideas are still important as we see later on.\\
Just in order to avoid possible confusion from the outset, we mention here that LQG comes in two versions. In the first version one solves the constraints of the theory, which arise due to the spacetime diffeomorphism invariance of Einstein's theory, in the quantum theory \cite{Thi96_1,Thi96_2}. In the second version one solves those constraints classically by gauge fixing the freedom to choose coordinates in terms of scalar matter fields (see e.g. \cite{GT12, DGKL10, HP11, HP13}). These two approaches are technically and conceptually very different, because in the first version the primary task is to solve the quantum constraints and to supply a Hilbert space structure on the resulting space of (distributional) solutions and it is a non-trivial task to find appropriate gauge invariant observables acting on it. There is no Hamiltonian in this first approach, because time translations are regarded as gauge transformations. In the second approach these tasks are already implemented classically. Furthermore, the classical construction automatically supplies a Hamiltonian that generates time evolution. In this paper we will therefore follow the second route, specifically the choice of scalar matter considered in \cite{KT89,GT07_4} as this brings us maximally close to the situation of pure Yang-Mills theory on Minkowski space.\\
The LQG Hilbert space, which was originally designed for the first approach, is necessarily non separable. This comes about because one considers the huge algebra of {\it all} fluxes and {\it all} holonomies, which in turn are needed if one wishes to implement the (spatial) diffeomorphism invariance of the theory in a (cyclic) representation of the holonomy - flux algebra \cite{LOST06,Flei06}. On the other hand, classically, far fewer functions on the phase space would suffice in order to separate all of its points, that is to say, much fewer paths and surfaces would suffice. In \cite{GT06_1,GT06_2,GT06_3,GT07_4} the observation was made, that - since in the second approach one has fixed the (spatial) diffeomorphism invariance of the theory - one may indeed restrict to a much smaller algebra. For instance, if the topology of spacetime is that of $\mathbb{R}^4$ then it suffices to consider rectangular paths and surfaces along the coordinate axes and planes, respectively. A further reduction of the number of degrees of freedom is obtained by passing to an abstract infinite graph and dual cell complex respectively, which have no information about their embedding into $\mathbb{R}^3$. The quantum theory is then formulated in terms of these abstract elementary holonomy and flux operators. The embedding scale reappears in the semiclassical limit in terms of coherent states \cite{STW01} for the gravitational degrees of freedom and can be chosen as small as one wishes.\\
In this paper we therefore consider the approach of \cite{GT07_4} to Einstein-Yang-Mills theory on the differential manifold $\mathbb{R}^4$ in the gauge fixed version of LQG\footnote{That is, the coordinate freedom is fixed but not the Yang-Mills like gauge freedom.} with scalar matter content and focus on the Yang-Mills contribution to the Hamiltonian, which then in the classical theory simply reads:

\begin{equation} \label{3}
H=\frac{1}{2Q^2} \;\int_{\mathbb{R}^3}\; d^3x\; \frac{q_{ab}}{\sqrt{\det(q)}}
[{\rm Tr}(\underline{E}^a \underline{E}^b)
+{\rm Tr}(\underline{B}^a \underline{B}^b)]
\end{equation}\\
Here $\underline{E},\underline{B}$ denote the electric and magnetic Yang-Mills field, $Q$ is the Yang-Mills coupling constant and $q_{ab}$ is the induced spatial metric on the Cauchy surface $\mathbb{R}^3$. The spatial indices are $a,b,c, ..=1,2,3$ and the traces are taken in the adjoint representation of the Lie algebra $\mathfrak{g}$ of the Yang-Mills gauge group $G$, e.g. su(N) for $G=SU(N)$. \\
\\
The architecture of this paper is as follows:\\
\\
In section \ref{c2} we will briefly review the quantisation of (\ref{3}), more details can be found in \cite{Thi98, GT07_4}. We also review the essentials of \cite{KS75} and compare these two theories.\\
Section \ref{c3} reviews useful facts about the representation theory of SU(3) (QCD gauge group) needed in sections \ref{c4} and \ref{c5}, while analogous knowledge for SU(2) (gravitational gauge group) are shifted to the appendix.\\
In section \ref{c4} we compute basic building blocks necessary in order to compute the {\it background spectrum} of (\ref{3}) with fixed Minkowski background metric, that is $q_{ab} = \delta_{ab}$, on a lattice of size $\epsilon$, i.e. we treat the Kogut \& Susskind situation.\\
In section \ref{c5} we do the same, but with $q_{ab}$ being a quantum operator on the LQG Hilbert space. The calculational steps performed here are the preparation for computing the {\it fundamental spectrum} of $H$ on the tensor product Hilbert space corresponding to both geometry and matter degrees of freedom.\\
In section \ref{c6} we summarize our findings and elucidate the necessary steps for our  future research.
\section{Review of Einstein-Yang-Mills-Theory}
\label{c2}
In this chapter we recap elements of the classical and quantum Einstein-Yang-Mills theories. In the first section we review the classical canonical formulation and in the second we formulate the quantum theory using the techniques of Loop Quantum Gravity (LQG). We also review the derivation of the Kogut-Susskind lattice Hamiltonian on Minkowski space. Notice that our quantisation makes use of the presence of additional scalar matter fields that do not explicitly appear in the Hamiltonian since they serve to fix the general coordinate freedom and therefore are ``Higgsed away''. See \cite{GT12} for all the details.
\subsection{Classical Einstein-Yang-Mills-Theory}
\label{c2.1}
The Yang-Mills action for a unitary gauge group $G$ in general relativity is:

\begin{equation}
S_{YM} = -\frac{1}{4Q^2}\underset{M}{\int} d^4x \sqrt{\left| det(g) \right|} g^{\mu\nu} g^{\rho\sigma} \underline{F}^I_{\mu\rho}\underline{F}^I_{\nu\sigma}
\end{equation}\\
where $\underline{F}$ is the curvature of the $G$-connection $\underline{A}$ and $Q$ is the coupling constant and $g^{\mu\nu}$ is the metric on the manifold $M$. The aim of this chapter is to cast this action into canonical form.
This is done using the ADM-formalism, the details of which can be found in \cite{Wal84}. The idea is to assume that $M$ may be splitted as $M=\mathbb{R}\times S$. This foliation into space-like hypersurfaces allows the replacement of the ten components of the spacetime metric by the six components of the induced Riemann metric $q_{ab}$ of $S$ and the three components of the shift vector $N_a$ and the lapse function $N$. Also the co-triad field $e^i_a$ is transformed to the densitized triad

\begin{equation}\label{E-field}
E^a_i = \frac{1}{2}\epsilon^{abc}\epsilon_{ijk}e^j_be^k_c = \sqrt{det(q)} e^a_i
\end{equation}\\
which serves as the canonical pair on the gravitational phase space together with the extrinsic curvature:

\begin{equation}\label{K-field}
K_{ab}=sgn\left(det(e^j_c)\right)K^i_ae^i_b
\end{equation}\\
\ref{E-field} and \ref{K-field}, together with the connection $A^i_a = \Gamma^i_a+K^i_a$ form the Asthekar-Barbero-Variables \cite{Ash91,Bar94,Bar95,Bar96}, where $\Gamma^i_a$ is the spin-connection of $e^i_a$.

In conjunction with the canonical pair from Yang-Mills-theory $\left(\underline{A}^i_a,\frac{1}{Q^2}\underline{E}^a_i\right)$, where the first is the above mentioned $G$-connection and the second the associated electric field, one is set up to start working on $SU(2)\times G$. Due to the gauge fixing dynamically induced by additional matter fields, lapse and shift get frozen to $N=1, \; N^a=0$ respectively. After performing the Legendre transformation, one finds \cite{Thi07}

\begin{equation} \label{2.4}
S_{YM} = \frac{1}{Q^2}\underset{\mathbb{R}}{\int}dt\underset{S}{\int}d^3x\left(
\underline{\dot{A}}_a^I\underline{E}^a_I-\left(
-\underline{A}^I_t\underline{D}_a\underline{E}^a_I+N^a\underline{F}^I_{ab}\underline{E}^b_I
+\frac{q_{ab}}{2\sqrt{det(q)}}\left(\underline{E}^a_I\underline{E}^b_I+\underline{B}^a_I\underline{B}^b_I\right)
\right)\right)
\end{equation}\\
where $\underline{B}^a_I=\epsilon^{abc}\underline{F}^I_{bc}$ and $\underline{D}_a$ acts like the Levi-Civita connection on tensor indices. The contributions to the spatial diffeomorphism constraint and the Hamiltonian can be directly read off: the Hamiltonian is

\begin{equation} \label{2.5}
H_{YM} = \frac{q_{ab}}{2Q^2\sqrt{det(q)}}\left(\underline{E}^a_I\underline{E}^b_I+\underline{B}^a_I\underline{B}^b_I\right)
\end{equation}
\subsection{Quantum Einstein-Yang-Mills-Theory}
\label{c2.2}
In this chapter one will construct a Hamiltonian for a Quantum Einstein-Yang-Mills-Theory. As already stated, the methods of quantisation (\ref{2.5}) will be those of Loop Quantum Gravity. We will present the construction separately for the Einstein-Term and the Yang-Mills-Term. Finally we show how the classical Kogut-Susskind-Hamiltonian emerges from the theory in the limit of a flat spacetime.

Let us stress again, that we are working in the framework of deparametrised models: a suitable gauge fixing leads to a reduced phase spacetime that (when quantised via the methods of LQG) provides a model where all the constraints are solved, all operators are spacetime diffeomorphism invariant and physical states respectively. In this formulation there is no Hamiltonian constraint, but a Hamiltonian operator \cite{GT12,GT06_2,GT07_4,GHTW10}.

Also the idea of Algebraic Quantum Gravity is used, where we work solely on abstract graphs, which do not care about their embedding. Instead only the nodes and their connection among themselves are of interest. In our case the graphs are of cubic topology (i.e. a general vertex will have six edges adjacent to it) which is very alike to the situation in Lattice Gauge Theory. In this manner we follow the proposal of \cite{GT06_1}, meaning that physics now happens on such a given graph leaving it invariant, a feature in which AQG differs from the first route of LQG, where there is no Hamiltonian but an infinite number of constraints which must commute with each other on the kernel of the diffeomorphism constraint. The only known way to achieve this without anomalies in this sense is to let the Hamiltonian constraint act by adding new edges. By contrast, with only one Hamiltonian, there is no anomaly to worry about anymore and the quantization of the Hamiltonian can be done in the way that is customary in lattice gauge theory. With every edge $e$ one associates an element $A(e)$ of SU(2) for the gravitational sector and an element $\underline{A}(e)$ of the Yang-Mills gauge group $G$, as well as elements $E(e)$, $\underline{E}(e)$, respectively, for the corresponding Lie algebra. Hence in both cases there are the following algebraic relations, with $Q$ being the coupling constant and $f_{jkl}$ the structure constant of SU(2) or $G$ respectively:

\begin{equation} \label{2.6}
\left[A(e),A(e')\right]=0
\end{equation}
\begin{equation} \label{2.7}
\left[E_j(e),A(e')\right] = i\hbar Q^2\delta_{e,e'}\tau_j/2 A(e)
\end{equation}
\begin{equation} \label{2.8}
\left[E_j(e), E_k(e')\right] = i \hbar	Q^2\delta_{e,e'}f_{jkl}E_l(e')
\end{equation}\\
A nice representation of this algebra is the Infinite Tensor Product Hilbert space $\mathcal{H}=\underset{e}{\bigotimes} \mathcal{H}_e$,  where on every edge $\mathcal{H}_e=L^2\left(G,d\mu_H(G)\right)\otimes L^2\left(SU(2),d\mu_H(SU(2)\right)$ \cite{GT07_4}. Here $A(e)$ is a unitary matrix valued operator and $E(e)$ an essential self-adjoint derivation operator. So e.g. the action of $E(e)$ on a function $f_e$ on $e$ is:

\begin{equation} \label{2.9}
E_j(e)f_e (h)=i\hbar Q^2 \frac{d}{ds}\left(f_e\left(e^{s\tau_j/2}h\right)\right)_{s=0}
\end{equation}\\
where $\tau_j$ are the generators of the corresponding Lie algebra. This choice gives a parallel to the concept of LQG. And although there is no strict derivation of an algebraic Hamiltonian, it appears sensible to take the quantum version of the operators derived in the LQG framework and use them in AQG. The derivation of those in LQG was first performed in \cite{Thi96_1, Thi96_2} for the gravitational sector and in \cite{Thi98} for the Yang-Mills sector).

Considering all this, the gravitational Hamiltonian is set to:

\begin{equation}\label{2.10}
\hat{H}_{Einstein}(v)= \hat{S}_E^{(1/2)}(v)-2(1+\gamma^2)\hat{T}(v)
\end{equation}
\\
with

\begin{equation}
\hat{S}_E^{(r)}(v)=\frac{1}{N_v}\underset{e_1\cap e_2\cap e_3 =v}{\sum}\frac{\epsilon(e_1,e_2,e_3)}{\left|L(v,e_1,e_2\right|}
\underset{\beta\in L(v,e_1,e_2)}{\sum}tr\left(
\left(\hat{A}(\beta)-\hat{A}(\beta)^{-1}\right)A(e_3)\left[A(e_3)^{-1},\hat{V}_v^r\right]
\right)
\end{equation}

\begin{equation}
\hat{T}(v)=\frac{1}{N_v}\underset{e_1\cap e_2\cap e_3 =v}{\sum}\epsilon(e_1,e_2,e_3) tr\left(
\hat{A}(e_1)\left[\hat{A}(e_1)^{-1},\hat{K}\right]\hat{A}(e_2)\left[\hat{A}(e_2)^{-1},\hat{K}\right]\hat{A}(e_3)\left[\hat{A}(e_3)^{-1},\sqrt{\hat{V}}\right]
\right)
\end{equation}
\\
where $\hat{K}=\left[\hat{S}^{(1)}_E,\hat{V}\right]$ and $\hat{S}^{(1)}_E=\underset{v}{\sum}\hat{S}_E^{(1)}$, $N_v$ is the number of unordered triples of mutually distinct edges incident at $v$ and $L(v,e,e')$ the set of minimal loops. These are all loops, which start at $v$ along $e$ and end at $v$ along $(e')^{-1}$ and are minimal in the sense that there are no other loops with the same restrictions and fewer edges traversed. In our case, where one is restricted to the once and for all fixed cubic graph, the elementary loops are the plaquettes, consisting of four edges. $\hat{V}$ is the algebraic quantum Volume Operator:

\begin{equation}
\hat{V}=\underset{N\rightarrow\infty}{\textrm{lim}}\underset{I=1}{\overset{N}{\sum}}\sqrt{\left|\frac{1}{3!}\epsilon\left(a,b,c\right)\hat{E}_{i}\left(S_{I}^{a}\right)\hat{E_{j}}\left(S_{I}^{b}\right)\hat{E}_{k}\left(S_{I}^{c}\right)\epsilon^{ijk}\right|}
\end{equation}
\\
where the skew function $\epsilon$ is chosen such that it matches that of the embedding dependent Ashtekar-Lewandowski-Volume operator of LQG \cite{AL97} when the algebraic graph is embedded in a generic way (see \cite{GT06_1} for further details). One can show that its spectrum has to be discrete and further analysis has been performed in greater detail in \cite{Brunnemann:2004xi}. Consequently, the action of the Hamiltonian on an algebraic graph or others is quite involved and the solution of eigenstates cannot be computed analytically, however it is numerically \cite{BR06} and semiclassically \cite{GT06_3} under good control. Some calculations have been done for the LQG Hamiltonian-constraint, which maybe could transfer directly to the algebraic version. For further reading see e.g. \cite{ATZ11,ALZ13}.

For the Yang-Mills Hamiltonian one sets:

\begin{equation}
\label{AYMHam}
\hat{H}_{YM}(v)=\frac{1}{2Q^2}\left(\hat{H}_E(v)+\hat{H}_B(v)\right)
\end{equation}
\\
with

\begin{equation}
\hat{H}_E(v)=\frac{1}{P_v}\underset{e_1\cap e_2=v}{\sum}tr\left(\hat{A}(e_1)\left[\hat{A}(e_1)^{-1},\sqrt{\hat{V}}\right]\hat{A}(e_2)\left[\hat{A}(e_2)^{-1},\sqrt{\hat{V}}\right]\right)\underline{\hat{E}}_J(e_1)\underline{\hat{E}}_J(e_2)
\end{equation}

\[
\hat{H}_B(v)=\frac{1}{T_v^2}\underset{e_1\cap e_2\cap e_3=v}{\sum}\underset{e_4\cap e_5\cap e_6=v}{\sum}\frac{\epsilon(e_1,e_2,e_3)}{\left|L(v,e_2,e_3)\right|}\frac{\epsilon(e_4,e_5,e_6)}{\left|L(v,e_5,e_6)\right|}\underset{\beta\in L(v,e_2,e_3)}{\sum}\underset{\beta'\in L(v,e_5,e_6}{\sum}\times
\]
\begin{equation}
\times
tr\left(\hat{\tau}_j\hat{A}(e_1)\left[\hat{A}(e_1)^{-1},\sqrt{\hat{V}}\right]\right)
tr\left(\hat{\tau}_j\hat{A}(e_4)\left[\hat{A}(e_4)^{-1},\sqrt{\hat{V}}\right]\right)
tr\left(\underline{\hat{\tau}}_J\underline{\hat{A}}(\beta)\right)tr\left(\underline{\hat{\tau}}_J\underline{\hat{A}}(\beta')\right)
\end{equation}
\\
where $P_v$ is the number of all pairs of edges incident at $v$, $T_v$ is the number of all non-trivial triples of edges incident at $v$ and the $\epsilon$-term is that of the Volume-operator. Note that as in the Kogut-Susskind case, while the Hamiltonian expressed in terms of lattice variables has the correct continuum limit when the lattice embedding becomes sufficiently fine, it is but one of infinitely many possible discretisations that have this property. For instance one could consider discretisations that also have next to next neighbor interaction terms.

For the moment one should also notice that the gravitational Gauss constraint as well as the Yang-Mills Gauss constraint have their algebraic quantum versions as well. Going over to the invariant subspace where these Gauss constraints are solved, leads (as in LQG) to the fact that one needs to introduce intertwiners $\pi$ of both gauge groups respectively on every vertex. The obtained subspace $\mathcal{H}^\mathcal{G}_{kin}$ is commonly referred to in the literature as the space of spin-network functions

\begin{equation}
T_{\gamma,{j_e},{\pi_v}} \left[A,\underline{A}\right] = \underset{v\subset \gamma}{\bigotimes} \hspace{2mm}\underline{\pi}_v\otimes \pi_v \hspace{2mm}\underset{e\subset\gamma}{\bigotimes}\hspace{2mm} h^{j_e}(e)\otimes \underline{h}^{\underline{j}_e}(e)
\end{equation}\\
where $h^{j_e}(e)=h^{j_e}(e)\left(A_e\right)$ corresponds to the irreducible representation of label $j_e$ of the holonomy of $SU(2)$ and $\underline{h}^{\underline{j}_e}(e)$ respectively of the Yang-Mills gauge group $G$. For more information on these see section \ref{c3}.

To compute the spectrum of the Hamiltonian one would have to compute its matrix elements and their calculation shall be done in chapter \ref{c5}. In the following the gauge group for the Gravitational spin-networks is of course $SU(2)$ and for the Yang-Mills gauge group we pick the case of QCD, i.e. $SU(3)$.
\newline
\\
This section finishes with a last remark on the Kogut-Susskind-Hamiltonian. While there are a lot of ways to derive it from the Wilson action (see e.g. \cite{KS75,Cr83}, having this Yang-Mills-Hamiltonian of Quantum Gravity at hand gives an easy derivation of the Kogut-Susskind, which should be seen as the classical limit of the theory. Hence we will replace the general metric with the flat Euclidean one and only quantise the Yang-Mills-Field. After embedding the graph in Minkowski space with a sufficiently small lattice length $\epsilon$, one arrives, still with only nearest neighbor interactions  (as in the case of the Wilson action), indeed at a version of the Kogut-Susskind Hamiltonian:

\begin{equation}\label{2.18}
\hat{H}_{KS}=\frac{1}{2Q^2\epsilon}\left(\underset{e\in \gamma}{\sum}\hat{E}_J(e)\hat{E}_J(e)+\underset{\beta,\beta'\in\gamma}{\sum}
tr\left(\tau_j\hat{A}(\beta)\right)tr\left(\tau_j\hat{A}(\beta')\right)
\right)
\end{equation}
\\
This is not the form generally found in the literature (e.g. \cite{KS75}), because for the derivation of the LQG version of (\ref{AYMHam}) a different approximation scheme for the curvature of the $G$-connection $F_{ab}$ is used. The approximation used in \cite{Thi96_1,Thi96_2} is  $\text{Im}\left(A(\beta)\right) \approx \epsilon^2F_{ab}^j\tau_j+\mathcal{O}\left(\epsilon^4\right)$, while the other one - which is in case of a flat background metric equivalent - is $\text{Re}\left(A(\beta)\right) \approx d_n +\epsilon^4 F_{ab}^iF_i^{ab} + \mathcal{O}\left(\epsilon^6\right)$. Kogut and Susskind used the latter one, however in the case of a non-trivial background it is not applicable. In any case this second approximation leads to the addition of a constant, the dimension of the group matrices $d_n$, which is treated in LQCD as a simple energy shift. Going along this road one obtains:

\begin{equation}
\hat{H}_{KS,\text{lit}}=\frac{1}{2Q^2\epsilon}\left(\underset{e\in \gamma}{\sum}\hat{E}_J(e)\hat{E}_J(e)+\underset{\beta\in\gamma}{\sum}
tr\left(\hat{A}(\beta)\right)+tr\left(\hat{A}(\beta)^{\dagger} \right) - 2d_n
\right)
\end{equation}
\section{Representation Theory and Graphical Calculus of $SU(3)$}
\label{c3}
Loop Quantum Gravity and Lattice Gauge theory both very heavily depend on the representation theory of the corresponding gauge group. ($SU(2)$ for the gravitational sector and for the purpose of this article we restrict ourselve to the $SU(3)$ for the Yang-Mills field). Brink and Satchler have introduced a formalism called graphical calculus \cite{BS68} for $SU(2)$, which simplifies the manipulations one wants to perform on the coupled representations of the spin-network by suppressing many of indices from the irreducible representations and makes the coupling of different links more obvious. There has also been a proposal for a graphical calculus in \cite{Cvi08} for any Lie Group but this works only in its defining representation, while for our purpose we want to combine different irreducible representations. The methods we will use throughout this paper regarding the computations of the gravitational degrees of freedom have been introduced in \cite{ATZ11}. With this framework it has been accomplished to evaluate the matrix elements of the Euclidian Part of the Hamiltonian constraint from \cite{Thi96_1, Thi96_2} and the matrix elements of its Lorentzian Part in \cite{ALZ13}. The matrix elements for the Euclidian and Lorentzian part have been found analytically modulo the matrix elements of the volume operator, which must be determined non analytically. To make this paper self-contained we provide a list of the most important identities of this $SU(2)$-related calculus to the appendix.
In this chapter we aim at the construction of a similar calculus for the gauge group of $SU(3)$. For this purpose we revisit the representation theory of $SU(3)$ in the following section. The familiar reader may jump forward to \ref{c3.2}.
\subsection{Representation Theory of $SU(3)$}
\label{c3.1}
In this chapter, we recall some general properties of the finite dimensional representations of the unitary, compact and semi-simple Lie-Group $SU(3)$ and we will construct its Clebsch-Gordan-Coefficients.
We start by choosing a suitable basis for the Lie algebra $su(3)$ as in \cite{Gri84}. This Lie algebra has a real form and we may pick a basis $\left\{A_{i,k}\right\}$ (where $i,k=1,2,3$), with the following commutation relations:

\begin{equation}
\left[ A_{i,k},A_{j,l}\right] = \delta_{k,j} A_{i,l} - \delta_{i,l} A_{j,k}
\end{equation}
\\
These are subject to the restriction $A_{11}+A_{22}+A_{33}=0$ and $A_{i,k}^{+}=A_{k,i}$, where the adjoint is taken in the respective representation. We will now consider representations of these commutation- and $\ast$-relations considered as an abstract Lie algebra.
Out of this set one can construct two (so-called) weight operators:

\begin{equation}
H_{1}=A_{11}-A_{22}
\end{equation}

\begin{equation}
H_{2}=A_{22}-A_{33}
\end{equation}
\\
Now given a finite dimensional representation $\left(D,V\right)$ over the vectorspace $V$ of $su(3)$ or equivalently $SU(3)$ (since any representation of $SU(3)$ corresponds to a unique one of $su(3)$ and vice versa, due to $SU(3)$ being simply connected), one can simultaneously diagonalize $D(H_{1})$ and $D(H_{2})$ as $\left[H_{1},H_{2}\right]=0$. A pair $j=(a,b)\in\mathbb{C}$ is called a weight for $D$ if there exists a $v\text{\ensuremath{\neq}}0$ in $V$ such that

\begin{equation}
D(H_{1})v=av
\end{equation}

\begin{equation}
D(H_{2})v=bv
\end{equation}
\\
Additionally $j$ is called highest weight, if for all weights $j'$
of $D$ and $\mu,\nu\geq0$ holds

\begin{equation}
j-j'=\mu\alpha_{1}+\nu\alpha_{2}
\end{equation}
\\
where the $\alpha_{i}$ are roots (a non-zero pair $(\alpha_{i,1},\alpha_{i,2})\in\mathbb{C}^{2}$, such that $\left[H_{j},Z_i\right]=\alpha_{i,j}Z_i$ with a non-zero $Z_i\in SU(3)$). In the following the irreducible representation of highest weight $j$ is denoted by $D^{(j)}$.

According to the Theorem of the highest weight \cite{Hall03} the following is true for
an irreducible representation $D$ of $SU(3)$
\begin{enumerate}
\item $D$ is the direct sum of weight spaces
\item $D$ has a unique highest weight $j=\left(a,b\right)$ with $a,b\in\mathbb{N}^{+}$
\item $D$ and $D'$ are equivalent $\Leftrightarrow$ $j=j'$
\end{enumerate}
From this we may can also deduce the following: The dimension of the irreducible representation
with highest weight $j=(a,b)$ is 

\begin{equation}
d_{j}=\frac{1}{2}\cdot(a+1)(b+1)(a+b+2)
\end{equation}
\\
A proof for this formula can be found e.g. in \cite{Hum72}. 

We work with finite dimensional representations of $SU(3)$, which is thus completely reducible \cite{VK95}. Consequently, the Tensor product of these representations can be rewritten as the sum of irreducible representations:

\begin{equation} \label{3.8}
D^{(j_{1})}\otimes D^{(j_{2})}=\underset{j}{\sum}\mu_{j}D^{(j)}
\end{equation}
\\
Let the vector-spaces on which these act be called $V_{j}$ and choose orthonormal bases in these spaces. Then a basis for $V_{j_{1}}\otimes V_{j_{2}}$ is

\[
\left\{ e_{m_{1}}^{j_{1}}\otimes e_{m_{2}}^{j_{2}}\right\} 
\]
\\
and equivalently for $V_{j}$ $\left\{ e_{m}^{j,s}\right\} $, where $j$ labels the weight and $s=1,...,\mu_{j}$ is used to distinguish the multiplicities. These bases can be connected by a unitary matrix:

\begin{equation} \label{3.9}
e_{m}^{j,s}=\underset{m_{1},m_{2}}{\sum}\left\langle e_{m_{1}}^{j_{1}},e_{m_{2}}^{j_{2}}\mid e_{m}^{j,s}\right\rangle e_{m_{1}}^{j_{1}}\otimes e_{m_{2}}^{j_{2}}
\end{equation}
\\
where the entries of the matrix are called the Clebsch-Gordan-Coefficients of the Tensor product. As they are elements of a unitary matrix, the following orthogonality relations hold:

\begin{equation} \label{3.10-bubble}
\underset{m_{1},m_{2}}{\sum}\bigl\langle e_{m}^{j,s}\mid e_{m_{1}}^{j_{1}},e_{m_{2}}^{j_{2}}\bigr\rangle\bigl\langle e_{m_{1}}^{j_{1}},e_{m_{2}}^{j_{2}}\mid e_{m'}^{j',s'}\bigr\rangle =\delta_{j,j'}\delta_{s,s'}\delta_{m,m'}
\end{equation}

\begin{equation} \label{3.11-2-1}
\underset{j,s,m}{\sum}\left\langle e_{m_{1}}^{j_{1}},e_{m_{2}}^{j_{2}}\mid e_{m}^{j,s}\right\rangle \bigl\langle e_{m}^{j,s}\mid e_{m_{1}'}^{j_{1}},e_{m_{2}'}^{j_{2}}\bigr\rangle =\delta_{m_{1},m_{1}'}\delta_{m_{2},m_{2}'}
\end{equation}
\\
To construct these Clebsch-Gordan-Coefficients explicitly, we follow the formalism developed by Pluha$\check{\mathrm{r}}$ et al. in \cite{PST86, PWH86}. It is useful to introduce additional linear combinations of the $A_{i,j}$. In addition to $H_1$ and $H_2$ one introduces the following operators: The two Casimir operators

\begin{equation}
F_2 = \frac{3}{2}\underset{i,k}{\sum}A_{i,j}A_{j,i}
\end{equation}

\begin{equation}
F_3= 9 \underset{i,j,k}{\sum}A_{i,j}A_{j,k}
\end{equation}
\\
which, in the $D^{(j)}$-representation, have the eigenvalues

\begin{equation}
f_{2}=\left(a+b+3\right)\left(a+b\right)-ab
\end{equation}

\begin{equation}
f_{3}=\left(a-b\right)\left(2a+b+3\right)\left(a+2b+3\right)
\end{equation}
\\
Also let us look at two sub-algebras, one isomorphic to $su(2)$:

\begin{equation}
I_{z}=\frac{1}{2}\left(A_{11}-A_{22}\right)\text{, {\color{white}.} }I_{+}=A_{12}\text{ and }I_{-}=A_{21}
\end{equation}
\\
There exist two eigenvalues for the group $SU(2)$, which we call isospin $i$ (from the total angular momentum operator $I^2$) and isospin projection $i_z$ (from the operator $I_z$).
Also there is a different sub-algebra isomorphic to $su(2)$:

\begin{equation}
\Lambda_{z}=A_{11}-A_{33}\text{, {\color{white}.} }\Lambda_{+}=\sqrt{2}\left(A_{12}-A_{23}\right)\text{ and }\Lambda_{-}=\sqrt{2}\left(A_{21}+A_{32}\right)
\end{equation}
\\
the eigenvalues of which are labeled $\lambda_0, \lambda_{0,z}$.

Both sub-algebras contain a linear combination of the weight operators. Thus their quantum numbers $i,\lambda_0$ can at most be $i_{0}=\frac{1}{2}a$ and $\lambda_{0}=a+b$, respectively \cite{PST86}. The eighth independent operator shall be:

\begin{equation}
Y=\frac{1}{3}\left(A_{11}+A_{22}-2A_{33}\right)
\end{equation}
\\
called the Hypercharge-Operator, whose Eigenvalues $y$ can be maximally $y_{0}=\frac{1}{3}\left(a+2b\right)$. This operator comes from particle physics where it unifies isospin and flavor into a single charge. $Y$ is just a linear combination of the $I_z$ and $\Lambda_z$ and thus the group, spanned from the latter operators, is, in principle, redundant. Hypercharge and isospin projection are weight components for SU(3).

Now one has to find how many quantum numbers are needed in general to describe a state in the vectorspace V of a irreducible highest weight representation $D^{(j)}$. With $su(n)$ being a complex, semisimple Lie algebra one can do a splitting in the cartan sub-algebra $\mathfrak{h}$, which is the maximal sub-lie-algebra of all abelian sub-algebras, consisting of semisimple elements. Thus

\begin{equation}
su(n)=\mathfrak{h}\oplus \mathfrak{g}_+ \oplus \mathfrak{g}_-
\end{equation}
\\
where $\mathfrak{g}_\pm$ are the sub-algebras to the corresponding to positive/negative roots with respect to a choice of simple positive roots. While $\mathfrak{h}$ has dimension $n-1$, $\mathfrak{g}_\pm$ have dimension $\frac{n(n-1)}{2}$. Every irreducible highest-weight representation is cyclic, i.e. there exists a non-trivial vector $v\in V$, which is a weight vector for $j$, with $D(\mathfrak{g}_+)v=0$ and the smallest subspace containing $v$ is all of $V$. The cyclic highest-weight representation depends on $r$ quantum numbers, where $r$ is the rank of the Lie algebra. These quantum numbers correspond to the highest weight vector eigenvalues of the Cartan sub-algebra generators. Moreover the ``occupation numbers'' are given by the generators of $\mathfrak{g}_-$, which are thus $\frac{n(n-1)}{2}$ many.

So now for $n=3$ one may see, that an additional quantum number next to the two weights $i_z$ and $y$ from the Cartan generators $I_z$ and $Y$ is needed. As the Casimir of the $su(2)$-subgroup $I^2$ commutates with both, it is convenient to use it.

Moreover, for a general rank $r$ semisimple Lie algebra the highest weight labels (here $a,b$) are in one-to-one correspondence with the eigenvalues of the $r$ algebraically independent Casimirs of rank $2,..,r+1$ (here $F_2,F_3$), hence $F_2,F_3,I_z,Y,I^2$ provides a maximally commuting set of self-adjoint operators characterising the irreducible representation completely.

Now one labels the basis states of $D^{(j)}$ with hypercharge $y$, isospin $i$ and isospin projection $i_{z}$ as $\left|\left(a,b\right),\left(y,i,i_{z}\right)\right\rangle \equiv\left|j,m\right\rangle $. To reduce the product $D^{(j_{1})}\otimes D^{(j_{2})}$ one has to deal with the multiplicity factors.  These contribute non-trivially here (in contrast to $SU(2)$), as can be seen very easily by looking at the corresponding sets of commutating operators. While there should be 10 commutating operators in the representation of $D^{(j_1)}\otimes D^{(j_2)}$, namely $\left(F_2,F_3,I_z,Y,I^2\right)^{(1)},\left(F_2,F_3,I_z,Y,I^2\right)^{(2)}$, after looking at the decomposition into irreducible representations there seem to be only 9 commutating ones: $\left(F_2,F_3,I_z,Y,I^2,F_2^{(1)},F_3^{(1)},F_2^{(2)},F_3^{(2)}\right)$. This strange occurrence is solved by introducing an additional operator $S$, which is a Casimir operator for the Lie algebra generated by $D^{(j_1)}(X)\otimes 1_{D^{(j_2)}}+ 1_{D^{(j_1)}}\otimes D^{(j_2)}(X),X\in su(3)$, and the s-classified reduced states, which are solutions to the eigenvalue problem

\begin{equation} \label{3.20}
S\left(\left\{ A\right\} _{1},\left\{ A\right\} _{2}\right)\left|(j_1,j_2),j,m,s\right\rangle =s\left|(j_1,j_2),j,m,s\right\rangle 
\end{equation}
\\
where we define

\begin{equation}
S\left(\left\{ A\right\} _{1},\left\{ A\right\} _{2}\right)=27\underset{i,j,k}{\sum}\left(A_{i,j;1}A_{j,k;2}A_{k,i;2}-A_{i,j;2}A_{j,k;1}A_{k,i;1}\right)-2F_{3;2}+2F_{3;1}
\end{equation}
\\
This operator is seen to fulfill some symmetry relations when acting on $D_{j_1}\otimes D_{j_2}\otimes D_{j_3}$

\begin{equation}
S\left(\left\{ A\right\} _{1},\left\{ A\right\} _{2}\right)=-S\left(\left\{ A\right\} _{2},\left\{ A\right\} _{1}\right)=-S\left(\left\{ A\right\} _{1},\left\{ A\right\} _{3}\right)=-S\left(\bar{\left\{ A\right\} }_{1},\bar{\left\{ A\right\} }_{2}\right)
\end{equation}
\\
where $D^{(j_{3})}$ stands for the coupled representation and the $\bar{A}_{ij}:=-A_{ij}$ define the generators of the conjugate (i.e. contragredient) representation. Finally these states have a phase ambiguity which can be resolved by setting:

\begin{equation}
\left\langle j_{1},j_{2}\lambda_{0;2},\lambda_{0,z;2}\mid j_{1},j_{2},j_{3},s\right\rangle >0
\end{equation}
\\
It should  be noted, however, that the $s$ are in general neither integral nor rational. Pluha$\mathrm{\check{r}}$ et al. \cite{PST86} have proposed a computational algorithm, where for a given set of highest weights the matrix $S\left(\left\{ A\right\} _{1},\left\{ A\right\} _{2}\right)$ is finite dimensional. With the last two equations it can be shown, that the Clebsch-Gordan-Coefficients $\langle j_{1},m_{1},j_{2},m_{2}\mid (j_1,j_2),j_{3},m_{3},s\rangle$, which couple the two representations $j_1,j_2$ to the resulting third $j_3$, while $m_1+m_2=m_3$, fulfil the following symmetry relations \cite{PST86}:

\begin{equation}\label{3.24}
\begin{aligned}
\langle j_{1},m_{1},j_{2},m_{2}\mid (j_1,j_2),\bar{j}_{3},\bar{m}_{3},s\rangle & =\langle j_{2},m_{2},j_{1},m_{1}\mid (j_1,j_2),\bar{j}_{3},\bar{m}_{3},\bar{s}\rangle\left(-\right)^{j_{1}+j_{2}+j_{3}}\tabularnewline
 & =\langle (j_{1},m_{1},j_{3},m_{3}\mid (j_1,j_2),\bar{j}_{2},\bar{m}_{2},\bar{s}\rangle\left(-\right)^{j_{1}+m_{1}}\sqrt{d_{j_2}/\d_{j_3}} \tabularnewline
 & =\langle\bar{j}_{1},\bar{m}_{1},\bar{j}_{2},\bar{m}_{2}\mid (j_1,j_2),j_{3},m_{3},\bar{s}\rangle\left(-\right)^{j_{1}+j_{2}+j_{3}}
 \tabularnewline
\end{aligned}
\end{equation}
\\
with $d_j=dim((a,b))$ the dimension of the space on which the irreducible representation corresponding to highest weight $(a,b)$ lives.
Also the following abbreviations have been introduced:

\[
\bar{j}=\left(b,a\right)\text{, {\color{white}.} }\bar{m}=\left(-y,i,-i_{z}\right)\text{ and }\bar{s}=-s
\]
\begin{equation}
\left(-\right)^{j}=\left(-1\right)^{a+b}\text{ and }\left(-\right)^{m}=\left(-1\right)^{\frac{3}{2}y+i_{z}}
\end{equation}
\subsection{Graphical Calculus of $SU(3)$}
\label{c3.2}
We will now develop a method to simplify computations involving the gauge group $SU(3)$. To the best of our knowledge, the graphical calculus developed here for $SU(3)$, while building on the one developed for $SU(2)$, is novel. We start by defining the so called s-classified 3j-Wigner-Symbol, an object, which represents the symmetry relations of the Clebsch-Gordan-Coeffecients in an easy way: \cite{PWH86}

\begin{equation}
\left(\begin{array}{cccc}
j_{1} & j_{2} & j_{3} & s\\
m_{1} & m_{2} & m_{3}
\end{array}\right)=\langle j_{1},m_{1},j_{2},m_{2}\mid (j_1,j_2),\bar{j}_{3},\bar{m}_{3},s\rangle\frac{\left(-\right)^{\bar{j}_{3}+\bar{m}_{3}}}{\sqrt{d_{\bar{j}_3}}}
\end{equation}
\\
The symmetry relations from the last chapter (\ref{3.24}) become:

\begin{equation}
\begin{aligned}
\left(\begin{array}{cccc}
j_{1} & j_{2} & j_{3} & s\\
m_{1} & m_{2} & m_{3}
\end{array}\right) & =\left(\begin{array}{cccc}
j_{2} & j_{\text{1}} & j_{3} & s\\
m_{2} & m_{1} & m_{3}
\end{array}\right)\left(-\right)^{j_{1}+j_{2}+j_{3}}\tabularnewline
 & =\left(\begin{array}{cccc}
j_{1} & j_{3} & j_{2} & s\\
m_{1} & m_{3} & m_{2}
\end{array}\right)\left(-\right)^{j_{1}+j_{2}+j_{3}}\tabularnewline
 & =\left(\begin{array}{cccc}
\bar{j}_{1} & \bar{j}_{2} & \bar{j}_{3} & \bar{s}\\
\bar{m}_{1} & \bar{m}_{2} & \bar{m}_{3}
\end{array}\right)\left(-\right)^{j_{1}+j_{2}+j_{3}}\tabularnewline
\end{aligned}
\end{equation}
\\
From this,it is apparent, that the s-classified 3j-symbols are invariant under even permutations and pick up a sign of $(-)^{j_1+j_2+j_3}$ for odd permutations. The usefulness of this Symbol lies in the fact, that any coupling of N representations can be expressed via 3j-symbols.  The aim now is to construct a graphical representation that allows one to represent multiple 3j-symbols and their distinct coupling (e.g. the s-classified 6j-symbols). We choose our notation such that it closely resembles the established calculus of \cite{BS68}. The graphical representation of the s-classified Wigner 3j-Symbol is a node, where the three representations are joined in, which are represented as lines

\begin{equation}
\left(\begin{array}{cccc}
j_{1} & j_{2} & j_{3} & s\\
m_{1} & m_{2} & m_{3}
\end{array}\right)=\begin{array}{c}\includegraphics[scale=0.8]{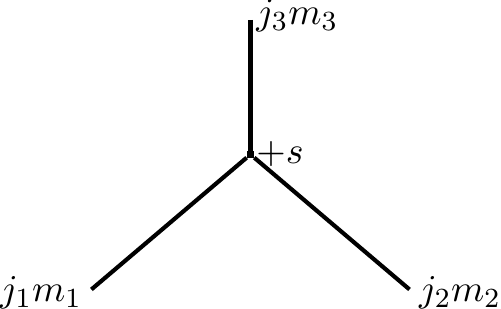}\end{array}
=\begin{array}{c}\includegraphics[scale=0.8]{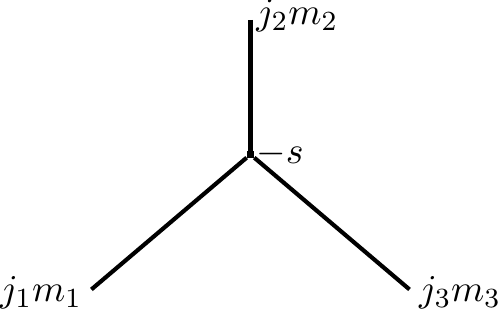}\end{array}
\end{equation}
\\
here the $+$ sign means that the elements of the 3j are ordered in an anti-clockwise orientation. Equivalently a $-$ sign indicates a clockwise orientation. E.g. a symmetry relation for the 3j is:

\begin{equation}
\begin{array}{c}\includegraphics[scale=0.8]{thesis-2_5pic01-node}\end{array}=\left(-\right)^{j_{1}+j_{2}+j_{3}}
\begin{array}{c}\includegraphics[scale=0.8]{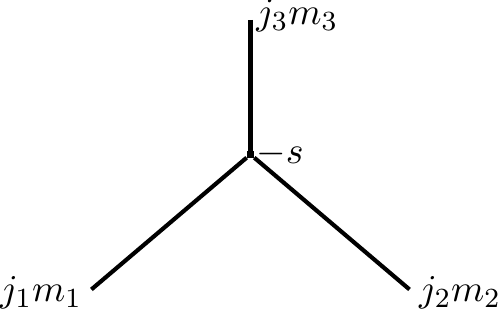}\end{array}
\end{equation}
\\
Additionally arrows will be introduced on the lines to indicate the ``metric tensor''. A line with no arrows means

\[
\begin{array}{c}\includegraphics{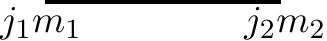}\end{array}
=\delta_{j_{1},j_{2}}\delta_{m_{1},m_{2}}
\]
\\
while a line with an arrow denotes the 1j-symbol:

\begin{equation}
\begin{array}{c}\includegraphics{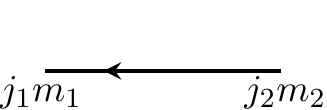}\end{array}
=\delta_{j_{1},\bar{j_{2}}}\left(\begin{array}{c}
j_{1}\\
m_{1},m_{2}
\end{array}\right)=\delta_{\bar{j}_{1},j_{2}}\delta_{\bar{m}_1,m_{2}}\left(-\right)^{j_{1}+m_{1}}
\end{equation}
\\
In the following we suppress the magnetic quantum numbers in the pictures. 
Having multiple arrows on one line, one can realize that (as well as for other orientations of the two arrows)

\begin{equation}
\begin{array}{c}\includegraphics{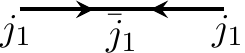}\end{array}
=
\begin{array}{c}\includegraphics{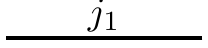}\end{array}
\end{equation}
\\
Given all of this we may calculate further: A contraction of 1j and
3j is:

\[
\begin{array}{c}\includegraphics[scale=0.8]{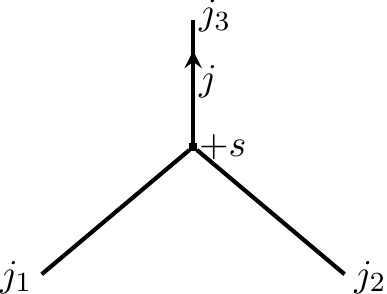}\end{array}
=\underset{m}{\sum}\left(\begin{array}{cccc}
j_{1} & j_{2} & j & s\\
m_{1} & m_{2} & m
\end{array}\right)\left(\begin{array}{c}
j_{3}\\
m_{3},m
\end{array}\right)\delta_{j_{3},\bar{j}}
\]
\[
=\underset{m}{\sum}\left(\begin{array}{cccc}
j_{1} & j_{2} & j & s\\
m_{1} & m_{2} & m
\end{array}\right)\delta_{m,\bar{m}_{3}}\delta_{j_{3},\bar{j}}\left(-\right)^{j_{3}+m_{3}}
\]
\begin{equation}
=\left(\begin{array}{cccc}
j_{1} & j_{2} & \bar{j}_{3} & s\\
m_{1} & m_{2} & \bar{m_3}
\end{array}\right)\left(-\right)^{j_{3}+m_{3}}\delta_{j_{3},\bar{j}}
\end{equation}
\\
Similarly we can write:

\[
\begin{array}{c}\includegraphics[scale=0.8]{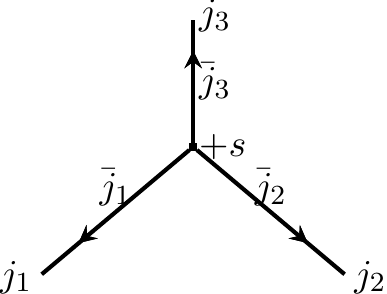}\end{array}
=\underset{m_{1}',m_{2}',m_{3}'}{\sum}\left(\begin{array}{cccc}
\bar{j}_{1} & \bar{j}_{2} & \bar{j}_{3} & s\\
m_{1}' & m_{2}' & m_{3}'
\end{array}\right)\left(\begin{array}{c}
j_{1}\\
m_{1},m_{1}'
\end{array}\right)\left(\begin{array}{c}
j_{2}\\
m_{2},m_{2}'
\end{array}\right)\left(\begin{array}{c}
j_{3}\\
m_{3},m_{3}'
\end{array}\right)
\]
\[
=\underset{m_{1}',m_{2}',m_{3}'}{\sum}\left(\begin{array}{cccc}
\bar{j}_{1} & \bar{j}_{2} & \bar{j}_{3} & s\\
m_{1}' & m_{2}' & m_{3}'
\end{array}\right)\delta_{m_{1}',\bar{m}_{1}}\delta_{m_{2}',\bar{m}_{2}}\delta_{m_{3}',\bar{m}_{3}}\left(-\right)^{\underset{i}{\sum}j_{i}+m_{i}}
\]
\[
=\left(\begin{array}{cccc}
\bar{j}_{1} & \bar{j}_{2} & \bar{j}_{3} & s\\
\bar{m}_{1} & \bar{m}_{2} & \bar{m}_{3}
\end{array}\right)\left(-\right)^{j_{1}+j_{2}+j_{3}}\left(-\right)^{m_{1}+m_{2}+m_{3}}
\]
\begin{equation}
=\left(\begin{array}{cccc}
j_{1} & j_{2} & j_{3} & \bar{s}\\
m_{1} & m_{2} & m_{3}
\end{array}\right)=
\begin{array}{c}\includegraphics[scale=0.8]{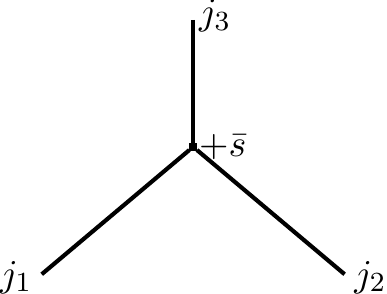}\end{array}
\end{equation}
\\
where we have used, that $(-)^{m_1+m_2+m_3}=0$.
In the following one uses the abbreviation:

\begin{equation}
\begin{array}{c}\includegraphics{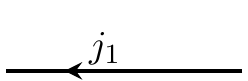}\end{array}
=
\begin{array}{c}\includegraphics{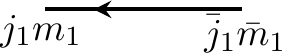}\end{array}
\end{equation}
\\
and thus only writes one index to each line from now on. For lines without arrow it  indicates the highest weights of its irreducible representation, and if the line has an arrow it indicates the highest weight of the representation where the arrow points towards.\\
Also the arrows can be changed by dualising the j.

\begin{equation}
\begin{array}{c}\includegraphics{3-34}\end{array}
=
\begin{array}{c}\includegraphics{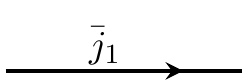}\end{array}
\end{equation}
\\
In order to represent more complex structures, lines can be joined as long as they carry the same highest weight. Note that the lines also carry a distinct group element.  Joining them means that the magnetic quantum numbers are set to equal and summed over. In the following these numbers are omitted in the graphs as already stated. With this definition one is, for example, able to represent the s-classified 6j-symbol, an object defined in the following way (similar to \cite{PWH86}):

\[
\left\{ \begin{array}{cccc}
j_{1} & j_{2} & j_{3}\\
j_{4} & j_{5} & j_{6}\\
s_{1} & s_{2} & s_{3} & s_{4}
\end{array}\right\}
 =\underset{\left\{ m\right\} }{\sum}\left(-\right)^{\underset{i}{\sum}j_{i}+m_{i}} 
 \left(\begin{array}
 {cccc}
j_{1} & j_{2} & j_{3} & s_{1}\\
m_{1} & m_{2} & m_{3}
\end{array}\right)
\left(\begin{array}{cccc}
\bar{j}_{1} & j_{5} & \bar{j}_{6} & s_{2}\\
\bar{m}_{1} & m_{5} & \bar{m}_{6}
\end{array}\right)
\cdot
\]
\begin{equation}
\cdot
\left(\begin{array}{cccc}
\bar{j}_{4} & \bar{j}_{2} & j_{6} & s_{3}\\
\bar{m}_{4} & \bar{m}_{2} & m_{6}
\end{array}\right)
\left(\begin{array}{cccc}
j_{4} & \bar{j}_{5} & \bar{j}_{3} & s_{4}\\
m_{4} & \bar{m}_{5} & \bar{m}_{3}
\end{array}\right)=\begin{array}{c}\includegraphics{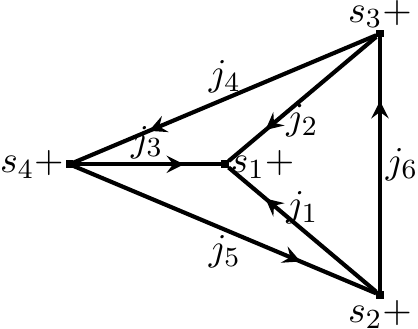}\end{array}
\end{equation}
\\
This object has a lot of symmetries at hand, so e.g., it holds

\[
\left\{ \begin{array}{cccc}
j_{1} & j_{2} & j_{3}\\
j_{4} & j_{5} & j_{6}\\
s_{1} & s_{2} & s_{3} & s_{4}
\end{array}\right\}=
\left\{ \begin{array}{cccc}
\bar{j}_{2} & \bar{j}_{1} & \bar{j}_{3}\\
j_{5} & j_{4} & j_{6}\\
s_{1} & s_{3} & s_{2} & s_{4}
\end{array}\right\}=
\left\{ \begin{array}{cccc}
\bar{j}_{1} & \bar{j}_{3} & \bar{j}_{2}\\
j_{4} & j_{6} & j_{5}\\
s_{1} & s_{2} & s_{4} & s_{3}
\end{array}\right\}=
\]

\begin{equation}
=\left\{ \begin{array}{cccc}
j_{4} & \bar{j}_{5} & \bar{j}_{3}\\
j_{1} &\bar{j}_{2} & \bar{j}_{6}\\
s_{4} & s_{3} & s_{2} & s_{1}
\end{array}\right\}=
\left\{ \begin{array}{cccc}
\bar{j}_{1} & \bar{j}_{2} & \bar{j}_{3}\\
\bar{j}_{4} & \bar{j}_{5} & \bar{j}_{6}\\
\bar{s}_{1} & \bar{s}_{2} & \bar{s}_{3} & \bar{s}_{4}
\end{array}\right\}
\end{equation}
\\
Also, for such a closed diagram (meaning that no open links remain) the object infers the invariance of the change of + $\leftrightarrow$ -, since every link obviously meets exactly two nodes, and $(-)^{2j}=1$, because - recalling the theorem of the highest weight - $j=(a,b)$ with $a,b\in\mathbb{N}$.

Important relations in the theory of group representations are the two orthogonality relations (\ref{3.10-bubble}) and (\ref{3.11-2-1}) Their form follows from the very definition of the  3j-symbols and the fact that they are real:

\[
\underset{m_1,m_2}{\sum} \left( \begin{array}{cccc}
j_{1} & j_2 & j_3 & s\\
m_1 & m_2 & m_3
\end{array}\right)
 \left( \begin{array}{cccc}
j_{1} & j_2 & j'_3 & s'\\
m_1 & m_2 & m'_3
\end{array}\right)
=\frac{1}{d_{\bar{j_3}}} \delta_{j_3,j'_3}\delta_{m_3,m'_3}\delta_{s,s'}
\]
\[
\underset{j_3,m_3,s}{\sum} \left( \begin{array}{cccc}
j_{1} & j_2 & j_3 & s\\
m_1 & m_2 & m_3
\end{array}\right)
\left( \begin{array}{cccc}
j_{1} & j_2 & j_3 & s\\
m_1' & m_2' & m_3
\end{array}\right)
=\frac{1}{d_{\bar{j}_3}} \delta_{m_1,m_1'}\delta_{m_2,m_2'}
\]
\\
Graphically these orthogonality relations can  be encoded as:

\begin{equation} \label{3.38-bubble}
\begin{array}{c}\includegraphics[scale=0.8]{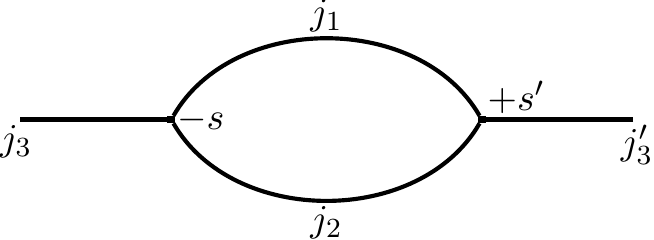}\end{array}
=
\frac{\delta_{s,s'}}{d_{j_3}}
\begin{array}{c}\includegraphics[scale=0.8]{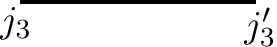}\end{array}
\end{equation}

\begin{equation} \label{3.39-2-1}
\underset{s,j_3}{\sum} d_{j_3}
\begin{array}{c}\includegraphics[scale=0.8]{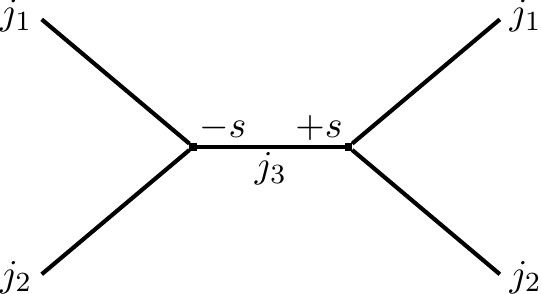}\end{array}
=\begin{array}{c}\includegraphics[scale=0.8]{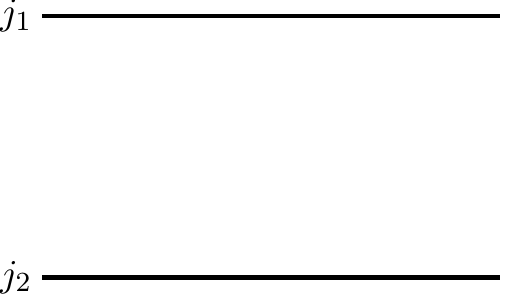}\end{array}
\end{equation}
\\
It should be noted at this point that the sum over $s$ goes over all the solutions from (\ref{3.20}) and is highly dependent on the coupled weights $j_1$, $j_2$ and $j_3$. While $j_3$ itself has to be chosen such that the three representations together form a triad (as for $SU(2)$) \cite{DS64,, VK91, VK95}, i.e. if $j_3$ is inside the set $\Pi_{j_1}+j_2$, with $\Pi_{j_1}$ denoting the set of all weights of the corresponding representation with heighest weight $j_1$.

One can immediately see that the expression of the second orthogonality with arrows on the links is stated as: 

\[
\underset{\bar{j}_3s}{\sum} d_{j_3}
\begin{array}{c}\includegraphics[scale=0.8]{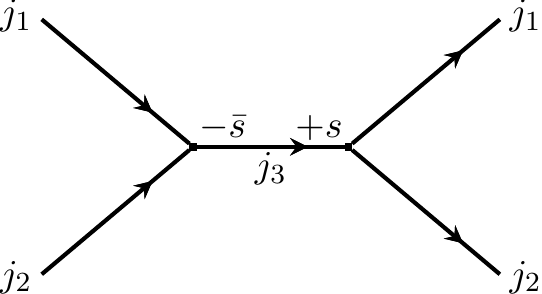}\end{array}
=
\begin{array}{c}\includegraphics[scale=0.8]{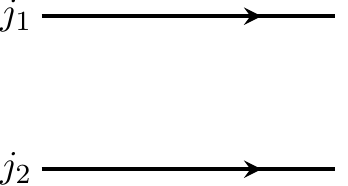}\end{array}
\]
\\
It is now obvious that transforming the algebraic expression of a graph alters its distinct representation, such that there also must exist some rules for transforming the graphs directly. We have already seen that e.g. the arrows can be changed in their direction, by going from weight $j=(a,b)$ to $\bar{j}=(b,a)$. Also: a line with two arrows is equivalent to a line with no arrows. Furthermore at a node one can add and remove arrows of the same direction on each line at the same time, while only changing the node internal index $s\rightarrow\bar{s}$.

Since one has for any general Lie group \cite{VK95}, that

\[
\underset{m_1'm_2'}{\sum}\langle e^{j_3,s}_{m_3} \mid e^{j_1}_{m_1} e^{j_2}_{m_2}\rangle D^{(j_1)}_{m_1'm_1}(g)D^{(j_2)}_{m_2'm_2}(g)=
\underset{m_3'}{\sum}\langle e^{j_3,s}_{m_3'} \mid e^{j_1}_{m_1}e^{j_2}_{m_2} \rangle D^{(j_3)}_{m_3'm_3}(g)
\]
\\
this translates as a transformation rule for our graphical calculus:

\begin{equation}\label{3.40}
\begin{array}{c}\includegraphics{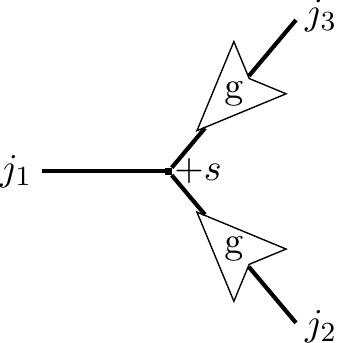}\end{array} =
\begin{array}{c}\includegraphics{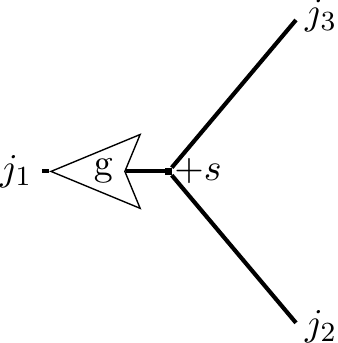}\end{array}
\end{equation}

We now look at further rules, which change the lines and their coupling itself. For this purpose we define objects equivalent to the $SU(2)$ $jm$-coefficients from \cite{YLV62}, which are blocks of connected nodes with an arrow on each line, whose explicit internal structure is of no importance. They have $n$ external lines with label $j_1 ... j_n$. Their graphical representation is:

\begin{equation}
F_n\left( \begin{array}{ccc}
j_1 & ... & j_n \\
m_1 & ... & m_n
\end{array}
\right)
=
\begin{array}{c}\includegraphics{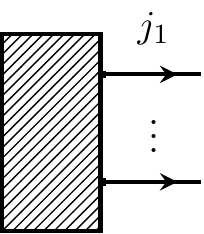}\end{array}
\end{equation}
\\
Using the orthogonality relations from above, a lot of manipulation on these external lines can be done. First one has to notice that a block with only one external line, i.e. $F_1\left(\begin{array}{c}
j\\
m
\end{array}\right)$, is equivalent to a scalar times a Clebsch-Gordan-Coefficient with two labels equal to zero and hence zero itself, if not $j=m=0$:

\begin{equation}
F_1\left(\begin{array}{c}
j\\
m
\end{array}\right)=F_1\left(\begin{array}{c}
0\\
0
\end{array}\right)\delta_{j,0}\delta_{m,0}=\left(\begin{array}{ccc}
0 & 0 & j\\
0 & 0 & m
\end{array}\right) \mathrm{const.}
\end{equation}
\\
This and the second orthogonality relation (\ref{3.39-2-1}) on an $F_2$ coefficient leads to:

\begin{equation}\label{3.43}
\begin{array}{c}\includegraphics{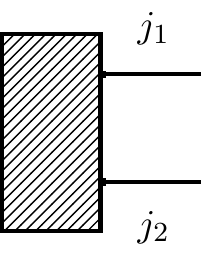}\end{array}
=
\underset{j,s}{\sum}
\begin{array}{c}\includegraphics{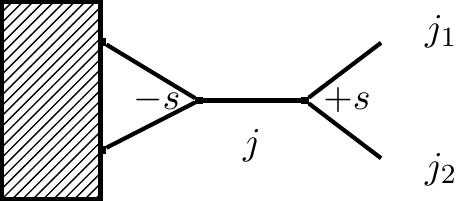}\end{array}
=
\delta_{j_1,j_2}
\begin{array}{c}\includegraphics{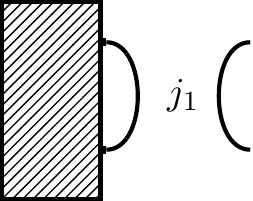}\end{array}
\end{equation}
\\
since the one connection link vanishes and the node reduces to a 1j-symbol and thus the sum over $s$ reduces to a $\delta_{s,2f_3(j_1)}$. With a similar calculation and using  (\ref{3.43}) we arrive at:

\begin{equation}
\begin{array}{c}\includegraphics{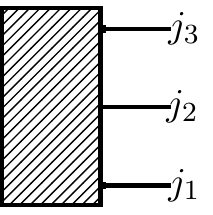}\end{array}
=\underset{s}{\sum}
\begin{array}{c}\includegraphics{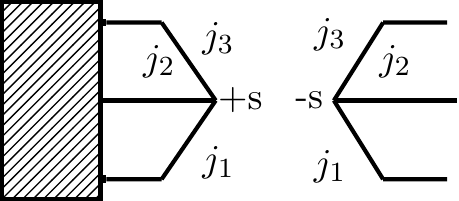}\end{array}
\end{equation}
\\
With this at hand, all the  tools of a graphical calculus necessary to simplify calculations involving the gauge group $SU(3)$ are provided. Before we dive into the computations of the matrix elements of the Quantum Yang-Mills Hamiltonian, we provide a final example: The following structure will be encountered numerous times in the remainder of this article:

\[
\begin{array}{c}\includegraphics{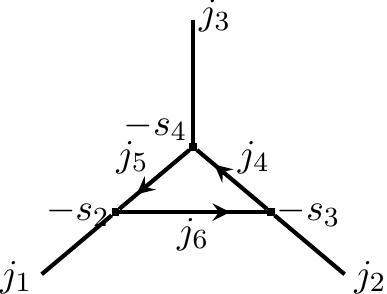}\end{array}
=\underset{s}{\sum}
\begin{array}{c}\includegraphics{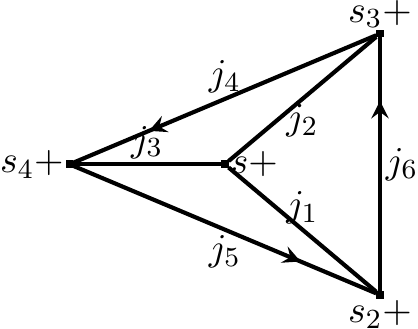}\end{array} \cdot 
\begin{array}{c}\includegraphics{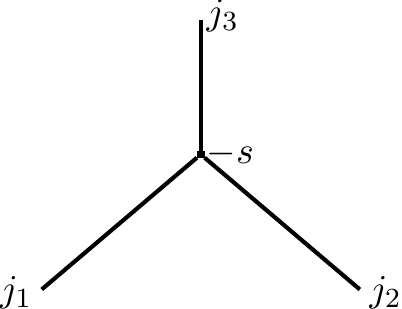}\end{array}
\]
\begin{equation}
=\underset{s}{\sum} \left\{\begin{array}{cccc}
j_1 & j_2 & j_3 \\
j_4 & j_5 & j_6 \\
s & s_2 & s_3 & s_4
\end{array}\right\} 
\begin{array}{c}\includegraphics{3-45-3}\end{array}
\end{equation}
\newpage
\section{Einstein-Yang-Mills-Theory in the Kogut-Susskind-Case}
\label{c4}
In this chapter we  present the results, when applying the developed methods in case of the background spectrum of the Kogut-Susskind-Hamiltonian in flat space. In this work we will not focus on any analytical solvable problem, e.g. the one-plaquette-graph, whose eigenstates are given in terms of Mathieu-functions \cite{MS54} in case of $U(1)$ or $SU(2)$ Gauge Theory\cite{RW80,RW81,RW82}. Instead we  concentrate on the physically interesting case of multiple-plaquette problems, which so far could be tackled using numerical investigations. A lot of work has been done on this, see e.g. \cite{BPMUV99, BPMUV00, Mat08, KHLK09, CDR88, DR85} and many more. The most promising approach up to today is still to calculate the matrix elements and continue afterwards with numerical simulations. For this reason this chapter will present the exact calculation of said matrix elements for further - yet to be done - computations.

The calculation is done in the notation of spin-networks, since this basis has certain advantages: e.g. the first Term, consisting of the Casimir Operators, diagonalzies here and gives the corresponding quadric Casimir $C_2(j)^2$ of the group \cite{BPMUV00}. Furthermore  (hence in the Kogut-Susskind-formalism one deals exclusively with it) a 3-dimensional spatial cubic lattice shall be considered. Thus at each vertex 6 links meet and the first question to answer is, how to choose the intertwiner at this node, which couples all six $j$'s to a resulting seventh which vanishes. There are multiple ways to do this and choosing one corresponds to the choice of a basis. Here we take the pairs of parallel edges (say e.g. in $\bar{e}_1$-direction) and couple these to a resulting third (e.g. $\pi_1$). At the end we couple all the three new representations $\pi_1,\pi_2, \pi_3$ to a vanishing fourth. This is independent of the gauge group and afterwards one single node looks as follows:

\[
\left|\nu\left(\left\{ \pi\right\} _{\bar{k}};\left\{ j\right\} _{\bar{k}};\left\{ s\right\} _{\bar{k}}\right)\right\rangle =
\begin{array}{c}\includegraphics[scale=1]{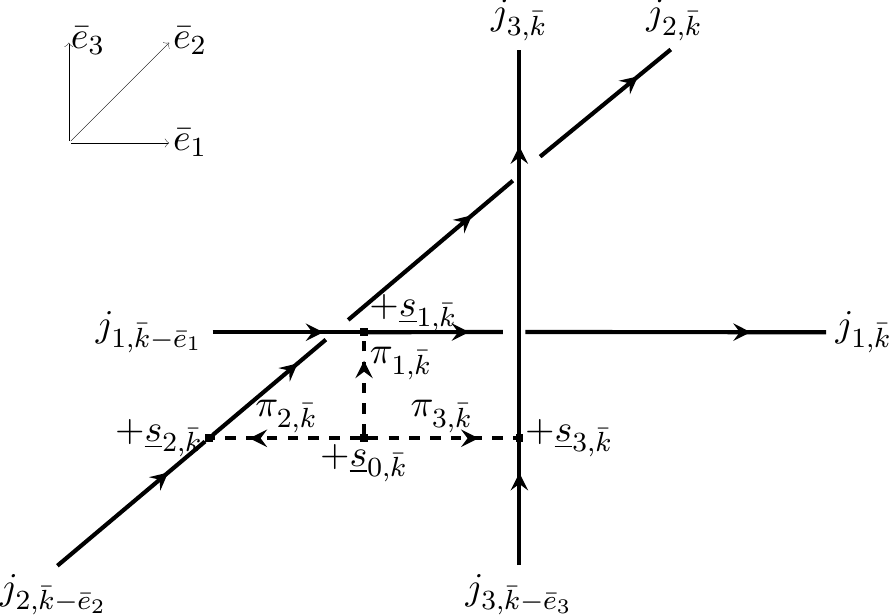}\end{array}
\]
\begin{equation}
=\left|\nu\left(\pi_{1,\bar{k}},\pi_{2,\bar{k}},\pi_{3,\bar{k}};j_{1,\bar{k}},j_{2,\bar{k}},j_{3,\bar{k}},j_{1,\bar{k}-\bar{e}_{1}},j_{1,\bar{k}-\bar{e}_{2}},j_{1,\bar{k}-\bar{e}_{3}};s_{0,\bar{k}},s_{1.\bar{k}}s_{2,\bar{k}},s_{3.\bar{k}}\right)\right\rangle 
\end{equation}
\\
For $SU(2)$ of course all the $s$ would vanish and thus be omitted. Out notation is chosen such that every edge is associated with its direction $\bar{e}_{i}$ and one point on the lattice $\bar{k}$. In total we write for the corresponding group element $\hat{A}_{i,\bar{k}}$. The group elements themselves however will not be written explicitly. If one recalls formula (\ref{3.40}) one sees, that when multiplies two representations of the same group element (as is done, when acting with the plaquette part of the Kogut-Susskind-Hamiltonian) one can shift it to the coupled representation. In this manner, one sees easily that one always ends up with the same lattice one started with (regarding the group elements), only its distinct irreducible representation will have changed. Since this concept translates to all the following calculations, all the corresponding group elements will obviously be omitted in the graphs.

Also, the lines, which are dashed in the picture, are those that are infinitesimally small (like those of $\pi_{i,\bar{k}}$), due to existing only at the vertex itself (and of course not carrying a group element).

To fix the orientation, we choose $\forall i\in\left\{ 1,2,3\right\} ,\forall\bar{k}\in\mathbb{Z}^{3}$

\begin{equation}
\begin{aligned}
\begin{array}{c}\includegraphics[scale=0.4]{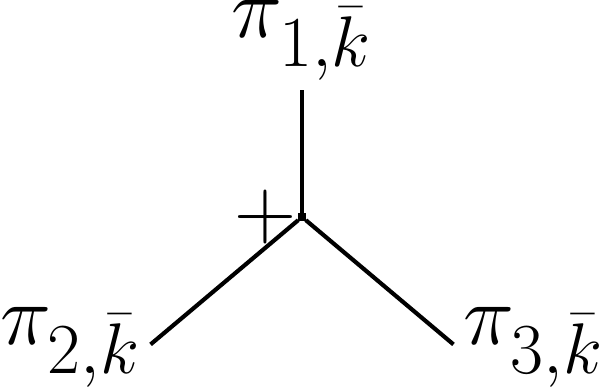}\end{array}\mathrm{ and }\begin{array}{c}\includegraphics[scale=0.8]{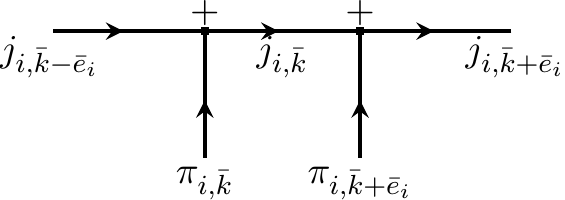}\end{array}
\end{aligned}
\end{equation}
\\
Let $\Psi$ be an arbitrary state of the lattice. As was already stated the Electric Term is diagonal, so we see immediately  that

\begin{equation}
2Q^2 \epsilon \hat{H}_{KS,lit} \left| \Psi \right\rangle = \underset{i,\bar{k}}{\sum} C_2(j_{i,\bar{k}})^2 + \underset{\beta}{\sum}  tr\left(\hat{A}(\beta)\right) + tr\left(\hat{A}(\beta)\right)^\dagger \left| \Psi \right\rangle
\end{equation}
\\
meaning we can restrict ourselves to the evaluation of the trace over all plaquettes. Even more: using that $\hat{A}+\hat{A}^\dagger = 2 Re(\hat{A})$ we focus only on the $tr(\hat{A}(\beta)$. Given the set $\left\{ k,m,n\right\} $ as an even permutation of
$\left\{ 1,2,3\right\} $, one can look w.l.o.g at the plaquette in
$\left(m,n\right)$-direction containing amongst others the vertex $\bar{k}$. In this notation the second term of the Hamiltonian is written:

\begin{equation}\label{trace}
\frac{1}{Q^2}\underset{\bar{k}}{\sum}\overset{3}{\underset{k=1}{\sum}}tr\left(\hat{A}_{m,\bar{k}}\hat{A}_{n,\bar{k}+\bar{e}_{m}}\hat{A}_{m,\bar{k}+\bar{e_{n}}}^{-1}\hat{A}_{n,\bar{k}}^{-1}\right)
\end{equation}
\\
We  first present the application of the graphical calculus to evaluate the matrix elements of (\ref{trace}) in case of the gauge group $SU(2)$ and later on state the corresponding results in case of the $SU(3)$ gauge group.
We note in passing that the Kogut.Susskind computation of the magnetic term performed here would be the same (for $SU(2)$) as the Euclidian piece of the gravitational contribution to the Hamiltonian, which also has not been done in the non graph changing setting before, although it was done for its semi classically valid $U(1)^3$ approximation \cite{GT06_2}. The action of the trace on a general graph $\left|\psi_{\bar{j},\bar{\pi}}\right\rangle $
is written as

\[
\begin{array}{c}\includegraphics[scale=0.8]{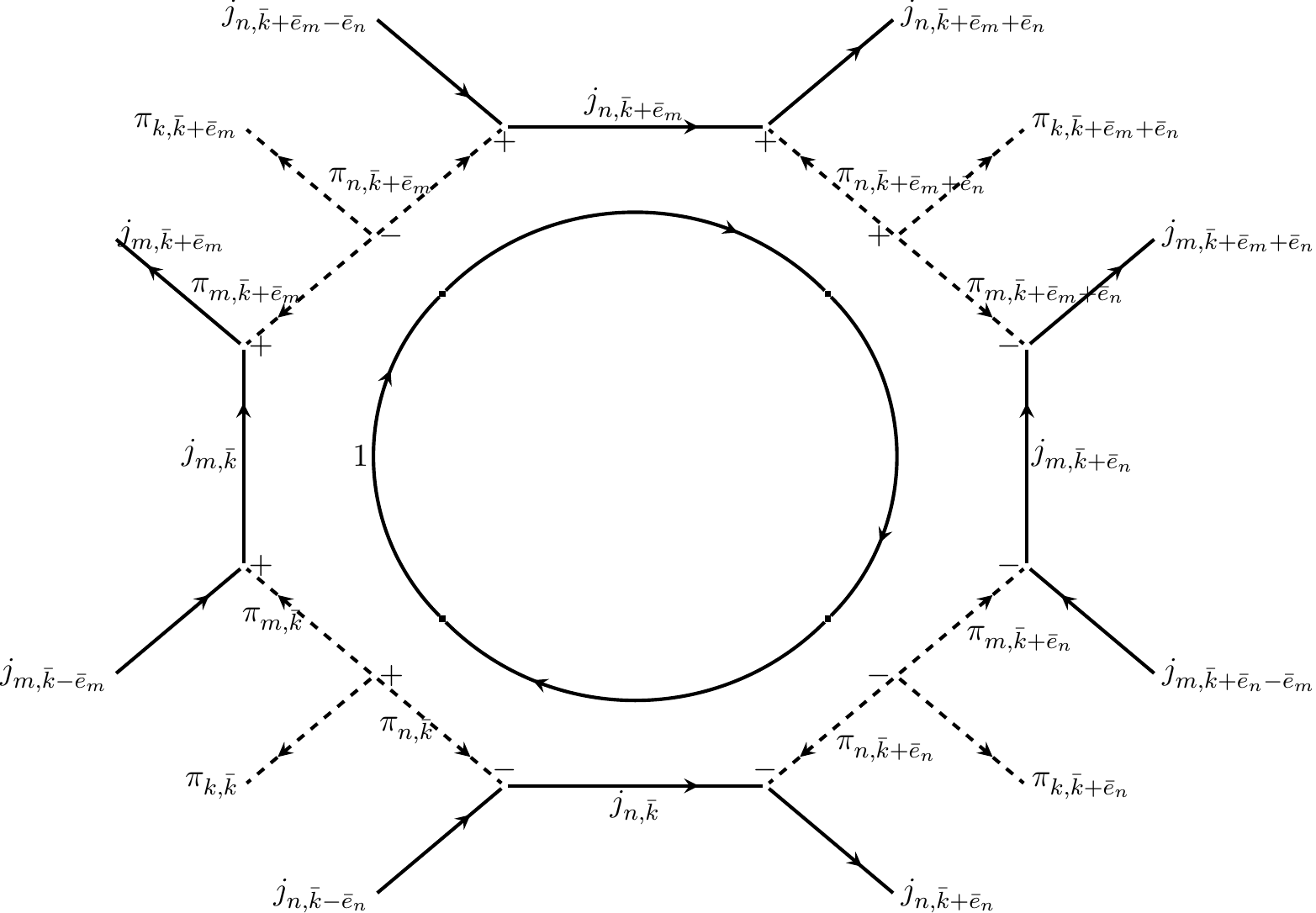}\end{array}
\]
\[
=\underset{\begin{array}{c}
j_{n,\bar{k}}'j_{m,\bar{k}}'\\
j{}_{n,\bar{k}+\bar{e}_{m}}',j_{m,\bar{k}+\bar{e}_{n}}'
\end{array}}{\sum}\underset{\begin{array}{c}
\pi_{n,\bar{k}}',\pi_{m,\bar{k}}',\pi_{n,\bar{k}+\bar{e}_{m}}',\pi_{n,\bar{k}+\bar{e}_{m}}'\\
\pi_{n,\bar{k}+\bar{e}_{n}}',\pi_{n,\bar{k}+\bar{e}_{n}}',\pi_{n,\bar{k}+\bar{e}_{m}+\bar{e}_{n}}',\pi_{n,\bar{k}+\bar{e}_{m}+\bar{e}_{n}}'
\end{array}}{\sum}
\]
\[
\cdot
\left(-\right)^{2\left(j_{m,\bar{k}}'+j_{n,\bar{k}}+j_{m,\bar{k}+\bar{e}_{n}}+j_{n,\bar{k}+\bar{e}_{m}}'+\pi_{m,\bar{k}}'+\pi_{n,\bar{k}}+\pi_{m,\bar{k}+\bar{e}_{m}}+\pi_{n,\bar{k}+\bar{e}_{m}}'+\pi_{m,\bar{k}+\bar{e}_{n}}+\pi_{n,\bar{k}+\bar{e}_{n}}'+\pi_{m,\bar{k}+\bar{e}_{m}+\bar{e}_{n}}'+\pi_{n,\bar{k}+\bar{e}_{m}+\bar{e}_{n}}\right)}\cdot
\]
\[
\cdot
\begin{array}{c}\includegraphics[scale=0.8]{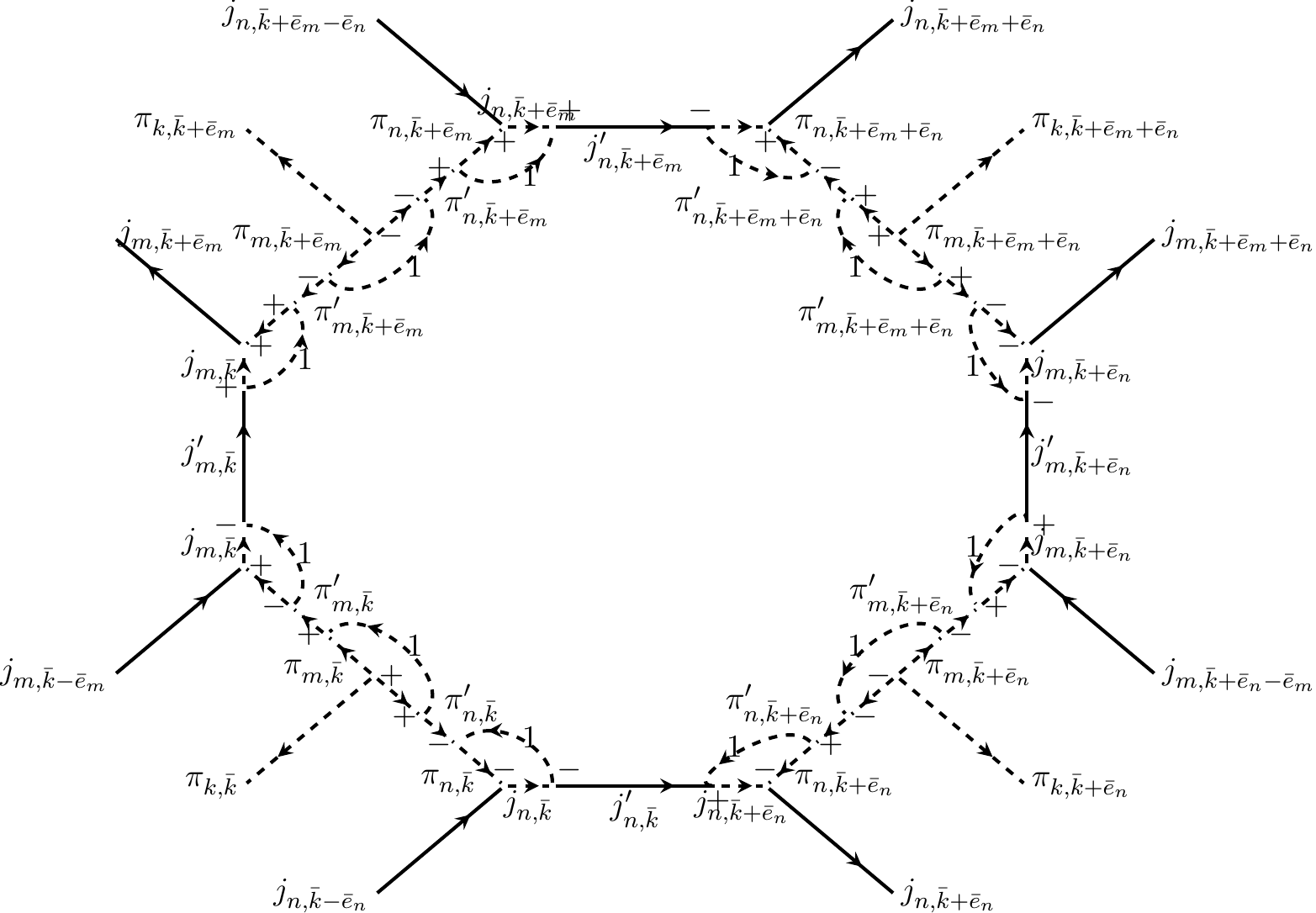}\end{array}
\]
\\
Now all the 6j-symbols have to be recoupled. One starts with the bottom left one in the figures (which is the easiest one with the 6j being exactly in the form as in equation (\ref{extraction}) and then one brings the orientation of the node back to normal order and continues clockwise. Finally if we define

\[
\mathfrak{P}_{SU(2)}\left(\left\{ \pi\right\} _{\bar{k}},\left\{ \pi\right\} _{\bar{k}+\bar{e}_{m}},\left\{ \pi\right\} _{\bar{k}+\bar{e}_{n}},\left\{ \pi\right\} _{\bar{k}+\bar{e}_{m}+\bar{e}_{n}};\left\{ j\right\} ;\pi_{n,\bar{k}}',\pi_{m,\bar{k}}',\pi_{m,\bar{k}+\bar{e}_{m}}',...;j_{n,\bar{k}}',j_{m,\bar{k}}',...\right)\equiv
\]
\[
\underset{i=0,1}{\prod}d_{j'_{m,\bar{k}+i\bar{e}_n}}d_{j'_{n,\bar{k}+i\bar{e}_m}}\underset{j=0,1}{\prod}d_{\pi'_{m,\bar{k}+i\bar{e}_m+j\bar{e}_n}}d_{\pi'_{n,\bar{k}+i\bar{e}_m+j\bar{e}_n}}\cdot
\]
\[
\left(-\right)^{2\left(j_{m,\bar{k}}'+j_{n,\bar{k}}+j_{m,\bar{k}+\bar{e}_{n}}+j_{n,\bar{k}+\bar{e}_{m}}'+\pi_{k,\bar{k}}+\pi_{k,\bar{k}+\bar{e}_{n}}+\pi_{k,\bar{k}+\bar{e}_{m}}+\pi_{k,\bar{k}+\bar{e}_{m}+\bar{e}_{n}}\right)+
j_{n,\bar{k}}'+j_{n,\bar{k}-\bar{e}_{n}}+\pi_{n,\bar{k}}'}
\left\{ \begin{array}{ccc}
j_{n,\bar{k}-\bar{e}_{n}} & j_{n,\bar{k}} & \pi_{n,\bar{k}}\\
1 & \pi_{n,\bar{k}}' & j_{n,\bar{k}}'
\end{array}\right\} \cdot
\]
\[
\cdot\left(-\right)^{2\pi_{n,\bar{k}}}\left\{ \begin{array}{ccc}
\pi_{k,\bar{k}} & \pi_{n,\bar{k}} & \pi_{m,\bar{k}}\\
1 & \pi_{m,\bar{k}}' & \pi_{n,\bar{k}}'
\end{array}\right\} \left(-\right)^{\pi_{k,\bar{k}}+\pi_{m,\bar{k}}'+\pi_{n,\bar{k}}'}\cdot\left(-\right)^{j_{m,\bar{k}}+j_{m,\bar{k}-\bar{e}_{m}}+\pi_{m,\bar{k}}+2j_{m,\bar{k}-\bar{e}_{m}}}\left\{ \begin{array}{ccc}
j_{m,\bar{k}-\bar{e}_{m}} & \pi_{m,\bar{k}} & j_{m,\bar{k}}\\
1 & j_{m,\bar{k}}' & \pi_{m,\bar{k}}'
\end{array}\right\} \cdot
\]
\[
\left(-\right)^{2\pi_{m,\bar{k}+\bar{e}_{m}}+j_{m,\bar{k}}'+j_{m,\bar{k}+\bar{e}_{m}}+\pi_{m,\bar{k}+\bar{e}_{m}}'}\left\{ \begin{array}{ccc}
j_{m,\bar{k}}' & 1 & j_{m,\bar{k}}\\
\pi_{m,\bar{k}+\bar{e}_{m}} & j_{m,\bar{k}+\bar{e}_{m}} & \pi_{m,\bar{k}+\bar{e}_m}'
\end{array}\right\} 
\cdot\left\{ \begin{array}{ccc}
\pi_{m,\bar{k}+\bar{e}_{m}}' & 1 & \pi_{m,\bar{k}+\bar{e}_{m}}\\
\pi_{n,\bar{k}+\bar{e}_{m}} & \pi_{k,\bar{k}+\bar{e}_{m}} & \pi_{n,\bar{k}+\bar{e}_{m}}'
\end{array}\right\}\cdot
\]
\[
\left(-\right)^{2\pi_{n,\bar{k}+\bar{e}_{m}}}
\left(-\right)^{\pi_{n,\bar{k}+\bar{e}_{m}}'+\pi_{k,\bar{k}+\bar{e}_{m}}+\pi_{m,\bar{k}+\bar{e}_{m}}'}
\cdot\left\{ \begin{array}{ccc}
\pi_{n,\bar{k}+\bar{e}_{m}}' & 1 & \pi_{n,\bar{k}+\bar{e}_{m}}\\
j_{n,\bar{k}+\bar{e}_{m}} & j_{n,\bar{k}+\bar{e}_{m}-\bar{e}_{n}} & j_{n,\bar{k}+\bar{e}_{m}}'
\end{array}\right\} \left(-\right)^{\pi_{n,\bar{k}+\bar{e}_{m}}'+j_{n,\bar{k}+\bar{e}_{m}-\bar{e}_{n}}+j_{n,\bar{k}+\bar{e}_{m}}'}\cdot
\]
\[
\left(-\right)^{j_{n,\bar{k}+\bar{e}_{m}}+j_{n,\bar{k}+\bar{e}_{m}+\bar{e}_{n}}+\pi_{n,\bar{k}+\bar{e}_{m}+\bar{e}_{n}}+2j_{n,\bar{k}+\bar{e}_{m}}}
\left\{ \begin{array}{ccc}
j_{n,\bar{k}+\bar{e}_{m}}' & 1 & j_{n,\bar{k}+\bar{e}_{m}}\\
\pi_{n,\bar{k}+\bar{e}_{m}+\bar{e}_{n}} & j_{n,\bar{k}+\bar{e}_{m}+\bar{e}_{n}} & \pi_{n,\bar{k}+\bar{e}_{m}+\bar{e}_{n}}'
\end{array}\right\}
\cdot\left(-\right)^{2\pi_{n,\bar{k}+\bar{e}_{m}+\bar{e}_{n}}}
\]
\[
\left\{ \begin{array}{ccc}
\pi_{n,\bar{k}+\bar{e}_{m}+\bar{e}_{n}}' & 1 & \pi_{n,\bar{k}+\bar{e}_{m}+\bar{e}_{n}}\\
\pi_{m,\bar{k}+\bar{e}_{m}+\bar{e}_{n}} & \pi_{n,\bar{k}+\bar{e}_{m}+\bar{e}_{n}} & \pi_{m,\bar{k}+\bar{e}_{m}+\bar{e}_{n}}'
\end{array}\right\} \left(-\right)^{\pi_{n,\bar{k}+\bar{e}_{m}+\bar{e}_{n}}'+\pi_{m,\bar{k}+\bar{e}_{m}+\bar{e}_{n}}'+\pi_{k,\bar{k}+\bar{e}_{m}+\bar{e}_{n}}}\cdot
\left(-\right)^{2\pi_{m,\bar{k}+\bar{e}_{m}+\bar{e}_{n}}}
\]
\[
\left\{ \begin{array}{ccc}
\pi_{m,\bar{k}+\bar{e}_{m}+\bar{e}_{n}}' & 1 & \pi_{m,\bar{k}+\bar{e}_{m}+\bar{e}_{n}}\\
j_{m,\bar{k}+\bar{e}_{n}} & j_{m,\bar{k}+\bar{e}_{m}+\bar{e}_{n}} & j_{m,\bar{k}+\bar{e}_{n}}'
\end{array}\right\} 
\left(-\right)^{j_{m,\bar{k}+\bar{e}_{n}}'+\pi_{m,\bar{k}+\bar{e}_{m}+\bar{e}_{n}}'+j_{m,\bar{k}+\bar{e}_{m}+\bar{e}_{n}}}\cdot\left(-\right)^{j_{m,\bar{k}+\bar{e}_{n}}+j_{m,\bar{k}-\bar{e}_{m}+\bar{e}_{n}}+\pi_{m,\bar{k}+\bar{e}_{n}}}
\]
\[
\left(-\right)^{2j_{m,\bar{k}+\bar{e}_{n}}+2\pi_{m,\bar{k}+\bar{e}_{n}}}
\cdot
\left\{ \begin{array}{ccc}
\pi_{m,\bar{k}+\bar{e}_{n}}' & \pi_{m,\bar{k}+\bar{e}_{n}} & 1\\
j_{m,\bar{k}+\bar{e}_{n}} & j_{m,\bar{k}+\bar{e}_{n}}' & j_{m,\bar{k}-\bar{e}_{m}+\bar{e}_{n}}
\end{array}\right\} \cdot
\left(-\right)^{2\pi_{n,\bar{k}+\bar{e}_{n}}}\left\{ \begin{array}{ccc}
\pi_{n,\bar{k}+\bar{e}_{n}}' & \pi_{n,\bar{k}+\bar{e}_{n}} & 1\\
\pi_{m,\bar{k}+\bar{e}_{n}} & \pi_{m,\bar{k}+\bar{e}_{n}}' & \pi_{k,\bar{k}+\bar{e}_{n}}
\end{array}\right\} 
\]
\[
\left(-\right)^{\pi_{k,\bar{k}+\bar{e}_{n}}+\pi_{m,\bar{k}+\bar{e}_{n}}'+\pi_{n,\bar{k}+\bar{e}_{n}}'}\cdot\left(-\right)^{j_{n,\bar{k}}+j_{n,\bar{k}+\bar{e}_{n}}+\pi_{n,\bar{k}+\bar{e}_{n}}+2j_{n,\bar{k}}}
\left\{ \begin{array}{ccc}
j_{n,\bar{k}}' & j_{n,\bar{k}} & 1\\
\pi_{n,\bar{k}+\bar{e}_{n}} & \pi_{n,\bar{k}+\bar{e}_{n}}' & j_{n,\bar{k}+\bar{e}_{n}}
\end{array}\right\}
\]
\\
we can write the complete Matrix element for the gauge group $SU(2)$:
\vspace{1cm}
\[
\left\langle \psi_{\bar{j}',\bar{\pi}'}\mid\hat{H}_{YM}\mid\psi_{\bar{j},\bar{\pi}}\right\rangle =\frac{1}{2Q^2}\underset{\bar{k}}{\sum}j_{\bar{k}}\left(j_{\bar{k}}+1\right)+\frac{1}{Q^2}\underset{\bar{k}}{\sum}\underset{m<n}{\sum}
\]
\begin{equation} \label{4.3.5}
\mathfrak{P}_{SU(2)}\left(\left\{ \pi\right\} _{\bar{k}},
 \left\{ \pi\right\}_{\bar{k}+\bar{e}_{m}},\left\{ \pi\right\} _{\bar{k}+\bar{e}_{n}},\left\{ \pi\right\} _{\bar{k}+\bar{e}_{m}+\bar{e}_{n}};\left\{ j\right\} ;\pi_{n,\bar{k}}',\pi_{m,\bar{k}}',\pi_{m,\bar{k}+\bar{e}_{m}}',...;j_{n,\bar{k}}',j_{m,\bar{k}}',...\right)
\end{equation}
\\
A similar calculation with the beforehand established calculus for $SU(3)$ gives us the new plaquette term with $\mathcal{S}_{int}:=\left\{ s_{j_m,\bar{k}}, s_{j_n,\bar{k}}, s_{j_n,\bar{k}+\bar{e}_m}, s_{j_m,\bar{k}+\bar{e}_n},
s_{\pi_m,\bar{k}}, s_{\pi_n,\bar{k}}, s_{\pi_n,\bar{k}+\bar{e}_m+\bar{e}_n}\right.$ 
, $s_{\pi_m,\bar{k}+\bar{e}_m+\bar{e}_n}, s_{\pi_m,\bar{k}+\bar{e}_m}$, 
$\left. s_{\pi_n,\bar{k}+\bar{e}_n}, s_{\pi_n,\bar{k}+\bar{e}_m}, s_{\pi_m, \bar{k}+\bar{e}_n} \right\}$ which denotes the internal set of multiplicities over which we have to sum this time (in contrast note the absence of an additional sign factor here)

\[
 \underset{\mathcal{S}_{int}}{\sum}
 \mathfrak{P}_{SU(3)}\left(\left\{ \pi,s\right\} _{\bar{k}},\left\{ \pi,s\right\} _{\bar{k}+\bar{e}_{m}},\left\{ \pi,s\right\} _{\bar{k}+\bar{e}_{n}},\left\{ \pi,s\right\} _{\bar{k}+\bar{e}_{m}+\bar{e}_{n}};\left\{ j\right\} ;\pi_{n,\bar{k}}',\pi_{m,\bar{k}}',\pi_{m,\bar{k}+\bar{e}_{m}}',...; \right.
\]
\[
\left.
s_{n,\bar{k}}',s_{0,\bar{k}}',s_{m,\bar{k}}',...;j_{n,\bar{k}}',j_{m,\bar{k}}',...\right)\equiv \underset{\mathcal{S}_{int}}{\sum}d_{j_{m,\bar{k}}}d_{j_{n,\bar{k}}}d_{j_{m,\bar{k}+\bar{e}_n}}d_{j_{n,\bar{k}+\bar{e}_m}}\left( \underset{i,j=0,1}{\prod}d_{\pi_{m,\bar{k}+i\bar{e}_n+j\bar{e}_m}} d_{\pi_{n,\bar{k}+i\bar{e}_n+j\bar{e}_m}}\right)\cdot
\]
\[
\left\{\begin{array}{cccc}
j_{n,\bar{k}-\bar{e}_n} & \bar{j}_{n,\bar{k}}' & \pi_{n,\bar{k}}' \\
1 &\pi_{n,\bar{k}}& j_{n,\bar{k}} \\
s_{n,\bar{k}}' & s_{n,\bar{k}} & \bar{s}_{j_n,\bar{k}} & s_{\pi_n,\bar{k}}
\end{array}
\right\}
\left\{\begin{array}{cccc}
\bar{\pi}_{k,\bar{k}} & \bar{\pi}_{n,\bar{k}}' & \bar{\pi}_{m,\bar{k}}'  \\
1 & \bar{\pi}_{m,\bar{k}} & \pi_{n,\bar{k}} \\
s_{0,\bar{k}}' & s_{0,\bar{k}} & \bar{s}_{\pi_m,\bar{k}} & \bar{s}_{\pi_n,\bar{k}} 
\end{array}
\right\}
\left\{\begin{array}{cccc}
j_{m,\bar{k}-\bar{e}_m} & \pi_{m,\bar{k}}' & \bar{j}'_{m,\bar{k}} \\
1 & \bar{j}_{m,\bar{k}} & \bar{\pi}_{m,\bar{k}} \\
s_{m,\bar{k}}' & s_{m,\bar{k}} & s_{\pi_m,\bar{k}} & s_{j_m,\bar{k}} 
\end{array}
\right\}
\]
\[
\left\{\begin{array}{cccc}
j_{m,\bar{k}}' & \pi_{m,\bar{k}+\bar{e}_m}' & \bar{j}_{m,\bar{k}+\bar{e}_m} \\
\pi_{m,\bar{k}+\bar{e}_m} & \bar{j}_{m,\bar{k}} & 1 \\
s'_{m,\bar{k}+\bar{e}_m} & s_{j_m,\bar{k}} & s_{\pi_m,\bar{k}+\bar{e}_m} &  s_{m,\bar{k}+\bar{e}_m}
\end{array}
\right\}
\left\{\begin{array}{cccc}
\bar{\pi}'_{m,\bar{k}+\bar{e}_m} & \bar{\pi}'_{n,\bar{k}+\bar{e}_m} & \bar{\pi}_{k,\bar{k}+\bar{e}_m} \\
\bar{\pi}_{n,\bar{k}+\bar{e}_m} & \pi_{m,\bar{k}+\bar{e}_m} & 1 \\
s'_{0,\bar{k}+\bar{e}_m} & \bar{s}_{\pi_m,\bar{k}+\bar{e}_m}& \bar{s}_{\pi_n,\bar{k}+\bar{e}_m} & s_{0,\bar{k}+\bar{e}_m}
\end{array}
\right\}
\]
\[
\left\{\begin{array}{cccc}
\pi'_{n,\bar{k}+\bar{e}_m} & \bar{j}'_{n,\bar{k}+\bar{e}_m} & j_{n,\bar{k}+\bar{e}_m-\bar{e}_n} \\
\bar{j}_{n,\bar{k}+\bar{e}_m} & \bar{\pi}_{n,\bar{k}+\bar{e}_m} & 1 \\
s'_{n,\bar{k}+\bar{e}_m} & s_{\pi_{n},\bar{k}+\bar{e}_m} & \bar{s}_{j_n,\bar{k}+\bar{e}_n} & s_{n,\bar{k}+\bar{e}_m}
\end{array}
\right\}
\left\{\begin{array}{cccc}
\pi'_{m,\bar{k}+\bar{e}_n} & j_{m,\bar{k}+\bar{e}_n-\bar{e}_m} & \bar{j}'_{m,\bar{k}+\bar{e}_n} \\
j_{m,\bar{k}+\bar{e}_n} & 1 & \pi'_{m,\bar{k}+\bar{e}_n} \\
s'_{m,\bar{k}+\bar{e}_n} & s_{\pi_m,\bar{k}+\bar{e}_n} & s_{m,\bar{k}+\bar{e}_n} & \bar{s}_{j_m,\bar{k}+\bar{e}_n}
\end{array}
\right\}
\]
\[
\left\{\begin{array}{cccc}
\bar{\pi}'_{n,\bar{k}+\bar{e}_n} & \bar{\pi}_{k,\bar{k}+ \bar{e}_n} & \bar{\pi}'_{m,\bar{k}+\bar{e}_n} \\
\pi_{m,\bar{k}+\bar{e}_n} & 1 & \bar{\pi}_{n,\bar{k}+\bar{e}_n} \\
s_{0,\bar{k}+\bar{e}_n}' & \bar{s}_{\pi_m,\bar{k}+\bar{e}_n} & s_{0,\bar{k}+\bar{e}_n} & \bar{s}_{\pi_m,\bar{k}+\bar{e}_n}
\end{array}
\right\}
\left\{\begin{array}{cccc}
j'_{n,\bar{k}+\bar{e}_m} & \pi'_{n,\bar{k}+\bar{e}_m+\bar{e}_n} & \bar{j}_{n,\bar{k}+\bar{e}_m+\bar{e}_n} \\
\pi_{n,\bar{k}+\bar{e}_m+\bar{e}_n} & \bar{j}_{n,\bar{k}+\bar{e}_m} & 1 \\
s'_{n,\bar{k}+\bar{e}_m+\bar{e}_n} & s_{j_m,\bar{k}+\bar{e}_m} & s_{\pi_n,\bar{k}+\bar{e}_m+\bar{e}_n} & s_{n,\bar{k}+\bar{e}_m+\bar{e}_n}
\end{array}
\right\}
\]
\[
\left\{\begin{array}{cccc}
j'_{n,k} & \bar{j}_{n,\bar{k}+\bar{e}_n} & \pi'_{n,\bar{k}+\bar{e}_n} \\
\pi_{n,\bar{k}+\bar{e}_n} & 1 & j_{n,\bar{k}} \\
s'_{n,\bar{k}+\bar{e}_n} & s_{j_n,\bar{k}} & s_{n,\bar{k}+\bar{e}_n} & s_{\pi_n,\bar{k}+\bar{e}_n}
\end{array}
\right\}
\left\{\begin{array}{cccc}
\bar{\pi}'_{n,\bar{k}+\bar{e}_m+\bar{e}_n} & \bar{\pi}'_{m,\bar{k}+\bar{e}_m+\bar{e}_n} & \bar{\pi}_{k,\bar{k}+\bar{e}_m+\bar{e}_n}  \\
\bar{\pi}_{m,\bar{k}+\bar{e}_m+\bar{e}_n} & \pi_{n,\bar{k}+\bar{e}_m+\bar{e}_n} & 1 \\
s'_{0,\bar{k}+\bar{e}_m+\bar{e}_n} & \bar{s}_{\pi_m,\bar{k}+\bar{e}_m+\bar{e}_n} & \bar{s}_{\pi_m,\bar{k}+\bar{e}_m+\bar{e}_n} & s_{0,\bar{k}+\bar{e}_m+\bar{e}_n}
\end{array}
\right\}
\]
\[
\left\{\begin{array}{cccc}
\pi_{m,\bar{k}+\bar{e}_m+\bar{e}_n} & j'_{m,\bar{k}+\bar{e}_n} & \bar{j}_{m,\bar{k}+\bar{e}_m+\bar{e}_n} \\
j_{m,\bar{k}+e_n} & \bar{\pi}_{m,\bar{k}+\bar{e}_m+\bar{e}_n} & 1 \\
s'_{m,\bar{k}+\bar{e}_m+\bar{e}_n} & s_{\pi_m,\bar{k}+\bar{e}_m+\bar{e}_n} & s_{j_m,\bar{k}+\bar{e}_n} & s_{m,\bar{k}+\bar{e}_m+\bar{e}_n}
\end{array}
\right\}
\]
\\
So the complete matrix-element is the same as in (\ref{4.3.5}) with this sum over the new plaquette term and the new Casimir. Note, that in the action of the Kogut-Susskind-Hamiltonian the group elements of the plaquette are in the defining representation. However the same calculation could be done for an arbitrary $m$-representation. Since this will be used later, there have been no simplifications in the above expressions, such that one can easily replace $1 \rightarrow m$ and denote the new plaquette term as $\mathfrak{P}\left(\ldots\mid m\right)$ to distinguish it from the QCD case.
\section{Einstein-Yang-Mills-Theory in Quantum Gravity}
\label{c5}
To compute the matrix elements of the full Quantum Gravity Yang-Mills-Hamiltonian, we adopt the same notation as in \ref{c2}, and denote the gravity-quantum-numbers with $j_i$ and the Yang-Mills-quantum-numbers with $\underline{j}_i$, whose gauge group will be set to $SU(3)$ for the remainder of this paper. The basis functions $\Psi$ on our cubic graph are labelled by

\begin{equation}
\left| \Psi\left(\{j\}\right) \underline{\Psi}\left(\{\underline{j}\},\{\underline{\pi}\};\{\underline{s}\}\right)\right\rangle = \underset{\bar{k}\in \mathbb{Z}^3}{\sum} \left| \nu\left(\{\pi\}_{\bar{k}},\{j\vphantom{\underline{j}}
\}_{\bar{k}}\right) \right\rangle \otimes \left| \nu\left(\{\underline{\pi}\}_{\bar{k}},\{\underline{j}\}_{\bar{k}}\right) \right\rangle
\end{equation}
\\
Due to the fact, that the result is quite lengthy and splits up into a lot of sub-cases, we split up this section. The Quantum Gravity Yang-Mills-Hamiltonian

\[
\hat{H}_{YM}(v)=\frac{1}{2Q^2}\left(\hat{H}_E(v)+\hat{H}_B(v)\right)
\]
\\
consists out of two big parts. The first being the Electric Term and the second one being the Magnetic Term. For both one can look separately at the gravitational degrees of freedom and at the Yang-Mills-degrees of freedom, i.e. the Electric Fluxes and the plaquette part respectively. Each of these four parts is calculated in its corresponding sub-chapter below.
\subsection{Gravity-Part of the Electric Term}
\label{c5.1}
The Gravity-Part of the YM-Hamiltonian is

\begin{equation}\label{5.2}
tr\left(\hat{A}_{j}\left[\hat{A}_{j}^{-1},\sqrt{\hat{V}}\right]\hat{A}_{m}\left[\hat{A}_{m}^{-1},\sqrt{\hat{V}}\right]\right)
\end{equation}
\\
Due to the commutators one gets four different parts. The first one is just the definition of the elements of the action of the Volume:

\[
\hat{V}\left|\nu\left(\left\{ \pi\right\} _{\bar{k}},\left\{ j\right\} _{\bar{k}}\right)\right\rangle \equiv\underset{\left\{ \pi\right\} _{\bar{k}}^{2}}{\tilde{\sum}}V_{\bar{k}}\left(\left\{ \pi\right\} _{\bar{k}},\left\{ \pi\right\} _{\bar{k}}^{2};\left\{ j_{\bar{k}}\right\} \right)\left|\nu\left(\left\{ \pi\right\} _{\bar{k}}^{2},\left\{ j\right\} _{\bar{k}}\right)\right\rangle 
\]
\\
The label $\bar{k}$ is purely of interest for the valency of the vertex (with $\bar{k}\in \mathbb{Z}^3$ there are six edges meeting at the node). Moreover one realizes that the Volume operator only changes the intertwiners, not the graph itself.
We have also introduced the weighted sum: $\underset{j}{\tilde{\sum}}=\underset{j}{\sum}d_j$.
\\
For the second one the action of the Volume on a non-gauge invariant node is needed. The notation here ($\sqrt{V}_{\bar{k}+\bar{e}_j}$) means that on the edge in $j$-direction a non gauge-invariant edge in $m$ representation is glued. The additional representation $j_j$ that changes to $j_j^2$, where one needs to sum over, is also displayed after the first semicolon: $\sqrt{V}_{\bar{k}+\bar{e}_j}\left(\ldots;j_j,j_j^2;\ldots\mid m\right)$. If $j=1,2,3$ only half of the edges are calculated. For the remaining ones, carrying the representation $\left(j_{j,\bar{k}-\bar{e}_{j}}\right)$, the calculation broadly remains exactly the same when replacing $j_{j,\bar{k}-\bar{e}_{j}}\Leftrightarrow j_{j,\bar{k}}$. However one wants to work on a vertex where all edges are outgoing to maximize the degree of symmetry, which explains the (temporary) additional sign in the second line of the computation. Moreover one also has to switch the orientation of the vertex itself, since the ``+''-sign would elsewhere become ``-''. To combine both cases in one in the following, we will introduce the parameter $p_{j}\in\left\{ \bar{0},\bar{e}_{j}\right\} $, which distinguishes the cases, using $j_{j,\bar{k}-\bar{p}_{j}}$ and $j_{j,\bar{k}-\bar{e}_{j}+\bar{p}_{j}}$.
So for one we will get a sign of ${\left|\bar{p}_{j}\right|\left(\pi_{j,\bar{k}}+j_{j,\bar{k}}+j_{j,\bar{k}+\bar{e}_{j}}\right)}$ to ensure that the sign at the vertex is always ``+''. With all of this the action for the second part is (where one also uses the $SU(2)$ version of the orthogonality relation (\ref{3.10-bubble}) in the last line after having coupled the last holonomy to the graph):

\[
tr\left(\sqrt{\hat{V}}\hat{A}_{m,\bar{p}_{m}}\sqrt{\hat{V}}\hat{A}_{m,\bar{p}_{m}}^{-1}\right)\left|\nu\left(\left\{ \pi\right\} _{\bar{k}};\left\{ j\right\} _{\bar{k}}\right)\right\rangle =
\]
\[
=tr\left(\sqrt{\hat{V}}\hat{A}_{m,\bar{p}_{m}}\sqrt{\hat{V}}\right)\left(-\right)^{2\left(j_{1,\bar{k}-\bar{e}_{1}}+j_{2,\bar{k}-\bar{e}_{2}}+j_{3,\bar{k}-\bar{e}_{3}}\right)}\left(-\right)^{\left|\bar{p}_{m}\right|\left(\pi_{m,\bar{k}}+j_{m,\bar{k}}+j_{m,\bar{k}+\bar{e}_{m}}\right)}
\]
\[
\underset{j_{m,\bar{k}-\bar{p}_{m}}^{2}}{\tilde{\sum}}\left(-\right)^{2j_{m,\bar{k}-\bar{p}_{m}}^{2}}\Biggl|
\begin{array}{c}\includegraphics[scale=0.5]{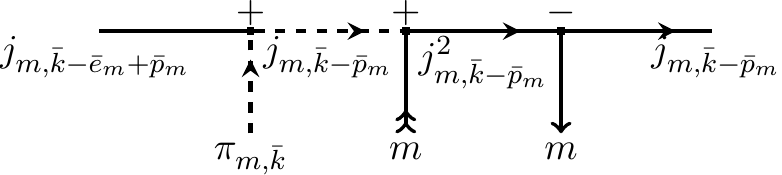}\end{array}
\Biggr\rangle =
\]
\[
=tr\left(\sqrt{\hat{V}}\hat{A}_{m,\bar{p}_{m}}\right)\sqrt{V}_{\bar{k}+\bar{e}_{m}-2\bar{p}_{m}}\left(\left\{ \pi\right\} _{\bar{k}},\left\{ \pi\right\} _{\bar{k}}^{2};j_{m,\bar{k}-\bar{p}_{m}},j_{m,\bar{k}-\bar{p}_{m}}^{3};\ldots j_{m,\bar{k}-\bar{p}_{m}}^{2}\ldots\mid m\right)
\]
\[
\left(-\right)^{2j_{m,\bar{k}-\bar{p}_{m}}^{2}}\Biggl|
\begin{array}{c}\includegraphics[scale=0.5]{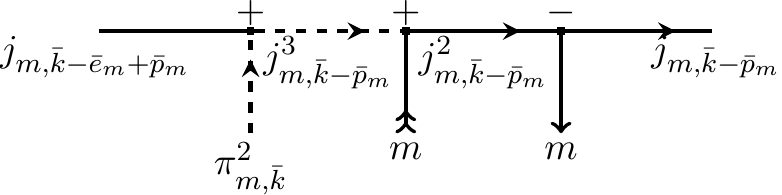}\end{array}
\Biggr\rangle =
\]
\[
=\underset{j_{m,\bar{k}+\bar{p_{m}}}^{2},\left\{ \pi\right\} _{\bar{k}}^{3},\left\{ \pi\right\} _{\bar{k}}^{2}}{\tilde{\sum}}\sqrt{V}_{\bar{k}+\bar{e}_{m}-2\bar{p}_{m}}\left(\left\{ \pi\right\} _{\bar{k}},\left\{ \pi\right\} _{\bar{k}}^{2};j_{m,\bar{k}-\bar{p}_{m}},j_{m,\bar{k}-\bar{p}_{m}};\ldots j_{m,\bar{k}-\bar{p}_{m}}^{2}\ldots\right)\cdot
\]
\[
\cdot\sqrt{V}_{\bar{k}}\left(\left\{ \pi\right\} _{\bar{k}}^{2},\left\{ \pi\right\} _{\bar{k}}^{3};\left\{ j\right\} _{\bar{k}}\right)
\cdot\left(-\right)^{\left|\bar{p}_{m}\right|\left(\pi_{m,\bar{k}}^{3}+\pi_{m,\bar{k}}\right)}\left|\nu\left(\left\{ \pi\right\} _{\bar{k}}^{3};\left\{ j\right\} _{\bar{k}}\right)\right\rangle 
\]
\\
And correspondingly the third part is:

\[
tr\left(\hat{A}_{j,\bar{p}_{j}}\sqrt{\hat{V}}\hat{A}_{j,\bar{p}_{j}}^{-1}\sqrt{\hat{V}}\right)\left|\nu\left(\left\{ \pi\right\} _{\bar{k}};\left\{ j\right\} _{\bar{k}}\right)\right\rangle =
\]
\[
=\underset{\left\{ \pi\right\} _{\bar{k}}^{2},\left\{ \pi\right\} _{\bar{k}}^{3},j_{j,\bar{k}-\bar{p}_{j}}^{2}}{\tilde{\sum}}\left(-\right)^{\left|\bar{p}_{j}\right|\left(\pi_{j,\bar{k}}-\pi_{j,\bar{k}}^{3}\right)}\sqrt{V_{\bar{k}}}\left(\left\{ \pi\right\} _{\bar{k}},\left\{ \pi\right\} _{\bar{k}}^{2};\left\{ j\right\} _{\bar{k}}\right)\cdot
\]
\[
\cdot\sqrt{V}_{\bar{k}+\bar{e}_{j}-2\bar{p}_{j}}\left(\left\{ \pi\right\} _{\bar{k}}^{2},\left\{ \pi\right\} _{\bar{k}}^{3};j_{j,\bar{k}-\bar{p}_{j}},j_{j,\bar{k}-\bar{p}_{j}};\ldots j_{j,\bar{k}-\bar{p}_{j}}^{2}\ldots\mid m\right)\left|\nu\left(\left\{ \pi\right\}^3_{\bar{k}};\left\{ j\right\} _{\bar{k}}\right)\right\rangle 
\]
\\
The fourth and last part of \ref{5.2} needs some more detailed treatment, since we deal now, with two holonomies that are glued to the graph, and that may go in different directions. The term of interest is $\hat{A}_j \sqrt{\hat{V}}\hat{A}_j^-1\hat{A}_m \sqrt{\hat{V}}\hat{A}_m^-1$, where $j,m$ denote the different directions of the glued edges. Summing over all possible combinations of choosing two (possibly the same) edges emanating from one vertex $\bar{k}$ we have 36 combinations, from which many due to symmetry reasons give the same result. In total we have thus only to distinguish three case: Both holonomies may
\begin{itemize}
\item[i)] lie on the same edge $\left(j_{m,\bar{k}}=j_{j,\bar{k}}\right)$
\item[ii)] lie on parallel edges $\left(j_{m,\bar{k}+\bar{e}_{m}}=j_{j,\bar{k}}\right)$
\item[iii)] go in different directions
\end{itemize}
For i) it is obvious that the holonomies in the middle cancel, leaving us with a rather simple expression:

\[
tr\left(\hat{A}_{j,\bar{p}_{j}}\hat{V}\hat{A}_{j,\bar{p}_j}^{-1}\right)\left|\nu\left(\left\{ \pi\right\} _{\bar{k}},\left\{ j\right\} _{\bar{k}}\right)\right\rangle=
\]
\[
=\underset{j_{j,\bar{k}-\bar{p}_{j}}^{2}\left\{ \pi\right\} _{\bar{k}}^{2}}{\tilde{\sum}}\left(-\right)^{\left|\bar{p}_{j}\right|\left(\pi_{j,\text{\c\ \ensuremath{\bar{k}}}}+\pi_{j,\bar{k}}^{2}\right)}V_{\bar{k}+\bar{e}_{j}-2\bar{p}_{j}}\left(\left\{ \pi\right\} _{\bar{k}},\left\{ \pi\right\} _{\bar{k}}^{2};j_{j,\bar{k}-\bar{p}_{j}},j_{j,\bar{k}-\bar{p}_{j}};\ldots j_{j,\bar{k}-\bar{p}_{j}}^{2}\ldots\mid m\right)\left|\nu\left(\left\{ \pi\right\} _{\bar{k}}^{2},\left\{ j\right\} _{\bar{k}}\right)\right\rangle 
\]
\\
The second part of course incorporates now a change from one link to the other and back to close the trace of the holonomies at the end. As one can easily see, the structures appearing again look similar to equation \ref{extraction} from the appendix and thus represent 6j-symbols. Note, moreover, that the open edges in the $m$-representation in the third line denotes the open ends of the holonomy. One is attached infinitesimal close to the vertex, hence the action of the volume elements also changes the link between these two, and the other open end (on the $j_{j,\bar{k}-\bar{p}_j}$-edge) is attached after the group element, which we have suppressed and trivially shifted to the $j^2_{j,\bar{k}-\bar{p}_j}$-edge.

\[tr\left(\hat{A}_{j,p_j}\sqrt{\hat{V}}\hat{A}_{j,p_j}^{-1}\hat{A}_{m,p_m}\sqrt{\hat{V}}\hat{A}_{m,p_m}^{-1}\right)\left|\nu(\left\{\pi\right\}_{\bar{k}},\left\{j\right\}_{\bar{k}})\right\rangle=\left(-\right)^{2\overset{3}{\underset{i=1}{\sum}}j_{i,\bar{k}-\bar{e}_{i}}+\left|\bar{p}_{j}\right|\left(j_{j,\bar{k}-\bar{e}_{j}}+j_{j,\bar{k}}+\pi_{j,\bar{k}}\right)}\cdot
\]
\[
\cdot
\underset{\begin{array}{c}
j_{j,\bar{k}-\bar{e}_{j}+\bar{p}_{j}}^{2},\left\{ \pi\right\} _{\bar{k}}^{2},\\j_{j,\bar{k}-\bar{e}_{j}+\bar{p}_{j}}^{3},\\
j_{j,\bar{k}-\bar{p}_{J}}^{2},j_{j,\bar{k}-\bar{e}_{j}+\bar{p}_{j}}^{4}\\
\end{array}}{\tilde{\sum}}
\left(-\right)^{2j_{j,\bar{k}-\bar{e}_{j}+\bar{p}_{j}}^{2}}
\sqrt{V}_{\bar{k}-\bar{e}_{j}+2\bar{p}_{j}}\left(\left\{ \pi\right\} _{\bar{k}},\left\{ \pi\right\} _{\bar{k}}^{2};j_{j,\bar{k}-\bar{e}_{j}+\bar{p}_{j}},j_{j,\bar{k}-\bar{e}_{j}+\bar{p}_{j}}^{3};\ldots j_{j,\bar{k}-\bar{e}_{j}+\bar{p}_{j}}^{2}\ldots\mid m\right)
\cdot\]
\[\cdot
\left(-\right)^{2j_{j,\bar{k}-\bar{p}_{j}}^{2}+2j_{j,\bar{k}-\bar{e}_{j}+\bar{p}_{j}}^{3}}
 tr\left(\hat{A}_{j,\bar{p}_{j}}\sqrt{\hat{V}}\right)
\Biggl|
\begin{array}{c}\includegraphics[scale=0.7]{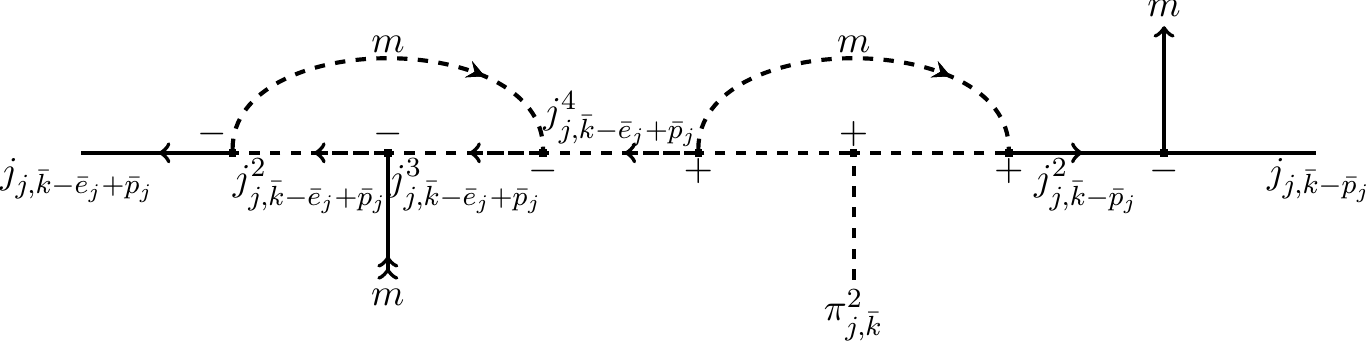}\end{array}
\Biggr\rangle =
\]
\[=\underset{\begin{array}{cc}
j_{j,\bar{k}-\bar{e}_{j}+\bar{p}_{j}}^{2\ldots5}, & j_{j,\bar{k}-\bar{p}_{J}}^{2},\\
\left\{ \pi\right\} _{\bar{k}}^{2}, & \left\{ \pi\right\} _{\bar{k}}^{3},
\end{array}}{\tilde{\sum}}
\sqrt{V}_{\bar{k}-\bar{e}_{j}+2\bar{p}_{j}}\left(\left\{ \pi\right\} _{\bar{k}},\left\{ \pi\right\} _{\bar{k}}^{2};j_{j,\bar{k}-\bar{e}_{j}+\bar{p}_{j}},j_{j,\bar{k}-\bar{e}_{j}+\bar{p}_{j}}^{3};\ldots j_{j,\bar{k}-\bar{e}_{j}+\bar{p}_{j}}^{2}\ldots\mid m\right)
\]
\[
\sqrt{V}_{\bar{k}-\bar{e}_{j}+2\bar{p}_{j}}\left(\left\{ \pi\right\} _{\bar{k}}^{2},\left\{ \pi\right\} _{\bar{k}}^{3};j_{j,\bar{k}-\bar{e}_{j}+\bar{p}_{j}}^{4},j_{j,\bar{k}-\bar{e}_{j}+\bar{p}_{j}}^{5};\ldots j_{j,\bar{k}-\bar{p}_{j}}^{2}\ldots\mid m\right)
\]
\[
\left(-\right)^{\left|\bar{p}_{j}\right|\left(\pi_{j,\bar{k}}+\pi_{j,\bar{k}}^{3}\right)}
(-)^{\pi^2_{\bar{k}}-\pi^3_{\bar{k}}}
(-)^{2j^5_{j,\bar{k}-\bar{e}_j}+\bar{p}_j}
\left\{ \begin{array}{ccc}
j_{j,\bar{k}-\bar{e}_j+\bar{p}_j} & j^2_{j,\bar{k}-\bar{e}_j+\bar{p}_j} & m \\
j^3_{j,\bar{k}-\bar{e}_j+\bar{p}_j} & j^4_{j,\bar{k}-\bar{e}_j+\bar{p}_j} & m
\end{array}\right\}
\]
\[
\left\{ \begin{array}{ccc}
j^4_{j,\bar{k}-\bar{e}_j+\bar{p}_j} & j^3_{j,\bar{k}-\bar{e}_j+\bar{p}_j} & m \\
j_{j,\bar{k}-\bar{p}_j} & j^2_{j,\bar{k}-\bar{p}_j} & \pi^2_{j\bar{k}}
\end{array}\right\}
\left\{ \begin{array}{ccc}
j_{j,\bar{k}-\bar{e}_j+\bar{p}_j} & j^5_{j,\bar{k}-\bar{e}_j+\bar{p}_j} & m \\
j^2_{j,\bar{k}-\bar{p}_j} & j_{j,\bar{k}-\bar{p}_j} & \pi^3_{j,\bar{k}}
\end{array}\right\}
\left|\nu\left(\{\pi\}^3_{\bar{k}},\{j\}_{\bar{k}}\right)\right\rangle
\]
\\
Note that the additional sign of $\pi^2_{\bar{k}}-\pi^3_{\bar{k}}$ stems from the fact, that one has to reorient the vertices in between to act with the second Volume-operator in the way it was defined on a node with given orientation.
For iii) things get again more complicated. We have to switch from one edge to another edge, which does not lie in the same direction. Explicitly, we are interested in the action of the holonomy $\hat{A}^{-1}_{j,\bar{p}_j}\hat{A}_{m,\bar{p}_m}$ on a vertex, which we will find useful to write in the following form, where $\sigma$ gives us the sign of the permutation of $m,j,q$:

\[
\left|\nu\left(\left\{ \pi\right\} _{\bar{k}},\left\{ j\right\} _{\bar{k}}\right)\right\rangle =\left(-\right)^{2\underset{i=1}{\overset{3}{\sum}}j_{i,\bar{k}-\bar{e}_{i}}}\left(-\right)^{\left|\bar{p}_{j}\right|\left(\pi_{j,\bar{k}}+j_{j,\bar{k}}+j_{j,\bar{k}-\bar{e}_{j}}\right)}\left(-\right)^{\left(1-\left|\bar{p}_{m}\right|\right)\left(\pi_{m,\bar{k}}+j_{m,\bar{k}}+j_{m,\bar{k}-\bar{e}_{m}}\right)}
\]
\[
\left(-\right)^{\sigma\left(m,j,q\right)\left(\pi_{j,\bar{k}}+\pi_{m,\bar{k}}+\pi_{q,\bar{k}}\right)}
\Biggl|
\begin{array}{c}\includegraphics[scale=0.8]{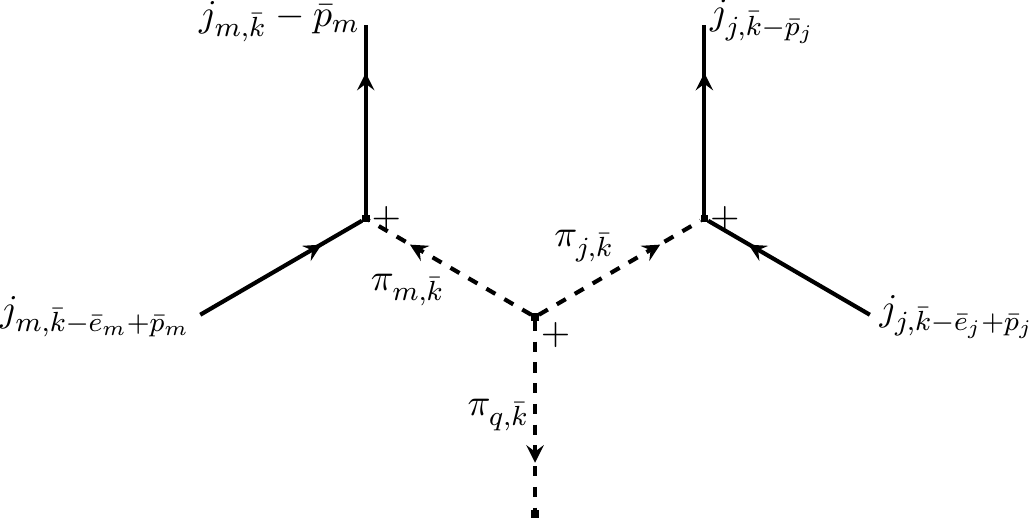}\end{array}
\Biggr\rangle 
\]
\\
Once our Hamiltonian acts on the state, we see that traversing the node results in a couple of 6j-symbols (four when going from $j_{m,\bar{k}-\bar{p}_m}$ to $j_{j,\bar{k}-\bar{p}_j}$ and three when going back. Remember that in between we have to bring the signs back into an orientation such that $\hat{V}$ can act and after its action we have to restore the given orientation, such that one can close the holonomies. In total one ends up with a fairly complicated expression:

\[
tr\left(\hat{A}_{j,\bar{p}_{j}}\sqrt{\hat{V}}\right)\underset{\begin{array}{c}
j_{m,\bar{k}-\bar{p}_{m}}^{2},j_{m,\bar{k}-\bar{p}_{m}}^{3}\\
\left\{ \pi\right\} _{\bar{k}}^{2}
\end{array}}{\tilde{\sum}}\left(-\right)^{2j_{m,\bar{k}-\bar{p}_{m}}^{2}}\left(-\right)^{2\underset{i=1}{\overset{3}{\sum}}j_{i,\bar{k}-\bar{e}_{i}}+\left|\bar{p}_{m}\right|\left(j_{m,\bar{k}}+j_{m,\bar{k}-\bar{e}_{m}}+\pi_{m,\bar{k}}\right)}
\left(-\right)^{\sigma\left(m,j,q\right)\left(\pi_{j,\bar{k}}^{2}+\pi_{m,\bar{k}}^{2}+\pi_{q,\bar{k}}^{2}\right)}
\]
\[\left(-\right)^{\left|\bar{p}_{j}\right|\left(j_{j,\bar{k}}+j_{j,\bar{k}-\bar{e}_{j}}+\pi_{j,\bar{k}}\right)}
\sqrt{V}_{\bar{k}+\bar{e}_{m}-2\bar{p}_{m}}\left(\left\{ \pi\right\} _{\bar{k}},\left\{ \pi\right\} _{\bar{k}}^{2};j_{m,\bar{k}-\bar{p}_{m}},j_{m,\bar{k}-\bar{p}_{m}}^{3};\ldots j_{m,\bar{k}-\bar{p}_{m}}^{2}\ldots\mid m\right)
\]
\[
\cdot
\underset{\begin{array}{c}
j_{m,\bar{k}-\bar{p}_{m}}^{4},j_{j,\bar{k}-\bar{p}_{j}}^{2}\\
\left\{ \pi\right\} _{\bar{k}}^{2},\pi_{m,\bar{k}}^{3},\pi_{j,\bar{k}}^{3}
\end{array}}{\tilde{\sum}}\left(-\right)^{2j_{m,\bar{k}-\bar{p}_{m}}^{3}+2\pi_{m,\bar{k}}^{2}+2\pi_{j,\bar{k}}^{3}+2j_{j,\bar{k}-\bar{p}_{j}}^{2}}\Biggl|
\begin{array}{c}\includegraphics[scale=0.8]{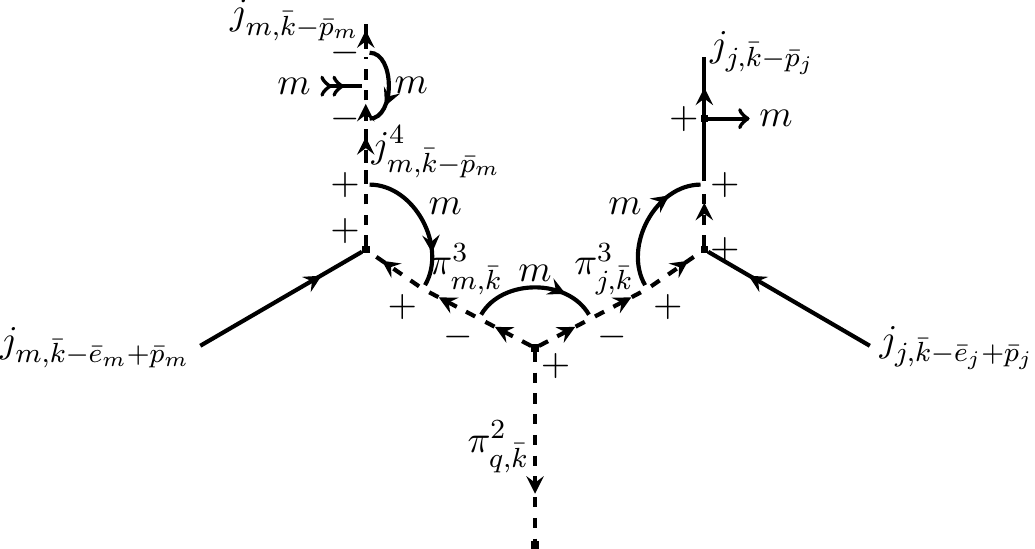}\end{array}
\Biggr\rangle =
\]
\[
=tr\left(\hat{A}_{j,\bar{p}_{j}}\right)\underset{\begin{array}{c}
j_{m,\bar{k}-\bar{p}_{m}}^{2\ldots4},j_{j,\bar{k}-\bar{p}_{j}}^{2}\\
\left\{ \pi\right\} _{\bar{k}}^{2},\pi_{m,\bar{k}}^{3},\pi_{j,\bar{k}}^{3}\\
\left\{ \pi\right\} _{\bar{k}}^{4},j_{m,\bar{k}-\bar{p}_{m}}^{5}
\end{array}}{\tilde{\sum}}\left(-\right)^{2\underset{i=1}{\overset{3}{\sum}}j_{i,\bar{k}-\bar{e}_{i}}}\left(-\right)^{\sigma\left(m,j,q\right)\left(\pi_{j,\bar{k}}^{2}+\pi_{m,\bar{k}}^{2}+2\pi_{q,\bar{k}}^{2}+\pi_{j,\bar{k}}^{3}+\pi_{m,\bar{k}}^{3}+\pi_{j,\bar{k}}^{4}+\pi_{m,\bar{k}}^{4}+\pi_{q,\bar{k}}^{4}\right)}
\]
\[
\left(-\right)^{\left|\bar{p}_{m}\right|\left(j_{m,\bar{k}}+j_{m,\bar{k}-\bar{e}_{m}}+\pi_{m,\bar{k}}+\pi_{m,\bar{k}}^{3}+j_{m,\bar{k}-\bar{p}_{m}}^{4}+2j_{m,\bar{k}-\bar{e}_{m}+\bar{p}_{m}}+j_{m,\bar{k}-\bar{p}_{m}}^{5}+\pi_{m,\bar{k}}^{4}\right)}
\]
\[
\left(-\right)^{\left|\bar{p}_{j}\right|\left(j_{j,\bar{k}}+j_{j,\bar{k}-\bar{e}_{j}}+\pi_{j,\bar{k}}+\pi_{j,\bar{k}}^{3}+2j_{j,\bar{k}-\bar{p}_{j}}^{2}+2j_{j,\bar{k}-\bar{e}_{j}+\bar{p}_{j}}+\pi_{j,\bar{k}}^{4}\right)}
(-)^{m+j_{m,\bar{k}-\bar{p}_m}+j_{m,\bar{k}-\bar{e}_m+\bar{p}_m}+j^2_{j,\bar{k}-\bar{p}_j}+j_{j,\bar{k}-\bar{e}_j+\bar{p}_j}}
\]
\[
(-)^{\pi^3_{m,\bar{k}}+\pi^3_{j,\bar{k}}+\pi^2_{m,\bar{k}}+\pi^2_{j,\bar{k}}+\pi^2_{q,\bar{k}}}
 \sqrt{V}_{\bar{k}+\bar{e}_{m}-2\bar{p}_{m}}\left(\left\{ \pi\right\} _{\bar{k}},\left\{ \pi\right\} _{\bar{k}}^{2};j_{m,\bar{k}-\bar{p}_{m}},j_{m,\bar{k}-\bar{p}_{m}}^{3};\ldots j_{m,\bar{k}-\bar{p}_{m}}^{2}\ldots\mid m\right)
\]
\[
 \sqrt{V}_{\bar{k}+\bar{e}_{m}-2\bar{p}_{m}}\left(\pi_{m,\bar{k}}^{3},\pi_{j,\bar{k}}^{3},\pi_{q,\bar{k}}^{2},\left\{ \pi\right\} _{\bar{k}}^{4};j_{m,\bar{k}-\bar{p}_{m}}^{4},j_{m,\bar{k}-\bar{p}_{m}}^{5};\ldots j_{j,\bar{k}-\bar{p}_{m}}^{2}\ldots\mid	m\right)
\]
\[
\left\{ \begin{array}{ccc}
\pi_{j,\bar{k}}^{3} & \pi_{j,\bar{k}}^{2} & m\\
j_{j,\bar{k}-\bar{p}_{j}} & j_{j,\bar{k}-\bar{p}_{j}}^{2} & j_{j,\bar{k}+\bar{e}_{j}-\bar{p}_{j}}
\end{array}\right\} 
\left\{ \begin{array}{ccc}
\pi_{q,\bar{k}}^{2} & \pi_{j,\bar{k}}^{2} & \pi_{m,\bar{k}}^{2}\\
m & \pi_{m,\bar{k}}^{3} & \pi_{j,\bar{k}}^{3}
\end{array}\right\}
\left\{ \begin{array}{ccc}
m & j_{m,\bar{k}-\bar{p}_{m}}^{3} & j_{m,\bar{k}-\bar{p}_{m}}\\
m & j_{m,\bar{k}-\bar{p}_{m}} & j_{m,\bar{k}-\bar{p}_{m}}^{4}
\end{array}\right\} 
\]
\[
\left\{ \begin{array}{ccc}
j_{m,\bar{k}-\bar{e}_{m}+\bar{p}_{m}} & \pi_{m,\bar{k}}^{2} & j_{m,\bar{k}-\bar{p}_{m}}^{3}\\
m & j_{m,\bar{k}-\bar{p}_{m}}^{4} & \pi_{m,\bar{k}}^{3}
\end{array}\right\}
\Biggl|
\begin{array}{c}\includegraphics[scale=0.7]{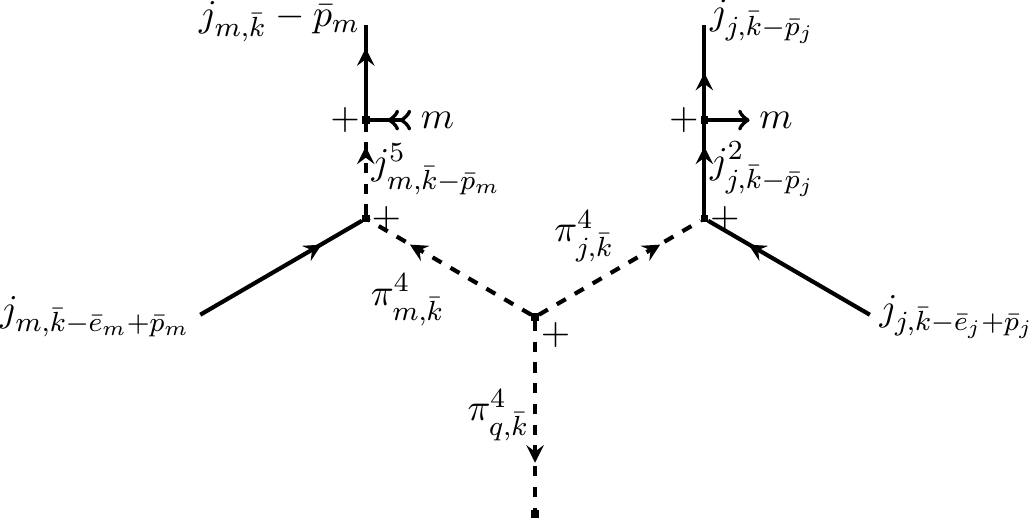}\end{array}
\Biggr\rangle =
\]
\[
=\underset{\begin{array}{c}
j_{m,\bar{k}-\bar{p}_{m}}^{2\ldots4},j_{j,\bar{k}-\bar{p}_{j}}^{2}\\
\left\{ \pi\right\} _{\bar{k}}^{2},\pi_{m,\bar{k}}^{3},\pi_{j,\bar{k}}^{3}, \left\{ \pi\right\} _{\bar{k}}^{4}\\
j_{m,\bar{k}-\bar{p}_{m}}^{5}, \pi^5_{m,\bar{k}},\pi^5_{j,\bar{k}}
\end{array}}{\tilde{\sum}}
\left(-\right)^{\sigma\left(m,j,q\right)\left(\pi_{j,\bar{k}}^{2}+\pi_{m,\bar{k}}^{2}+2\pi_{q,\bar{k}}^{2}+\pi_{j,\bar{k}}^{3}+\pi_{m,\bar{k}}^{3}+\pi_{j,\bar{k}}^{4}+\pi_{m,\bar{k}}^{4}+2\pi_{q,\bar{k}}^{4}+\pi_{m,\bar{k}}^{5}+\pi_{j,\bar{k}}^{5}\right)}
\]
\[
\left(-\right)^{\left|\bar{p}_{m}\right|\left(2j_{m,\bar{k}-\bar{p}_{m}}+j_{m,\bar{k}-\bar{p}_{m}}^{4}+j_{m,\bar{k}-\bar{p}_{m}}^{5}+\pi_{m,\bar{k}}+\pi_{m,\bar{k}}^{3}+\pi_{m,\bar{k}}^{4}+\pi_{m,\bar{k}}^{5}\right)}\left(-\right)^{\left|\bar{p_{j}}\right|\left(\pi_{j,\bar{k}}+\pi_{j,\bar{k}}^{3}+\pi_{j,\bar{k}}^{4}+\pi_{j,\bar{k}}^{5}+2m\right)}\]
\[
(-)^{2\pi^4_{q,\bar{k}}+\pi^5_{j,\bar{k}}+\pi^4_{m,\bar{k}}+\pi^4_m,\bar{k}+\pi^4_{j,\bar{k}}+\pi^4_{q,\bar{k}}+\pi^3_{m,\bar{k}}+\pi^3_{j,\bar{k}}+\pi^2_{m,\bar{k}}+\pi^2_{m,\bar{k}}+\pi^2_{j,\bar{k}}+\pi^2_{q,\bar{k}}}
\]
\[
 \sqrt{V}_{\bar{k}+\bar{e}_{m}-2\bar{p}_{m}}\left(\left\{ \pi\right\} _{\bar{k}},\left\{ \pi\right\} _{\bar{k}}^{2};j_{m,\bar{k}-\bar{p}_{m}},j_{m,\bar{k}-\bar{p}_{m}}^{3};\ldots j_{m,\bar{k}-\bar{p}_{m}}^{2}\ldots\mid m\right)
\]
\[
 \sqrt{V}_{\bar{k}+\bar{e}_{m}-2\bar{p}_{m}}\left(\pi_{m,\bar{k}}^{3},\pi_{j,\bar{k}}^{3},\pi_{q,\bar{k}}^{2},\left\{ \pi\right\} _{\bar{k}}^{4};j_{m,\bar{k}-\bar{p}_{m}}^{4},j_{m,\bar{k}-\bar{p}_{m}}^{5};\ldots j_{j,\bar{k}-\bar{p}_{m}}^{2}\ldots\mid	m\right)
\]
\[
\left\{ \begin{array}{ccc}
\pi_{j,\bar{k}}^{3} & \pi_{j,\bar{k}}^{2} & m\\
j_{j,\bar{k}-\bar{p}_{j}} & j_{j,\bar{k}-\bar{p}_{j}}^{2} & j_{j,\bar{k}+\bar{e}_{j}-\bar{p}_{j}}
\end{array}\right\} 
\left\{ \begin{array}{ccc}
\pi_{q,\bar{k}}^{2} & \pi_{j,\bar{k}}^{2} & \pi_{m,\bar{k}}^{2}\\
m & \pi_{m,\bar{k}}^{3} & \pi_{j,\bar{k}}^{3}
\end{array}\right\}
\left\{ \begin{array}{ccc}
m & j_{m,\bar{k}-\bar{p}_{m}}^{3} & j_{m,\bar{k}-\bar{p}_{m}}\\
m & j_{m,\bar{k}-\bar{p}_{m}} & j_{m,\bar{k}-\bar{p}_{m}}^{4}
\end{array}\right\} 
\]
\[\left\{ \begin{array}{ccc}
j_{m,\bar{k}-\bar{e}_{m}+\bar{p}_{m}} & \pi_{m,\bar{k}}^{2} & j_{m,\bar{k}-\bar{p}_{m}}^{3}\\
m & j_{m,\bar{k}-\bar{p}_{m}}^{4} & \pi_{m,\bar{k}}^{3}
\end{array}\right\}
\left\{ \begin{array}{ccc}
j_{m,\bar{k}-\bar{e}_{m}+\bar{p}_{m}} & \pi_{m,\bar{k}}^{4} & j_{m,\bar{k}-\bar{p}_{m}}^{5}\\
m & j_{m,\bar{k}-\bar{p}_{m}} & \pi_{m,\bar{k}}^{5}
\end{array}\right\}
\left\{ \begin{array}{ccc}
\pi_{m,\bar{k}}^{5} & \pi_{m,\bar{k}}^{4} & m\\
\pi_{j,\bar{k}}^{4} & \pi_{j,\bar{k}}^{5} & \pi_{q,\bar{k}}^{4}
\end{array}\right\}
\]
\[
\left\{ \begin{array}{ccc}
\pi_{j,\bar{k}}^{5} & \pi_{j,\bar{k}}^{4} & m\\
j_{j,\bar{k}-\bar{p}_{j}}^{2} & j_{j,\bar{k}-\bar{p}_{j}} & j_{j,\bar{k}-\bar{e}_{j}+\bar{p}_{j}}
\end{array}\right\}
\left|\nu\left(\pi_{m,\bar{k}}^{5},\pi_{j,\bar{k}}^{5},\pi_{q,\bar{k}}^{4};\left\{ j\right\} _{\bar{k}}\right)\right\rangle 
\]
\subsection{Gluon electric Fluxes of the Electric Term}
\label{c5.2}
The Electric Part of the Hamiltonian is

\[
\underline{\hat{E}}_I(e_1)\underline{\hat{E}}_I(e_2)
\]
\\
where $e_1$ and $e_2$ correspond again to all possible tuples of edges incident at a vertex $v$.  The Electric Fluxes $\underline{\hat{E}}_{I}(j)$ themselves are the grasping operators, whose action on a group element has been defined in (\ref{2.9}). The operator adds a generator of the Lie-Algebra, which can be viewed as a new intertwiner on the holonomy in the defining (i.e. $\underline{j}=1$) representation. Hence the action is determined up to a normalization factor, which depends on the gauge group and possibly also on the multiplicity-factor corresponding to the chosen intertwiner. However it is easy to check, that when choosing an arbitrary $\underline{s}_I$ multiplicity everywhere, the normalization does not depend on it and becomes $N^{(j)}=\sqrt{C_2(j)d_j}$ (the computation for this is, in principle, the same as in \cite{AR07}). Writing everything down in our graphical calculus:

\begin{equation}
\underline{\hat{E}}(j)
	\begin{array}{c}\includegraphics[scale=1]{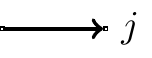}\end{array}
=i\sqrt{C_{2}\left(j\right)d_{j}}
\begin{array}{c}\includegraphics[scale=1]{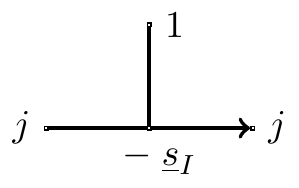}\end{array}
\end{equation}
\\
With this at hand we turn again to the three cases i)-iii) from \ref{c5.2}:
However, due to the nature of the SU(3) gauge group, one cannot obtain a node with all edges outgoing  by simply multiplying it with a sign factor. Instead, one now has to take care of the fact, that the switched edges carry the dual representation. So one works in the following with an oriented graph, denoted the following way:

\[
\left|\nu_{orient}\left(\underline{j}_{1,\bar{k}},\underline{j}_{2,\bar{k}},\underline{j}_{3,\bar{k}},\bar{\underline{j}}_{1,\bar{k}-\bar{e}_{1}}\underline{\bar{j}}_{2,\bar{k}-\bar{e}_{2}}\underline{\bar{j}}_{3,\bar{k}-\bar{e}_{3}},\ldots\right)\right\rangle =\left|\nu\left(\underline{j}_{1,\bar{k}},\underline{j}_{2,\bar{k}},\underline{j}_{3,\bar{k}},\underline{j}_{1,\bar{k}-\bar{e}_{1}}\underline{j}_{2,\bar{k}-\bar{e}_{2}}\underline{j}_{3,\bar{k}-\bar{e}_{3}},\ldots\right)\right\rangle
\]
\\
The first case i) $\left(\underline{j}_{j,\bar{k}}=\underline{j}_{m,\bar{k}}\right)$ means, that both grasping operators act on the same edge, hence we get twice the square root of the corresponding quadric Casimir and using the orthogonality relation (\ref{3.38-bubble}) one calculates:

\[
\underline{\hat{E}}(\underline{j}_j)^I\underline{\hat{E}}_(\underline{j}_j)^{I}\left|\nu_{orient}\left(\left\{ \underline{\pi}\right\} _{\bar{k}};\left\{ \underline{j}\right\} _{\bar{k}};\left\{ \underline{s}\right\} _{\bar{k}}\right)\right\rangle =-C_{2}\left(\underline{j}_{j,\bar{k}-\bar{p}_{j}}\right)\Biggl|
\begin{array}{c}\includegraphics[scale=0.6]{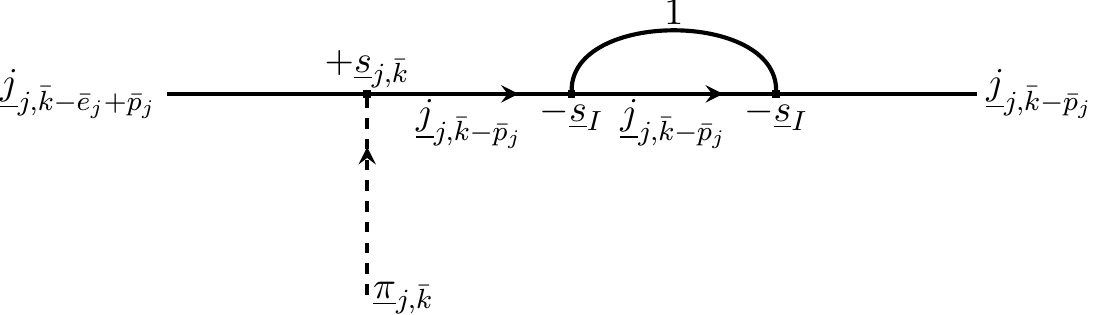}\end{array}
\Biggr\rangle 
\]
\[
=C_{2}\left(\underline{j}_{j,\bar{k}-\bar{p}_{j}}\right)\left|\nu_{orient}\left(\left\{ \underline{\pi}\right\} _{\bar{k}};\left\{ \underline{j}\right\} _{\bar{k}};\left\{ \underline{s}\right\} _{\bar{k}}\right)\right\rangle 
\]
\\
The second case ii), where  the edges in question lie in parallel direction $\left(\underline{j}_{j,\bar{k}}=\underline{j}_{m,\bar{k}-\bar{e}_{m}}\right)$
uses again the extraction of the s-classified 3j-symbol and thus one gets:

\[
\underline{\hat{E}}(\underline{j}_{j,\bar{k}-\bar{e}_{j}+\bar{p}_{j}})^{I}\underline{\hat{E}}(\underline{j}_{j,\bar{k}-\bar{p}_{j}})^{I}\left|\nu_{orient}\left(\left\{ \underline{\pi}\right\} _{\bar{k}};\left\{ \underline{j}\right\} _{\bar{k}};\left\{ \underline{s}\right\} _{\bar{k}}\right)\right\rangle =
\sqrt{C_{2}\left(\underline{j}_{j,\bar{k}-\bar{p}_{j}}\right)C_{2}\left(\underline{j}_{j,\bar{k}-\bar{e}_{j}+\bar{p}_{j}}\right)} \cdot
\]
\[
\cdot\underset{\underline{s}_{j,\bar{k}}'}{\sum}
\left\{ \begin{array}{cccc}
\underline{\bar{j}}_{j,\bar{k}-\bar{e}_i+\bar{p}_j} & \underline{\pi}_{j,\bar{k}} & \underline{\bar{j}}_{j,\bar{k}-\bar{p}_j} \\
\underline{j}_{j,\bar{k}-\bar{p}_j} & 1 & \underline{\bar{j}}_{j,\bar{k}-\bar{e}_j+\bar{p}_j} \\
\underline{s}'_{j,\bar{k}} & \underline{s}_I & \underline{s}_{j,\bar{k}} & \underline{s}_I
\end{array}\right\} \left|\nu_{orient}\left(\left\{ \underline{\pi}\right\} _{\bar{k}},\left\{ \underline{j}\right\} _{\bar{k}},\ldots\underline{s}_{j,\bar{k}}'\ldots\right)\right\rangle 
\]
\\
Lastly we look at iii), where both holonomies go in different directions. With the same strategy as before, we see:

\[
\underline{\hat{E}}_{j}^{I}\underline{\hat{E}}_{m}^{I}\left|\nu_{out}\left(\left\{ \underline{\pi}\right\} _{\bar{k}};\left\{ \underline{j}\right\} _{\bar{k}};\left\{ \underline{s}\right\} _{\bar{k}}\right)\right\rangle
=
\underset{\begin{array}{c}
\underline{\bar{\pi}}_{m,\bar{k}}^{2}\underline{\bar{\pi}}_{j,\bar{k}}^{2}\\
\underline{s}_{\pi_{j,\bar{k}}}\underline{s}_{\pi_{m,\bar{k}}}
\end{array}}{\tilde{\sum}}
\left(-\right)^{
\left(1-\left|\bar{p}_{j}\right|\right)\left(\underline{\pi}_{j,\bar{k}}+\underline{j}_{j,\bar{k}}+\underline{j}_{j,\bar{k}-\bar{e}_{j}}\right)+\left|\bar{p}_{m}\right|\left(\underline{\pi}_{m,\bar{k}}+\underline{j}_{m,\bar{k}}+\underline{j}_{m,\bar{k}-\bar{e}_{m}}\right)}\cdot
\]
\[
\left(-\right)^{\sigma\left(m,j,q\right)\left(\underline{\pi}_{j,\bar{k}}+\underline{\pi}_{m,\bar{k}}+\underline{\pi}_{q,\bar{k}}\right)}
\sqrt{C_{2}\left(\underline{j}_{m,\bar{k}-\bar{p}_{m}}\right)C_{2}\left(\underline{j}_{j,\bar{k}-\bar{p}_{j}}\right)}
\Biggl|
\begin{array}{c}\includegraphics[scale=0.7]{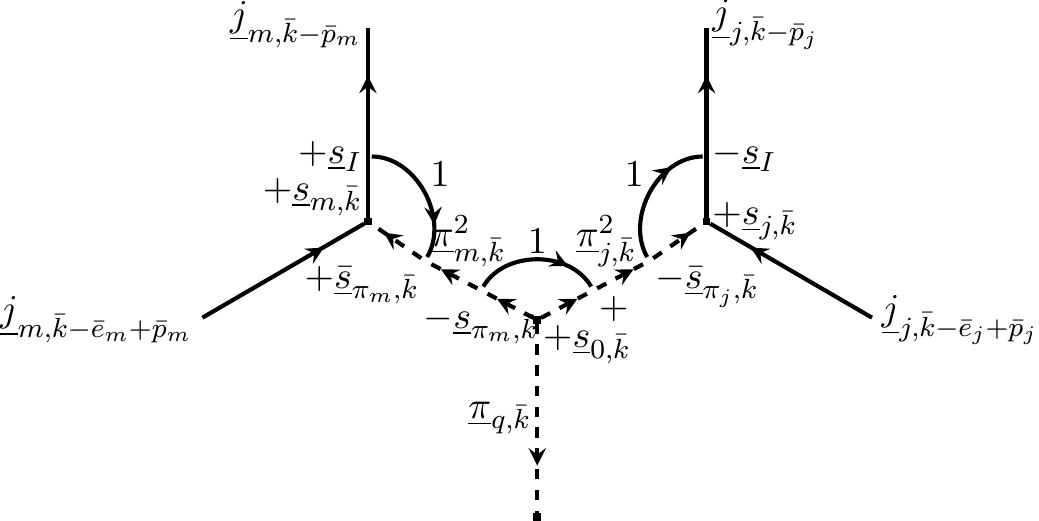}\end{array}
\Biggr\rangle 
\]
\[
=\underset{\begin{array}{c}
\underline{\pi}_{m,\bar{k}}^{2}\underline{\pi}_{j,\bar{k}}^{2}\\
\underline{s}_{\pi_{j,\bar{k}}}\underline{s}_{\pi_{m,\bar{k}}}\\
\underline{s}_{0,\bar{k}}'\underline{s}_{m,\bar{k}}'\underline{s}_{j,\bar{k}}'
\end{array}}{\sum}\left(-\right)^{\sigma\left(m,j,q\right)\left(\underline{\pi}_{m,\bar{k}}^{2}+\underline{\pi}_{j,\bar{k}}^{2}+\underline{\pi}_{m,\bar{k}}+\underline{\pi}_{j,\bar{k}}\right)+\left(1-\left|\bar{p}_{j}\right|\right)\left(\underline{\pi}_{j,\bar{k}}+\underline{\pi}_{j,\bar{k}}^{2}\right)+\left|\bar{p}_{m}\right|\left(\underline{\pi}_{m,\bar{k}}+\underline{\pi}_{m,\bar{k}}^{2}\right)}
(-)^{\pi^3_{j,\bar{k}}+\pi_{m,\bar{k}}+1}
\]
\[
\sqrt{C_{2}\left(\underline{j}_{m,\bar{k}-\bar{p}_{m}}\right)C_{2}\left(\underline{j}_{j,\bar{k}-\bar{p}_{j}}\right)}
\left\{\begin{array}{cccc}
\underline{\bar{j}_{m,\bar{k}-\bar{e}_m+\bar{p}_m}} & \underline{\pi}^2_{m,\bar{k}} & \underline{\bar{j}}_{m,\bar{k}-\bar{p}_m} \\
1 & \underline{\bar{j}}_{m,\bar{k}-\bar{p}_m} & \underline{\bar{\pi}}_{m,\bar{k}} \\
\underline{s}'_{m,\bar{k}} & s_{m,\bar{k}} & \underline{s}_{\pi_m,\bar{k}} & \underline{s}_I
\end{array}\right\}
\left\{\begin{array}{cccc}
\underline{\bar{\pi}}^2_{m,\bar{k}} & \underline{\bar{\pi}}_{q,\bar{k}} & \underline{\bar{\pi}}^2_{j,\bar{k}} \\
\underline{\pi}_{j,\bar{k}} & 1 & \underline{\pi}_{m,\bar{k}} \\
\underline{s}'_{0,\bar{k}} & \underline{s}_{\pi_m,\bar{k}} & s_{0,\bar{k}} & s_{\pi_j,\bar{k}}
\end{array}\right\}
\]
\[
\left\{\begin{array}{cccc}
\underline{\pi}^2_{j,\bar{k}} & \underline{\bar{j}}_{j,\bar{k}-\bar{e}_j+\bar{p}_j} & \underline{\bar{j}}_{\bar{k}-\bar{p}_j} \\
\underline{j}_{j,\bar{k}-\bar{p}_j} & 1 & \underline{\pi}_{j,\bar{k}} \\
\underline{s}'_{j,\bar{k}} & \underline{s}_{\pi_j,\bar{k}} & s_{j,\bar{k}} & \underline{s}_I
\end{array}\right\}
\left|\nu_{orient}\left(\{\underline{\pi}\}_{\bar{k}};\{\underline{j}\};\underline{s}'_{0,\bar{k}},\underline{s}'_{j,\bar{k}},\underline{s}'_{m,\bar{k}},\underline{s}_{q,\bar{k}}\right)\right\rangle
\]
\subsection{Gravity-Part of the Magnetic Term}
The Gravity-Part of the Magnetic Term is

\[
tr\left(\hat{\tau}_{i}\hat{A}_{l}\left[\hat{A}_{l}^{-1},\sqrt{\hat{V}}\right]\right)tr\left(\hat{\tau}_{i}\hat{A}_{p}\left[\hat{A}_{p}^{-1},\sqrt{\hat{V}}\right]\right)
\]
\\
Since there are again two commutators we have, in principle, four different terms to look at. However three of them vanish trivially. For example look at the expression, where the $\hat{A}_{p}$ cancel:

\[
tr\left(\hat{\tau}_{i}\hat{A}_{l}\left[\hat{A}_{l}^{-1},\sqrt{\hat{V}}\right]\right)tr\left(\hat{\tau}_{i}\sqrt{\hat{V}}\right)\left|\nu_{out}\left(\left\{ \pi\right\} _{\bar{k}};\left\{ j\right\} _{\bar{k}}\right)\right\rangle =
\]
\[
tr\left(\hat{\tau}_{i}\hat{A}_{l}\left[\hat{A}_{l}^{-1},\sqrt{\hat{V}}\right]\right)tr\left(\hat{\tau}_{i}\right)\underset{\left\{ \pi\right\} _{\bar{k}}^{2}}{\tilde{\sum}}\sqrt{V}_{\bar{k}}\left(\left\{ \pi\right\} _{\bar{k}},\left\{ \pi\right\} _{\bar{k}}^{2};\left\{ j\right\} _{\bar{k}}\right)\left|\nu_{out}\left(\left\{ \pi\right\} _{\bar{k}}^{2};\left\{ j\right\} _{\bar{k}}\right)\right\rangle =0
\]
\\
since $tr\left(\hat{\tau}_{i}\right)=0$ for $\tau_{i}\in SU(2)$. The same argument is of course also true, in case of the $A_l$ canceling.

Thus only the term with both Volume operators nested remains.  Again we distinguish on which edges the holonomies lie (cases i)-iii) from section \ref{c5.1}).
Since one has seen that the orientation of the arrows of the edges does not change the result, we will suppress this temporary sign from now on and just assume the vertex has been brought in a form such that all links are outgoing.
 If i) $\left(j_{p,\bar{k}}=j_{l,\bar{k}}\right)$
then one gets from the first trace an 6j-symbol and the inserted $\hat{\tau}_i$ acts like adding an intertwiner in the defining representation, which hence remains open, after closing the first trace. To close the second one, one uses again (\ref{3.38-bubble}) twice. In total one obtains:

\[
tr\left(\hat{\tau}_{i}\hat{A}_{p}\sqrt{\hat{V}}\hat{A}_{p}^{-1}\right)tr\left(\hat{\tau}_{i}\hat{A}_{p}\sqrt{\hat{V}}\hat{A}_{p}^{-1}\right)\left|\nu\left(\left\{ \pi\right\} _{\bar{k}};\left\{ j\right\} _{\bar{k}}\right)\right\rangle =tr\left(\hat{\tau}_{i}\hat{A}_{p}\sqrt{\hat{V}}\hat{A}_{p}^{-1}\right)\cdot
\]
\[
\underset{\begin{array}{c}
\left\{ \pi\right\} _{\bar{k}}^{2}j_{p,\bar{k}-\bar{p}_{p}}^{2}\\
j_{p,\bar{k}-\bar{p}_{p}}^{3}
\end{array}}{\tilde{\sum}}\sqrt{V}_{\bar{k}-\bar{e}_{p}+2\bar{p}_{p}}\left(\left\{ \pi\right\} _{\bar{k}},\left\{ \pi\right\} _{\bar{k}}^{2};j_{p,\bar{k}-\bar{p}_{p}},j_{p,\bar{k}-\bar{p}_{p}}^{3};\ldots j_{p,\bar{k}-\bar{p}_{p}}^{2}\dots\right)
\left\{ \begin{array}{ccc}
j_{p,\bar{k}-\bar{p}_{p}}^{3} & j_{p,\bar{k}-\bar{p}_{p}}^{2} & m\\
m & 1 & j_{p,\bar{k}-\bar{p}_{p}}
\end{array}\right\} 
\]
\[
\left(-\right)^{\left|\bar{p_{p}}\right|\left(\pi_{p,\bar{k}}^{2}+j_{p,\bar{k}-\bar{p}_{p}}^{3}+j_{p,\bar{k}-\bar{e}_{p}+\bar{p}_{p}}\right)+2m+j_{p,\bar{k}-\bar{p}_{p}}^{2}+j_{p,\bar{k}-\bar{p}_{p}}+m+1}\Biggl|
\begin{array}{c}\includegraphics[scale=0.8]{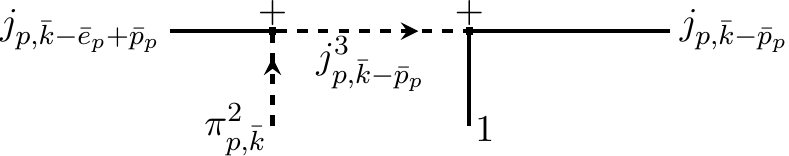}\end{array}
\Biggr\rangle 
\]
\[
=\underset{\begin{array}{c}
\left\{ \pi\right\} _{\bar{k}}^{2}j_{p,\bar{k}-\bar{p}_{p},m_{1}}^{2}\\
\left\{ \pi\right\} _{\bar{k}}^{3}j_{p,\bar{k}-\bar{p}_{p}}^{3}\\
j_{p,\bar{k}-\bar{p}_{p}}^{4}
\end{array}}{\tilde{\sum}}\sqrt{V}_{\bar{k}-\bar{e}_{p}+2\bar{p}_{p}}\left(\left\{ \pi\right\} _{\bar{k}},\left\{ \pi\right\} _{\bar{k}}^{2};j_{p,\bar{k}-\bar{p}_{p}},j_{p,\bar{k}-\bar{p}_{p}}^{3};\ldots j_{p,\bar{k}-\bar{p}_{p}}^{2}\dots\mid	m\right)
\]
\[
\sqrt{V}_{\bar{k}-\bar{e}_{p}+2\bar{p}_{p}}\left(\left\{ \pi\right\} _{\bar{k}}^{2},\left\{ \pi\right\} _{\bar{k}}^{3};j_{p,\bar{k}-\bar{p}_{p}}^{3},j_{p,\bar{k}-\bar{p}_{p}};\ldots j_{p,\bar{k}-\bar{p}_{p}}^{4}\dots\mid m\right)
(-)^{2m+1+j^2_{p,\bar{k}-\bar{p}_p}+j^4_{p,\bar{k}-\bar{p}_p}}
\]
\[
\left\{ \begin{array}{ccc}
j_{p,\bar{k}-\bar{p}_{p}}^{3} & m & j_{p,\bar{k}-\bar{p}_{p}}^{2}\\
m & j_{p,\bar{k}-\bar{p}_{p}}^{2} & 1
\end{array}\right\} \left\{ \begin{array}{ccc}
j_{p,\bar{k}-\bar{p}_{p}}^{3} & 1 & j_{p,\bar{k}-\bar{p}_p} \\
m & j^4_{,\bar{k}-\bar{p}_p} & m
\end{array}\right\}
\left|\nu\left(\left\{ \pi\right\} _{\bar{k}}^{3};\left\{ j\right\} _{\bar{k}}\right)\right\rangle 
\]
\\
With the same methods as established before, we get for case ii), meaning both links go in parallel direction  $\left(j_{p,\bar{k}}=j_{l,\bar{k}-\bar{e}_{l}}\right)$, that:

\[
tr\left(\hat{\tau}_{i}\hat{A}_{p,\bar{p}_{p}}\sqrt{\hat{V}}\hat{A}_{p,\bar{p}_{p}}^{-1}\right)tr\left(\hat{\tau}_{i}\hat{A}_{p,\bar{p}_{p}-\bar{e}_{p}}\sqrt{\hat{V}}\hat{A}_{p,\bar{p}_{p}-\bar{e}_{p}}^{-1}\right)\left|
\nu\left(\left\{ \pi\right\} _{\bar{k}};\left\{ j\right\} _{\bar{k}}\right)\right\rangle=
\underset{\begin{array}{c}
\left\{ \pi\right\} _{\bar{k}}^{2}\left\{ \pi\right\} _{\bar{k}}^{3}\\
j_{p,\bar{k}-\bar{e}_{p}+\bar{p}_{p}}^{2}j_{p,\bar{k}-\bar{e}_{p}+\bar{p}_{p}}^{3}\\
j_{p,\bar{k}-\bar{p}_{p}}^{2}j_{p,\bar{k}-\bar{p}_{p}}^{3}
\end{array}}{\tilde{\sum}}
\]
\[
\left(-\right)^{\left|\bar{p}_{p}\right|\left(2\pi_{p,\bar{k}}^{3}+j_{p,\bar{k}-\bar{p}_{p}}^{3}+j_{p,\bar{k}-\bar{p}_{p}}^{5}+j_{p,\bar{k}-\bar{e}_{p}+\bar{p}}+j_{p,\bar{k}-\bar{e_{p}+}\bar{p}}^{3}\right)}
(-)^{j^2_{p,\bar{k}-\bar{p}_p}+j_{p,\bar{k}-\bar{e}_p+\bar{p}_p}+j^2_{p,\bar{k}-\bar{e}_p+\bar{p}_p}+j^3_{p,\bar{k}-\bar{e}_p+\bar{p}_p}+\pi^2_{p,\bar{k}}+1}\]
\[
\left\{ \begin{array}{ccc}
j_{p,\bar{k}-\bar{p}_{p}}^{3} & j_{p,\bar{k}-\bar{p}_{p}} & 1\\
m & m & j_{p,\bar{k}-\bar{p}_{p}}^{2}\end{array}\right\}
\left\{ \begin{array}{ccc}
j_{p,\bar{k}-\bar{e}_{p}+\bar{p}_{p}} & j_{p,\bar{k}-\bar{e}_{p}+\bar{p}_{p}}^{3} & 1\\
j_{p,\bar{k}-\bar{p}_{p}} & j_{p,\bar{k}-\bar{p}_{p}}^{3} & \pi_{p,\bar{k}}^{2}
\end{array}\right\}
\left\{\begin{array}{ccc}
j_{p,\bar{k}-\bar{e}_p+\bar{p}_p} & j^2_{p,\bar{k}-\bar{e}_p+\bar{p}_p} & m \\
m & 1 & j^3_{p,\bar{k}-\bar{e}_p+\bar{p}_p}
\end{array}\right\}
 \]
\[
\sqrt{V}_{\bar{k}-2\bar{p}_{p}}\left(\left\{ \pi\right\} _{\bar{k}},\left\{ \pi\right\} _{\bar{k}}^{2};j_{p,\bar{k}-\bar{e}_{p}+\bar{p}_{p}},j_{p,\bar{k}-\bar{e}_{p}+\bar{p}_{p}}^{3};\ldots j_{p,\bar{k}-\bar{e}_{p}+\bar{p}_{p}}^{2}\ldots\right)
\]
\[
\sqrt{V}_{\bar{k}+\bar{e}_{p}-2\bar{p}_{p}}\left(\left\{ \pi\right\} _{\bar{k}}^{2},\left\{ \pi\right\} _{\bar{k}}^{3};j_{p,\bar{k}-\bar{p}_{p}}^{3},j_{p,\bar{k}-\bar{p}_{p}}^{4};\ldots j_{p,\bar{k}-\bar{p}_{p}}^{2}\ldots\right)
 \left|\nu\left(\{\pi\}^3_{\bar{k}},\{j\}_{\bar{k}}\right)\right\rangle
\]
\\
And finally with more suppressed calculation, it follows iii)  (both holonomies go into different directions):

\[
tr\left(\hat{\tau}_{i}\hat{A}_{l}\sqrt{\hat{V}}\hat{A}_{l}^{-1}\right)tr\left(\hat{\tau}_{i}\hat{A}_{p}\sqrt{\hat{V}}\hat{A}_{p}^{-1}\right)\left|\nu\left(\left\{ \pi\right\} _{\bar{k}};\left\{ j\right\} _{\bar{k}}\right)\right\rangle =
\]
\[
=\underset{\begin{array}{c}
\left\{ \pi\right\} _{\bar{k}}^{2}\pi_{p,\bar{k}}^{3}\pi_{l,\bar{k}}^{3}\left\{ \pi\right\} _{\bar{k}}^{4}\\
j_{p,\bar{k}-\bar{p}_{p}}^{2}j_{p,\bar{k}-\bar{p}_{p}}^{3}j_{l,\bar{k}-\bar{p}_{l}}^{2}j_{l,\bar{k}-\bar{p}_{l}}^{3}
\end{array}}{\tilde{\sum}}\sqrt{V}_{\bar{k}+\bar{e}_{p}-2\bar{p}_{p}}\left(\left\{ \pi\right\} _{\bar{k}},\left\{ \pi\right\} _{\bar{k}}^{2};j_{p,\bar{k}-\bar{p}_{p}},j_{p,\bar{k}-\bar{p}_{p}}^{3};\ldots j_{p,\bar{k}-\bar{p}_{p}}^{2}\ldots\right)
\]
\[\sqrt{V}_{\bar{k}-\bar{e}_{l}-2\bar{p}_{l}}\left(\pi_{l,\bar{k}}^{3},\pi_{p,\bar{k}}^{3},\pi_{q,\bar{k}}^{2},\left\{ \pi\right\} _{\bar{k}}^{4};j_{l,\bar{k}-\bar{p}_{l}}^{3},j_{l,\bar{k}-\bar{p}_{l}};\ldots j_{l,\bar{k}-\bar{p}_{l}}^{2}\ldots\right)
\left(-\right)^{\sigma\left(p,l,q\right)\left(\pi_{p,\bar{k}}^{2}+\pi_{l,\bar{k}}^{2}+2\pi_{q,\bar{k}}^{2}+\pi_{p,\bar{k}}^{3}+\pi_{l,\bar{k}}^{3}\right)}
\]
\[
(-)^{\left|\bar{p}_{p}\right|\left(\pi^2_{p,\bar{k}}+j^2_{p,\bar{k}-\bar{p}_p}+2j_{p,\bar{k}-\bar{e}_p+\bar{p}_p}+j_{l,\bar{k}-\bar{p}_l}+\pi^3_{l,\bar{k}}\right)+\left(1-\left|\bar{p}_l\right|\right)\left(\pi^2_{l,\bar{k}}+j_{l,\bar{k}-\bar{p}_l}+2j_{l,\bar{k}-\bar{e}_l+\bar{p}_l}+\pi^3_{l,\bar{k}}+j^3_{l,\bar{k}-\bar{p}_l}+\pi^4_{l,\bar{k}}+j_{l,\bar{k}-\bar{p}_l}+j_{l,\bar{k}-\bar{e}_l+\bar{p}_l}\right)}
\]
\[
(-)^{2m+m+j_{p,\bar{k}-\bar{p}_p}+j^2_{p,\bar{k}-\bar{p}_p}+j^3_{p,\bar{k}-\bar{p}_p}+j_{p,\bar{k}-\bar{e}_p+\bar{p}_p}-j_{l,\bar{k}-\bar{e}_l+\bar{p}_l}+j^3_{l,\bar{k}-\bar{p}_l}+\pi^2_{p,\bar{k}}+\pi^3_{p,\bar{k}}+\pi^2_{q,\bar{k}}+\pi^2_{l,\bar{k}}+\pi^3_{l,\bar{k}}}
\]
\[
\left\{ \begin{array}{ccc}
1 & m & m\\
j_{p,\bar{k}-\bar{p}_{p}}^{2} & j_{p,\bar{k}-\bar{p}_{p}} & j_{p,\bar{k}-\bar{p}_{p}}^{3}
\end{array}\right\}
\left\{ \begin{array}{ccc}
j_{p,\bar{k}-\bar{e}_{p}+\bar{p}_{p}} & \pi_{p,\bar{k}}^{2} & j_{p,\bar{k}-\bar{p}_{p}}^{3}\\
1 & j_{p,\bar{k}-\bar{p}_{p}} & \pi_{p,\bar{k}}^{3}
\end{array}\right\}
\]
\[
 \left\{ \begin{array}{ccc}
\pi_{p,\bar{k}}^{3} & \pi_{p,\bar{k}}^{2} & 1\\
\pi_{l,\bar{k}}^{2} & \pi_{l,\bar{k}}^{3} & \pi_{q,\bar{k}}^{2}
\end{array}\right\} 
\left\{ \begin{array}{ccc}
\pi_{l,\bar{k}}^{3} & \pi_{l,\bar{k}}^{2} & 1\\
j_{l,\bar{k}-\bar{p}_{l}} & j_{l,\bar{k}-\bar{p}_{l}}^{3} & j_{l,\bar{k}-\bar{e}_{l}+\bar{p}_{l}}
\end{array}\right\} \left\{ \begin{array}{ccc}
m_{1} & 1 & m\\
j_{l,\bar{k}-\bar{p}_{l}} & j_{l,\bar{k}-\bar{p}_{l}}^{2} & j_{l,\bar{k}-\bar{p}_{l}}^{3}
\end{array}\right\} \left|\nu\left(\left\{ \pi\right\} _{\bar{k}}^{4},\left\{ j\right\} _{\bar{k}}\right)\right\rangle 
\]
\subsection{Gluons plaquette of the Magnetic Term}
The plaquette part is given by

\[
tr\left(\underline{\hat{\tau}}_I\underline{\hat{A}}_{jk}\right)
tr\left(\underline{\hat{\tau}}\underline{\hat{A}}_{mn}\right)
\]
\\
which again acts only on the magnetic graph. Each of these two plaquettes, which we add, looks very similar in its structure to chapter \ref{c4}. Using this resemblance and inserting again the corresponding plaquette terms $\mathfrak{P}_{SU(3)}$ will simplify the task at hand. Again one has to distinguish different cases, i.e. determined by the possible combinations of $j,k,m$ and $n$.

The most simple one is $j=m$ and $k=n$:

\[
tr\left(\underline{\hat{\tau}}_{I}\underline{\hat{A}}_{mn}\right)tr\left(\underline{\hat{\tau}}_{I}\underline{\hat{A}}_{mn}\right)\left|\nu_{orient}\left(\left\{ \underline{\pi}\right\} _{\bar{k}};\left\{ \underline{j}\right\} \right)\right\rangle =\left(-\right)^{\sigma\left(n,m,p\right)\left(\underline{\pi}_{m,\bar{k}}+\underline{\pi}_{n,\bar{k}}+\underline{\pi}_{p,\bar{k}}\right)}
\]
\[
\left(-\right)^{\left|\bar{p}_{n}\right|\left(\underline{\pi}_{n,\bar{k}}+\underline{j}_{n,\bar{k}-\bar{p}_{n}}+\underline{j}_{n,\bar{k}-\bar{e}_{n}+\bar{p}_{n}}\right)+\left(1-\left|\bar{p}_{m}\right|\right)\left(\underline{\pi}_{m,\bar{k}}+\underline{j}_{m,\bar{k}}+\underline{j}_{m,\bar{k}-\bar{e}_{m}}\right)}\Biggl|
\begin{array}{c}\includegraphics[scale=0.7]{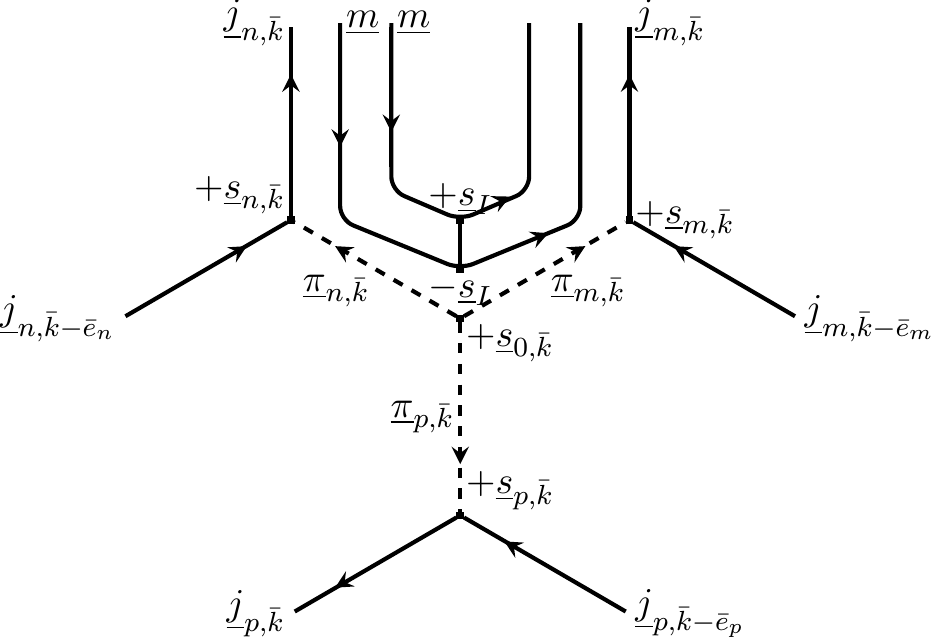}\end{array}
\Biggr\rangle 
\]
\[
=\left(-\right)^{\sigma\left(n,m,p\right)\left(\ldots\right)+\left|\bar{p}_{n}\right|\left(\ldots\right)+\left(1-\left|\bar{p}_{m}\right|\right)\left(\dots\right)}
\]
\[
\underset{\underline{m}_{1},\underline{s}}{\sum}\left(-\right)^{1}\left\{ \begin{array}{cccc}
\underline{m} & \underline{m} & \underline{\bar{m}}_1\\
1 & \underline{m} & \underline{\bar{m}} \\
\underline{\bar{s}} & \underline{s}_I & \underline{s}_I & \underline{s}
\end{array}\right\} \Biggl|
\begin{array}{c}\includegraphics[scale=0.7]{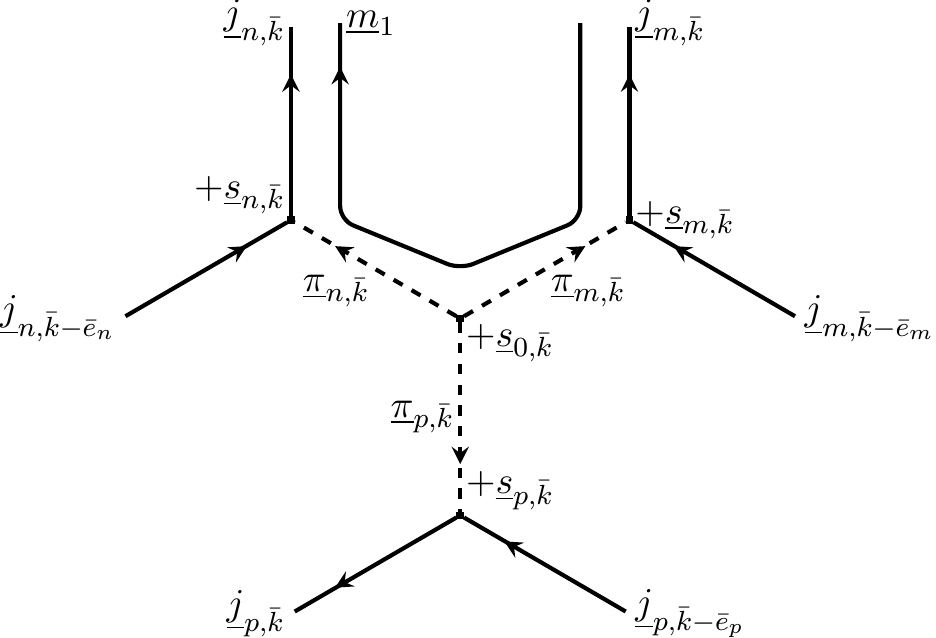}\end{array}
\Biggr\rangle 
\]
\\
Now we have exactly the same plaquette we inserted in the Kogut Susskind case. To extract exactly the same term again we have to bring the graph in an ordered form, which means we have to take care of the fact, that the Loop also touches four other nodes. In contrast to the Kogut-Susskind case these signs of the intertwiners now only depend on the chosen permutation of $n,m,p$ which means that we get a somewhat more complicated sign factor in front:

\[
=\underset{\underline{\bar{m}}_{1},\mathcal{S}}{\sum}\left(-\right)^{1+\underline{m}_{1}}\left\{ \begin{array}{cccc}
\underline{m} & \underline{m} & \underline{\bar{m}}_1\\
1 & \underline{m} & \underline{\bar{m}} \\
\underline{\bar{s}} & \underline{s}_I & \underline{s}_I & \underline{s}
\end{array}\right\}\cdot
\]
\[
(-)^{\sigma(n,m,p)\left(\underset{i,j=0,1}{\sum}\pi_{m,\bar{k}+i\bar{e}_m+j\bar{e}_n}+\pi_{n,\bar{k}+i\bar{e}_m+j\bar{e}_n}+2\pi_{p,\bar{k}+i\bar{e}_m+j\bar{e}_n}+\pi^2_{m,\bar{k}+i\bar{e}_m+j\bar{e}_n}+\pi^2_{n,\bar{k}+i\bar{e}_m+j\bar{e}_n}\right)}
\]
\[
(-)^{|\bar{p}_n|\left(\underset{i,j=0,1}{\sum}\pi_{n,\bar{k}+i\bar{e}_n+j\bar{e}_m}+j_{n,\bar{k}-\bar{p}_n+i\bar{e}_m}+2j_{n,\bar{k}-\bar{e}_n+\bar{p}_n+i\bar{e}_m+2j\bar{e}_n}+j^2_{n,\bar{k}-\bar{p}_n+i\bar{e}_m}+\pi^2_{n,\bar{k}+i\bar{e}_m+j\bar{e}_n}
\right)}
\]
\[
(-)^{(1-|\bar{p}_m|)\left(\underset{i,j=0,1}{\sum}\pi_{m,\bar{k}+i\bar{e}_m+j\bar{e}_m}+j_{m,\bar{k}-\bar{p}_m+i\bar{e}_n}+2j_{m,\bar{k}-\bar{e}_m+\bar{p}_m+i\bar{e}_n+2j\bar{e}_m}+j^2_{m,\bar{k}-\bar{p}_m+i\bar{e}_n}+\pi^2_{m,\bar{k}+\bar{e}_i+\bar{e}_j}
\right)}
\]
\[
\mathfrak{P}_{SU(3)}\left(\left\{ \underline{\pi}\right\} _{\bar{k}}\ldots;\left\{ \underline{j}\right\} _{\bar{k}};\left\{ \underline{s}\right\} _{\bar{k}}\ldots;\underline{\pi}_{n,\bar{k}}^{2},\ldots;\underline{j}_{n,\bar{k}}^{2}\ldots;\underline{s}_{0,\bar{k}}^{2}\ldots\mid\bar{m}\right)
\]
\[
\left|\nu_{orient}\left(\underline{\pi}_{p,\bar{k}},\underline{\pi}_{m,\bar{k}}^{2},\underline{\pi}_{n,\bar{k}}^{2};\underline{j}_{n,\bar{k}}^{2},\underline{j}_{m,\bar{k}}^{2},\ldots;\underline{s}_{0,\bar{k}}^{2},\underline{s}_{m,\bar{k}}^{2},\underline{s}_{n,\bar{k}}^{2},\underline{s}_{p,\bar{k}}\right)\right\rangle 
\]
\\
where $\mathcal{S}$ is the set of all new appearing labels in the state, which are the ones one has to sum over.

There are now four different cases, one has to look at, left:
\begin{description}
\item [{i)}] $j=m$ $\left(p_{j}=p_{m}\right)$ and $k=n$ $\left(p_{k}\neq p_{n}\right)$
\item [{ii)}] $j=m$ $\left(p_{j}=p_{m}\right)$ and $k\neq n$
\item [{iii)}] $j=m$ $\left(p_{j}\ne p_{m}\right)$ and $k=n$ $\left(p_{k}\ne p_{n}\right)$
\item [{iv)}] $j=m$ $\left(p_{j}\neq p_{m}\right)$ and $k\neq n$
\end{description}
Everything else is (up to a relabelling or switching the orientation
of the loop) one of theses cases. We could draw them as seen in Figure 5.2.
\begin{figure}[h]
\centering
\subfigure[Case i)]{\includegraphics[width=0.4\textwidth]{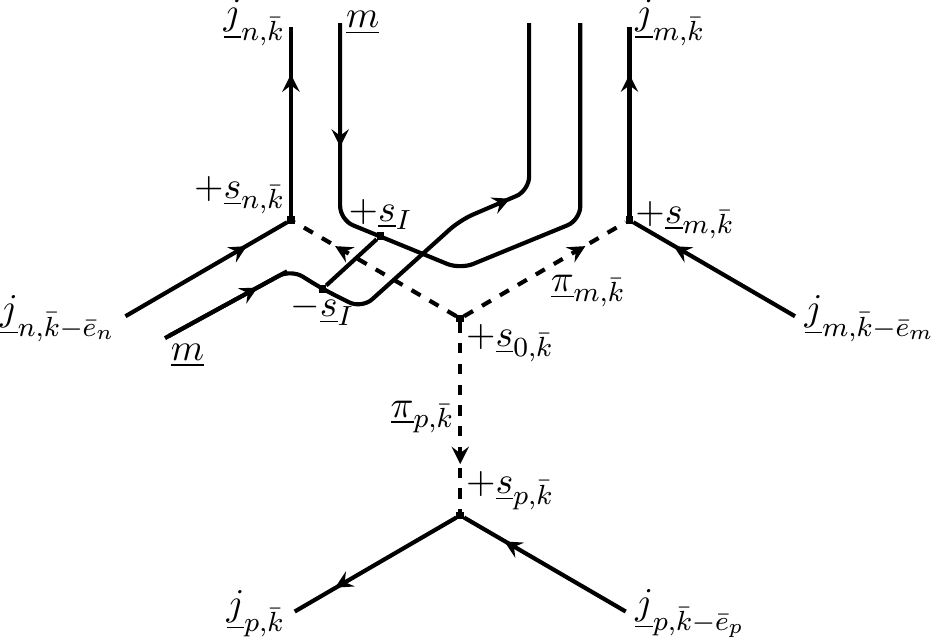}}
\subfigure[Case ii)]{\includegraphics[width=0.4\textwidth]{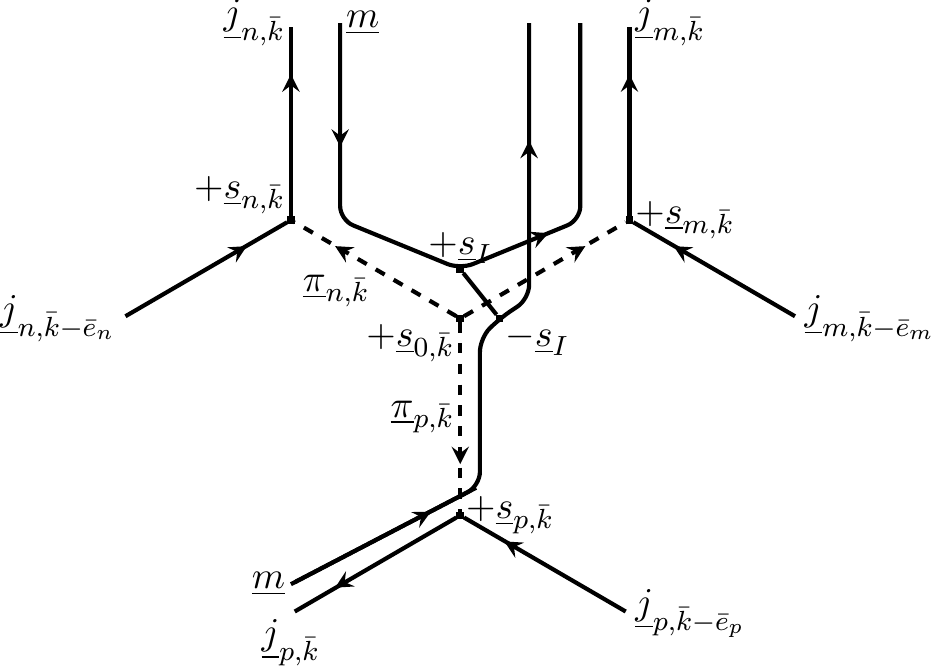}}
\subfigure[Case iii)]{\includegraphics[width=0.4\textwidth]{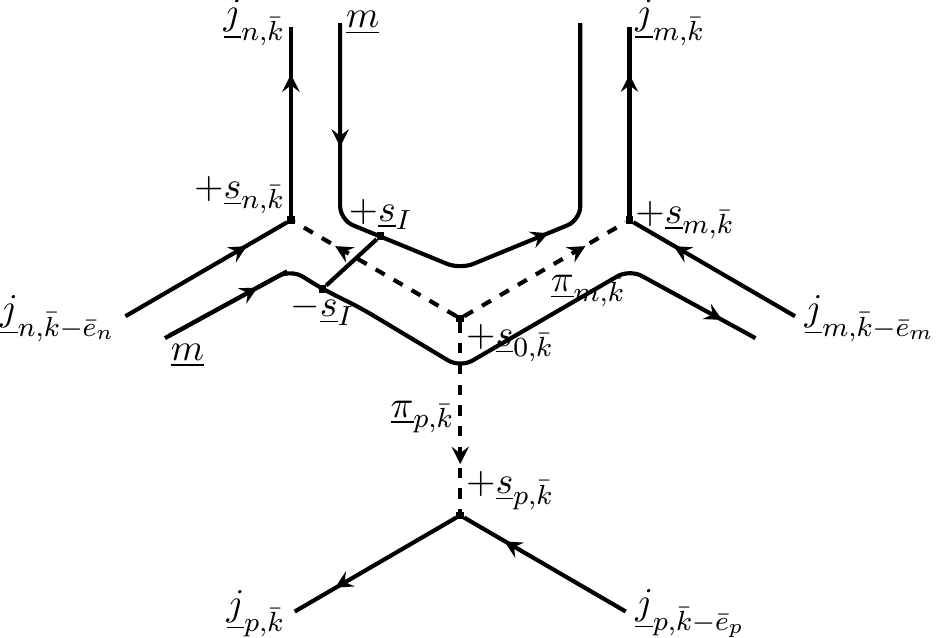}}
\subfigure[Case iv)]{\includegraphics[width=0.4\textwidth]{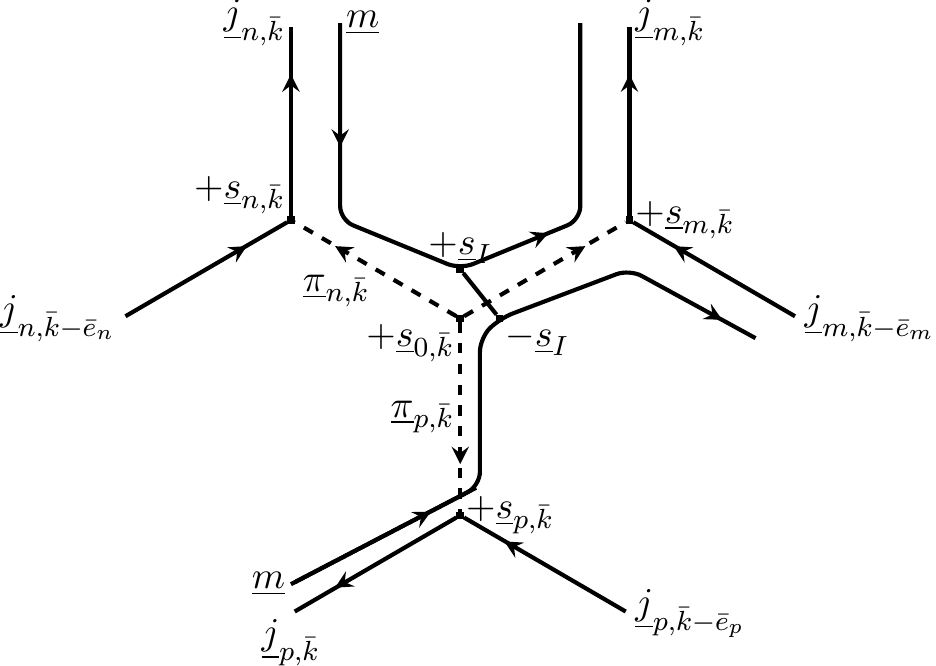}}
        \caption{Different cases of how the holonomies can be oriented. The first plaquette is fixed to be between the $m$ and $n$ direction and the second one can then have four different placements}
\end{figure}
Each loop can be recoupled with the previous techniques, giving a $\mathfrak{P}_{SU(3)}\left(\ldots\right)$-term up to one 6j each, which is due to the coupled $\hat{\tau}_j$. Instead one will get a 12j-symbol, which is defined in the following way:

\begin{equation}
\left\{ \begin{array}{ccccc}
j_{1} & j_{2} & j_{3} & j_{4}\\
l_{1} & l_{2} & l_{3} & l_{4}\\
k_{1} & k_{2} & k_{3} & k_{4}\\
 & s_{1} & s_{2} & s_{3} & s_{4}\\
 & s_{5} & s_{6} & s_{7} & s_{8}
\end{array}\right\} =
\begin{array}{c}\includegraphics[scale=0.7]{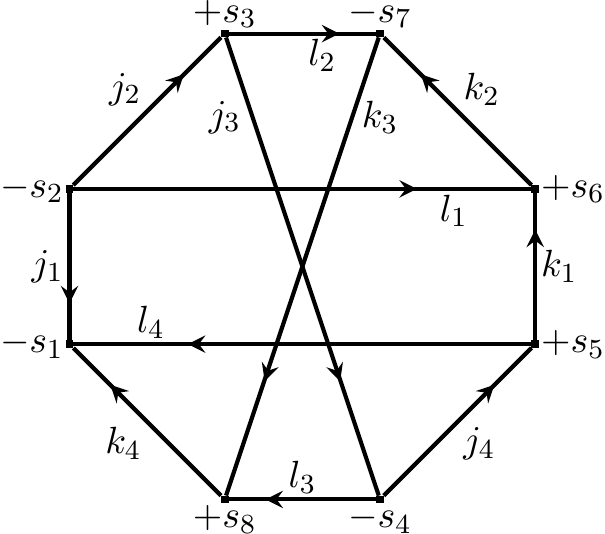}\end{array}
\end{equation}
\\
For instance it can be used to recouple the following object:

\[
\Biggl|
\begin{array}{c}
\includegraphics[scale=0.8]{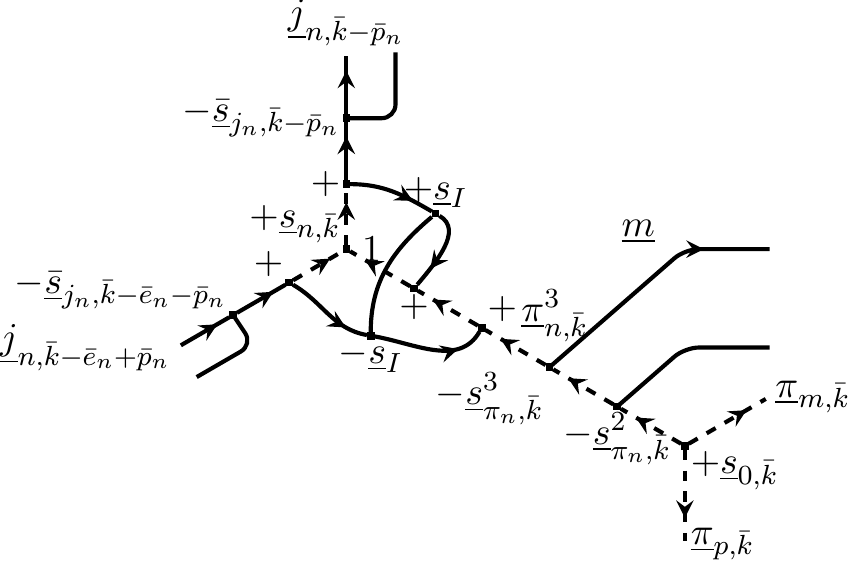}
\end{array}
\Biggr\rangle 
\]
\[
=\underset{\underline{s}_{n,\bar{k}}^{3}}{\sum}
\begin{array}{c}\includegraphics[scale=0.7]{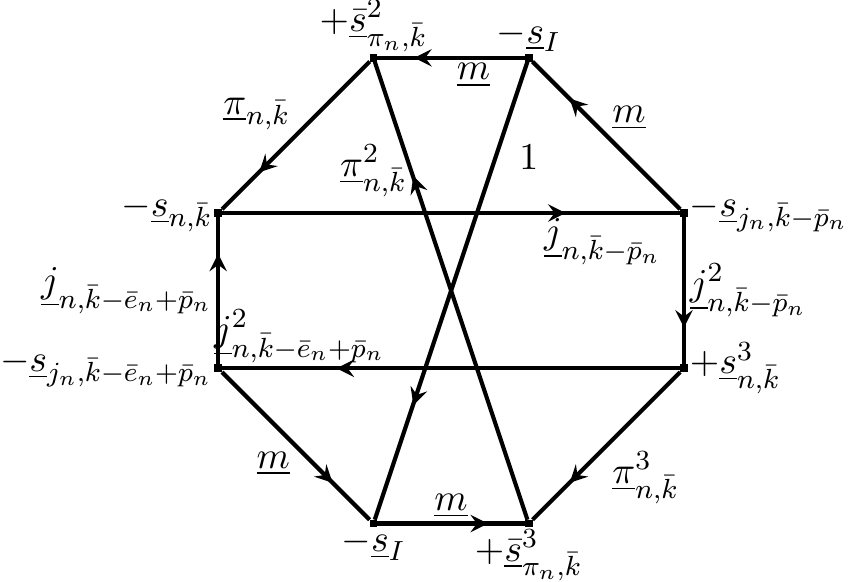}\end{array}
\Biggl|
\begin{array}{c}\includegraphics[scale=0.8]{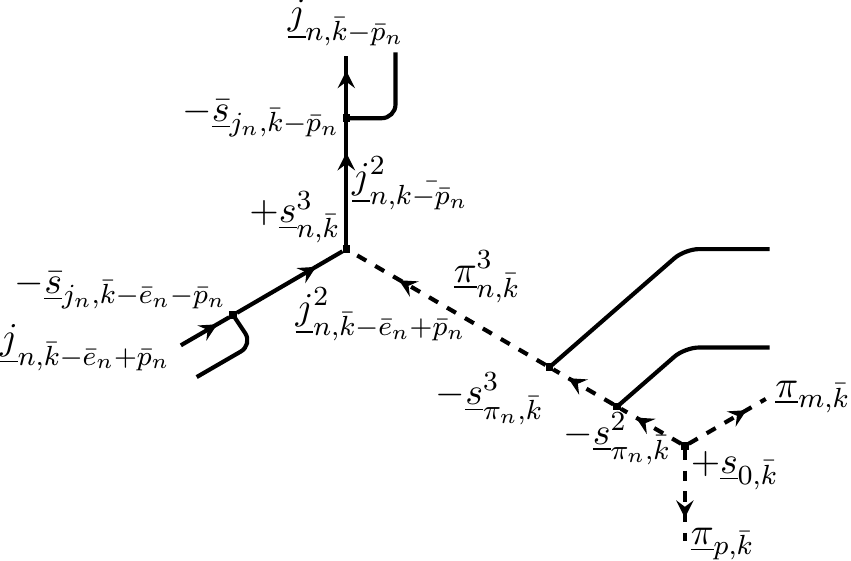}\end{array}
\Biggr\rangle\
\]
\\
And this is exactly the non-trivial operation for case i). So using it, we obtain:

\[
tr\left(\underline{\hat{\tau}}_{I}\underline{\hat{A}}_{m,n,p_{m},1-p_{n}}\right)tr\left(\underline{\hat{\tau}}_{I}\underline{\hat{A}}_{m,n}\right)\left|\nu_{orient}\left(\left\{ \underline{\pi}\right\} _{\bar{k}};\left\{ \underline{j}\right\} _{\bar{k}};\left\{ \underline{s}\right\} _{\bar{k}}\right)\right\rangle =
\]
\[
\underset{\mathcal{S}}{\sum} (-)^{(...)}
(-)^{\underline{\pi}^3_{n,\bar{k}}+\underline{\pi}^2_{n,\bar{k}}+\underline{m}+j_{n,\bar{k}-\bar{p}_n}+\underline{j}^2_{n,\bar{k}-\bar{p}_n}+\underline{\pi}^3_{n,\bar{k}}+\underline{\pi}^2_{n,\bar{k}}+1}
\left\{ \begin{array}{ccccc}
\underline{\bar{j}}_{n,\bar{k}-\bar{e}_{n}+\bar{p}_{n}} & \underline{\bar{\pi}}_{n,\bar{k}} & \underline{\bar{\pi}}_{n,\bar{k}}^{2} & \underline{\bar{\pi}}_{n,\bar{k}}^{3}\\
\underline{j}_{n,\bar{k}-\bar{p}_{n}} & \underline{\bar{m}} & \underline{\bar{m}} & \underline{j}_{n,\bar{k}-\bar{e}_{n}+\bar{p}_{n}}^{2}\\
\underline{\bar{j}}_{n,\bar{k}-\bar{p}_{n}}^{2} & \underline{m} & 1 & \underline{\bar{m}}\\
 & \underline{s}_{j_{n},\bar{k}-\bar{e}_{n}+\bar{p}_{n}} & \underline{s}_{n,\bar{k}} & \underline{\bar{s}}_{\pi_{n},\bar{k}}^{2} & \underline{\bar{s}}_{\pi_{n},\bar{k}}^{3}\\
 & \underline{s}_{n,\bar{k}}^{3} & \underline{s}_{j_{n},\bar{k}-\bar{p}_{n}} & \underline{s}_{I} & \underline{s}_{I}
\end{array}\right\} 
\]
\[
\underset{\underline{s}^2_{n,\bar{k}}}{\sum}
\left\{\begin{array}{cccc}
\underline{\bar{j}}_{n,\bar{k}-\bar{e}_n}+\bar{p}_n & \underline{\pi}^2_{n,\bar{k}} & \underline{\bar{j}}^2_{n,\bar{k}-\bar{p}_n} \\
\underline{\bar{m}} & \underline{\bar{j}}_{n,\bar{k}-\bar{p}_n} & \underline{\bar{\pi}}_{n,\bar{k}} \\
\underline{s}^2_{n,\bar{k}} & \underline{s}_{n,\bar{k}} & \underline{s}^2_{\pi_n,\bar{k}} & \underline{s}_{j_n,\bar{k}-\bar{p}_n}
\end{array}\right\}^{-1}
\left\{\begin{array}{cccc}
\underline{j}^2_{n,\bar{k}-\bar{e}_n+\bar{p}_n} & \underline{\pi}^3_{n,\bar{k}} & \underline{j}^2_{n,\bar{k}-\bar{p}_n} \\
\underline{\pi}^2_{n,\bar{k}} & \underline{\bar{j}}_{n,\bar{k}-\bar{e}_n+\bar{p}_n} & \underline{m} \\
\underline{s}^3_{n,\bar{k}} & \underline{s}_{j_n,\bar{k}-\bar{e}_n+\bar{p}_n} & \underline{s}^3_{\pi_n,\bar{k}} & \underline{s}^2_{n,\bar{k}}
\end{array}\right\}^{-1}
\]
\[
\mathfrak{P}_{SU(3)}\left(\underline{\pi}_{n,\bar{k}},\underline{\pi}_{p,\bar{k}},\underline{\pi}_{m,\bar{k}},\left\{ \underline{\pi}\right\} _{\bar{k}+\bar{e}_{n}},\ldots;\left\{ \underline{j}\right\} ;\underline{s}_{0,\bar{k}},\underline{s}_{m,\bar{k}},\underline{s}_{n,\bar{k}},\underline{s}_{p,\bar{k}},\left\{ \underline{s}\right\} _{\bar{k}+\bar{e}_{n}},\ldots;\right.
\]
\[
\left.
\underline{\pi}_{n,\bar{k}}^{2},\underline{\pi}_{m,\bar{k}}^{2},\underline{\pi}_{n,\bar{k}+\bar{e}_{n}}^{2},\ldots;\underline{j}_{n,\bar{k}-\bar{p}_{n}}^{2},\underline{j}_{m,\bar{k}-\bar{p}_{m}}^{2}\ldots;\underline{s}_{0,\bar{k}}^{2},\underline{s}_{m,\bar{k}}^{2},\underline{s}_{n,\bar{k}}^{2},\ldots\mid\underline{m}\right)
\]
\[
\mathfrak{P}_{SU(3)}\left(\underline{\pi}_{n,\bar{k}}^{2},\underline{\pi}_{m,\bar{k}}^{2},\underline{\pi}_{p,\bar{k}}\ldots;\underline{j}_{n,\bar{k}-\bar{p}_{n}}^{2},\underline{j}_{m,\bar{k}-\bar{p}_{m}}^{2},\ldots;\underline{s}_{0,\bar{k}}^{2},\underline{s}_{m,\bar{k}},\underline{s}_{n,\bar{k}}^{2}\ldots;
\right.
\]
\[
\left.
\underline{\pi}_{n,\bar{k}}^{3},\underline{\pi}_{m,\bar{k}}^{3},\ldots;\underline{j}_{n,\bar{k}-\bar{e}_{n}+\bar{p}_{n}}^{2},\underline{j}_{m,\bar{k}-\bar{p}_{m}}^{3};\underline{s}_{0,\bar{k}}^{3},\underline{s}_{m,\bar{k}}^{3},\underline{s}_{n,.\bar{k}}^{3},\ldots\mid\underline{m}\right)
\]
\[
\left|\nu_{orient}\left(\underline{\pi}_{n,\bar{k}}^{3},\underline{\pi}_{m,\bar{k}}^{3}\underline{\pi}_{p,\bar{k}},\ldots;\underline{j}_{n,\bar{k}}^{2},\underline{j}_{n,\bar{k}-\bar{e}_{n}}^{2},\underline{j}_{m,\bar{k}-\bar{p}_{m}}^{3},\underline{j}_{m,\bar{k}-\bar{e}_{m}+\bar{p}_{m}},\ldots;\underline{s}_{0,\bar{k}}^{3},\underline{s}_{n,\bar{k}}^{3},\underline{s}_{m,\bar{k}}^{3}\underline{s}_{p,\bar{k}}\right)\right\rangle 
\]
\\
The additional sign $(-)^{(...)}$ contains again the resulting sign, which stems from the permutation of $m,n,p$ and the choices of $\bar{p}_n,\bar{p}_m$. Since its construction is the same as before we refrain from writing it down explicitly. The inverse $s$-classified 6j-symbols are chosen in such a way that they cancel the corresponding elements in both $\mathcal{P}_{SU(3)}$ expressions.

For case ii) we get:

\[
tr\left(\underline{\hat{\tau}}_{I}\underline{\hat{h}}_{m,p}\right)tr\left(\underline{\hat{\tau}}_{I}\underline{\hat{h}}_{m,n}\right)\left|\nu_{orient}\left(\left\{ \underline{\pi}\right\} _{\bar{k}};\left\{ \underline{j}\right\} _{\bar{k}};\left\{ \underline{s}\right\} _{\bar{k}}\right)\right\rangle =
\]
\[
=\underset{\mathcal{S}}{\sum}\left(-\right)^{\left(\ldots\right)}\left\{ \begin{array}{ccccc}
\underline{\pi}_{p,\bar{k}} & \underline{\pi}_{m,\bar{k}} & \underline{\pi}_{m,\bar{k}}^{2} & \underline{\pi}_{m,\bar{k}}^{3}\\
\underline{\pi}_{n,\bar{k}} & \underline{\bar{m}} & \underline{\bar{m}} & \underline{\bar{\pi}}_{p,\bar{k}}^{2}\\
\underline{\bar{\pi}}_{n,\bar{k}}^{2} & \underline{m} & 1 & \underline{\bar{m}}\\
 & \underline{s}^2_{\pi_{p},\bar{k}} & \underline{s}_{0,\bar{k}} & \underline{\bar{s}}^2_{\pi_{m},\bar{k}} & \underline{\bar{s}}^3_{\pi_{m},\bar{k}}\\
 & \underline{s}^3_{0,\bar{k}} & \underline{s}^2_{\pi_{n},\bar{k}} & \underline{s}_{I} & \underline{s}_{I}
\end{array}\right\}
 \left(-\right)^{\underline{\pi}^2_{m,\bar{k}}+\underline{\pi}_{m,\bar{k}}^{3}+1}
\]
\[
\left\{\begin{array}{cccc}
\underline{\bar{\pi}}_{p,\bar{k}} & \underline{\bar{\pi}}^2_{m,\bar{k}} & \underline{\bar{\pi}}^2_{n,\bar{k}} \\
\underline{\bar{m}} & \underline{\bar{\pi}}_{n,\bar{k}} & \underline{\pi}_{m,\bar{k}}\\
\underline{s}^2_{0,\bar{k}} & \underline{s}_{0,\bar{k}} & \underline{s}^2_{\pi_m,\bar{k}} & \underline{s}^2_{\pi_n,\bar{k}}
\end{array}\right\}
\left\{\begin{array}{cccc}
\underline{\bar{\pi}}^2_{p,\bar{k}} & \underline{\bar{\pi}}^3_{m,\bar{k}} & \underline{\bar{\pi}}^2_{n,\bar{k}}\\
\underline{\bar{\pi}}^2_{m,\bar{k}} & \underline{\pi}_{p,\bar{k}} & m\\
\underline{s}^3_{0,\bar{k}} & \underline{s}^2_{\pi_p,\bar{k}} & \underline{s}^3_{\pi_m,\bar{k}} & \underline{s}^2_{0,\bar{k}}
\end{array}\right\}
\]
\[
\mathfrak{P}_{SU(3)}\left(\underline{\pi}_{n,\bar{k}},\underline{\pi}_{p,\bar{k}},\underline{\pi}_{m,\bar{k}},\left\{ \underline{\pi}\right\} _{\bar{k}+\bar{e}_{n}},\ldots;\left\{ \underline{j}\right\} ;\underline{s}_{n,\bar{k}},\underline{s}_{m,\bar{k}},\underline{s}_{0,\bar{k}},\left\{ \underline{s}\right\} _{\bar{k}+\bar{e}_{n}},\ldots;
\right.
\]
\[
\left.
\underline{\pi}_{n,\bar{k}},\underline{\pi}_{m,\bar{k}}^{4},\underline{\pi}_{n,\bar{k}+\bar{e}_{n}}^{2},\ldots;\underline{j}_{n,\bar{k}}^{2}\ldots;\underline{s}_{0,\bar{k}}^{2},\underline{s}_{n,\bar{k}}^{2},\underline{s}_{m,\bar{k}}^{2},\ldots\mid\bar{m}\right)
\]
\[
\mathfrak{P}_{SU(3)}\left(\underline{\pi}_{n,\bar{k}}^{2},\underline{\pi}_{p,\bar{k}},\underline{\pi}_{m,\bar{k}}^{2},\left\{ \underline{\pi}\right\} _{\bar{k}+\bar{e}_{p}},\ldots;\ldots\underline{j}_{p,\bar{k}},\underline{j}_{m,\bar{k}}^{2},\ldots;\underline{s}_{m,\bar{k}}^{2},\underline{s}_{p,\bar{k}},\underline{s}_{0,\bar{k}}^{2},\left\{ \underline{s}\right\} _{\bar{k}+\bar{e}_{n}},\ldots;
\right.
\]
\[
\left.
\underline{\pi}_{n,\bar{k}}^{3},\underline{\pi}_{p,\bar{k}}^{2},\ldots;\underline{j}_{p,\bar{k}}^{2},\underline{j}_{m,\bar{k}}^{3}\ldots;\underline{s}_{0,\bar{k}}^{3},\underline{s}_{m,\bar{k}}^{3},\underline{s}_{p,\bar{k}}^{2},\ldots\mid\bar{m}\right)
\]
\[
\left|
\nu_{out}\left(\underline{\pi}_{n,\bar{k}}^{2},\underline{\pi}_{m,\bar{k}}^{3},\underline{\pi}_{p,\bar{k}}^{2}\ldots;\underline{j}_{n,\bar{k}}^{2},\underline{j}_{m,\bar{k}}^{3},\underline{j}_{p,\bar{k}}^{2},\ldots;\underline{s}_{0,\bar{k}}^{3},\underline{s}_{m,\bar{k}}^{3},\underline{s}_{n,\bar{k}}^{2},\underline{s}_{p,\bar{k}}^{2}\right)
\right\rangle
\]
\\
For iii) one gets almost the same as for i):

\[
tr\left(\underline{\hat{\tau}}_{I}\hat{A}_{m,n,1-\bar{p}_{m},1-\bar{p}_{n}}\right)tr\left(\underline{\hat{\tau}}_{I}\underline{\hat{A}}_{m,n}\right)\left|\nu_{orient}\left(\left\{ \underline{\pi}\right\} _{\bar{k}},\left\{ \underline{j}\right\} ;\left\{ \underline{s}\right\} _{\bar{k}}\right)\right\rangle =
\]
\[
\underset{\mathcal{S}}{\sum} (-)^{(...)}
(-)^{\underline{\pi}^3_{n,\bar{k}}+\underline{\pi}^2_{n,\bar{k}}+\underline{m}+j_{n,\bar{k}-\bar{p}_n}+\underline{j}^2_{n,\bar{k}-\bar{p}_n}+\underline{\pi}^3_{n,\bar{k}}+\underline{\pi}^2_{n,\bar{k}}+1}
\left\{ \begin{array}{ccccc}
\underline{\bar{j}}_{n,\bar{k}-\bar{e}_{n}+\bar{p}_{n}} & \underline{\bar{\pi}}_{n,\bar{k}} & \underline{\bar{\pi}}_{n,\bar{k}}^{2} & \underline{\bar{\pi}}_{n,\bar{k}}^{3}\\
\underline{j}_{n,\bar{k}-\bar{p}_{n}} & \underline{\bar{m}} & \underline{\bar{m}} & \underline{j}_{n,\bar{k}-\bar{e}_{n}+\bar{p}_{n}}^{2}\\
\underline{\bar{j}}_{n,\bar{k}-\bar{p}_{n}}^{2} & \underline{m} & 1 & \underline{\bar{m}}\\
 & \underline{s}_{j_{n},\bar{k}-\bar{e}_{n}+\bar{p}_{n}} & \underline{s}_{n,\bar{k}} & \underline{\bar{s}}_{\pi_{n},\bar{k}}^{2} & \underline{\bar{s}}_{\pi_{n},\bar{k}}^{3}\\
 & \underline{s}_{n,\bar{k}}^{3} & \underline{s}_{j_{n},\bar{k}-\bar{p}_{n}} & \underline{s}_{I} & \underline{s}_{I}
\end{array}\right\} 
\]
\[
\underset{\underline{s}^2_{n,\bar{k}}}{\sum}
\left\{\begin{array}{cccc}
\underline{\bar{j}}_{n,\bar{k}-\bar{e}_n}+\bar{p}_n & \underline{\pi}^2_{n,\bar{k}} & \underline{\bar{j}}^2_{n,\bar{k}-\bar{p}_n} \\
\underline{\bar{m}} & \underline{\bar{j}}_{n,\bar{k}-\bar{p}_n} & \underline{\bar{\pi}}_{n,\bar{k}} \\
\underline{s}^2_{n,\bar{k}} & \underline{s}_{n,\bar{k}} & \underline{s}^2_{\pi_n,\bar{k}} & \underline{s}_{j_n,\bar{k}-\bar{p}_n}
\end{array}\right\}^{-1}
\left\{\begin{array}{cccc}
\underline{j}^2_{n,\bar{k}-\bar{e}_n+\bar{p}_n} & \underline{\pi}^3_{n,\bar{k}} & \underline{j}^2_{n,\bar{k}-\bar{p}_n} \\
\underline{\pi}^2_{n,\bar{k}} & \underline{\bar{j}}_{n,\bar{k}-\bar{e}_n+\bar{p}_n} & \underline{m} \\
\underline{s}^3_{n,\bar{k}} & \underline{s}_{j_n,\bar{k}-\bar{e}_n+\bar{p}_n} & \underline{s}^3_{\pi_n,\bar{k}} & \underline{s}^2_{n,\bar{k}}
\end{array}\right\}^{-1}
\]
\[
\mathfrak{P}_{SU(3)}\left(\underline{\pi}_{n,\bar{k}},\underline{\pi}_{p,\bar{k}},\underline{\pi}_{m,\bar{k}},\left\{ \underline{\pi}\right\} _{\bar{k}+\bar{e}_{n}},\ldots;\left\{ \underline{j}\right\} ;\underline{s}_{m,\bar{k}},\underline{s}_{0,\bar{k}},\underline{s}_{n,\bar{k}},\left\{ \underline{s}\right\} _{\bar{k}+\bar{e}_{n}}\ldots;
\right.
\]
\[
\left.
\underline{\pi}_{n,\bar{k}}^2,\underline{\pi}_{m,\bar{k}}^{2},\underline{\pi}_{n,\bar{k}+\bar{e}_{n}}^{2},\ldots;\underline{j}_{n,\bar{k}-\bar{p}_{n}}^{2},\ldots;\underline{s}_{0,\bar{k}}^{2},\underline{s}_{m,\bar{k}}^{2},\underline{s}_{n,\bar{k}}^{2},\ldots\mid\bar{m}\right)
\]
\[
\mathfrak{P}_{SU(3)}\left(\underline{\pi}_{n,\bar{k}}^2,\underline{\pi}_{p,\bar{k}},\underline{\pi}_{m,\bar{k}}^{2},\left\{ \underline{\pi}\right\} _{\bar{k}+\bar{e}_{n}}^{2},\ldots;\ldots\underline{j}_{m,\bar{k}-\bar{p}_{m}}^{2},\underline{j}_{n,\bar{k}-\bar{p}_{n}}^{2},\ldots;\underline{s}_{m,\bar{k}}^{2},\underline{s}_{0,\bar{k}}^{2},\underline{s}_{n,\bar{k}}^{2},\left\{ \underline{s}\right\} _{\bar{k}+\bar{e}_{n}}^{2}\ldots;
\right.
\]
\[
\left.
\underline{\pi}_{n,\bar{k}}^{3},\underline{\pi}_{m,\bar{k}}^{3},\ldots;\underline{j}_{n,\bar{k}-\bar{e}_{n}+\bar{p}_{n}}^{2},\underline{j}_{m,\bar{k}-\bar{e}_{m}+\bar{p}_{m}}^{3},\ldots;\underline{s}_{0,\bar{k}}^{3},\underline{s}_{m,\bar{k}}^{3},\underline{s}_{n,\bar{k}}^{3},\ldots\mid\bar{m}\right)
\]
\[
\left|\nu_{orient}\left(\underline{\pi}_{n,\bar{k}}^{3},\underline{\pi}_{m,\bar{k}}^{3}\underline{\pi}_{p,\bar{k}},\ldots;\underline{j}_{n,\bar{k}}^{2},\underline{j}_{n,\bar{k}-\bar{e}_{n}}^{2},\underline{j}_{m,\bar{k}-\bar{p}_{m}}^{3},\underline{j}^2_{m,\bar{k}-\bar{e}_{m}+\bar{p}_{m}},\ldots;\underline{s}_{0,\bar{k}}^{3},\underline{s}_{n,\bar{k}}^{3},\underline{s}_{m,\bar{k}}^{3}\underline{s}_{p,\bar{k}}\right)\right\rangle 
\]
\\
For iv) finally (compare to (ii)):

\[
tr\left(\underline{\hat{\tau}}_{I}\hat{A}_{m,p,1-\bar{p}_{m},\bar{p}_{p}}\right)tr\left(\underline{\hat{\tau}}_{I}\underline{\hat{A}}_{m,n}\right)\left|\nu_{orient}\left(\left\{ \underline{\pi}\right\} _{\bar{k}},\left\{ \underline{j}\right\} ;\left\{ \underline{s}\right\} _{\bar{k}}\right)\right\rangle =
\]
\[=
\underset{\mathcal{S}}{\sum}\left(-\right)^{\left(\ldots\right)}\left\{ \begin{array}{ccccc}
\underline{\pi}_{p,\bar{k}} & \underline{\pi}_{m,\bar{k}} & \underline{\pi}_{m,\bar{k}}^{2} & \underline{\pi}_{m,\bar{k}}^{3}\\
\underline{\pi}_{n,\bar{k}} & \underline{\bar{m}} & \underline{\bar{m}} & \underline{\bar{\pi}}_{p,\bar{k}}^{2}\\
\underline{\bar{\pi}}_{n,\bar{k}}^{2} & \underline{m} & 1 & \underline{\bar{m}}\\
 & \underline{s}^2_{\pi_{p},\bar{k}} & \underline{s}_{0,\bar{k}} & \underline{\bar{s}}^2_{\pi_{m},\bar{k}} & \underline{\bar{s}}^3_{\pi_{m},\bar{k}}\\
 & \underline{s}^3_{0,\bar{k}} & \underline{s}^2_{\pi_{n},\bar{k}} & \underline{s}_{I} & \underline{s}_{I}
\end{array}\right\}
 \left(-\right)^{\underline{\pi}^2_{m,\bar{k}}+\underline{\pi}_{m,\bar{k}}^{3}+1}
\]
\[
\left\{\begin{array}{cccc}
\underline{\bar{\pi}}_{p,\bar{k}} & \underline{\bar{\pi}}^2_{m,\bar{k}} & \underline{\bar{\pi}}^2_{n,\bar{k}} \\
\underline{\bar{m}} & \underline{\bar{\pi}}_{n,\bar{k}} & \underline{\pi}_{m,\bar{k}}\\
\underline{s}^2_{0,\bar{k}} & \underline{s}_{0,\bar{k}} & \underline{s}^2_{\pi_m,\bar{k}} & \underline{s}^2_{\pi_n,\bar{k}}
\end{array}\right\}
\left\{\begin{array}{cccc}
\underline{\bar{\pi}}^2_{p,\bar{k}} & \underline{\bar{\pi}}^3_{m,\bar{k}} & \underline{\bar{\pi}}^2_{n,\bar{k}}\\
\underline{\bar{\pi}}^2_{m,\bar{k}} & \underline{\pi}_{p,\bar{k}} & m\\
\underline{s}^3_{0,\bar{k}} & \underline{s}^2_{\pi_p,\bar{k}} & \underline{s}^3_{\pi_m,\bar{k}} & \underline{s}^2_{0,\bar{k}}
\end{array}\right\}
\]
\[
\mathfrak{P}_{SU(3)}\left(\underline{\pi}_{n,\bar{k}},\underline{\pi}_{p,\bar{k}},\underline{\pi}_{m,\bar{k}},\left\{ \underline{\pi}\right\} _{\bar{k}+\bar{e}_{n}},\ldots;\left\{ \underline{j}\right\} ;\underline{s}_{n,\bar{k}},\underline{s}_{m,\bar{k}},\underline{s}_{0,\bar{k}},\left\{ \underline{s}\right\} _{\bar{k}+\bar{e}_{n}},\ldots;
\right.
\]
\[
\left.
\underline{\pi}_{n,\bar{k}},\underline{\pi}_{m,\bar{k}}^{4},\underline{\pi}_{n,\bar{k}+\bar{e}_{n}}^{2},\ldots;\underline{j}_{n,\bar{k}}^{2}\ldots;\underline{s}_{0,\bar{k}}^{2},\underline{s}_{n,\bar{k}}^{2},\underline{s}_{m,\bar{k}}^{2},\ldots\mid\bar{m}\right)
\]
\[
\mathfrak{P}_{SU(3)}\left(\underline{\pi}_{n,\bar{k}}^{2},\underline{\pi}_{p,\bar{k}},\underline{\pi}_{m,\bar{k}}^{2},\left\{ \underline{\pi}\right\} _{\bar{k}+\bar{e}_{p}-2\bar{p}_{p}}\ldots;\ldots\underline{j}_{p,\bar{k}},\underline{j}_{m,\bar{k}}^{2},\ldots;\underline{s}_{m,\bar{k}}^{2},\underline{s}_{p,\bar{k}},\underline{s}_{0,\bar{k}},\left\{ \underline{s}\right\} _{\bar{k}+\bar{e}_{p}-2\bar{p}_{p}},\ldots;
\right.
\]
\[
\left.
\underline{\pi}_{m,\bar{k}}^{3},\underline{\pi}_{p,\bar{k}}^{2},\ldots;\underline{j}_{p,\bar{k}}^{2},\underline{j}_{m,\bar{k}-\bar{e}_{m}+\bar{p}_{m}}^{3},\ldots;\underline{s}_{0,\bar{k}}^{3},\underline{s}_{m,\bar{k}}^{3},\underline{s}_{p,\bar{k}}^{2},\ldots\mid\bar{m}\right)
\]
\[
\left|
\nu_{out}\left(\underline{\pi}_{n,\bar{k}}^{2},\underline{\pi}_{m,\bar{k}}^{3},\underline{\pi}_{p,\bar{k}}^{2}\ldots;\underline{j}_{n,\bar{k}}^{2},\underline{j}_{m,\bar{k}}^{3},\underline{j}_{p,\bar{k}}^{2},\underline{j}_{m,\bar{k}-\bar{e}_{m}}^{2}\ldots;\underline{s}_{0,\bar{k}}^{3},\underline{s}_{m,\bar{k}}^{3},\underline{s}_{n,\bar{k}}^{2},\underline{s}_{p,\bar{k}}^{2}\right)
\right\rangle
\]
\section{Conclusion}
\label{c6}
In this paper we have taken the first steps towards the computation of the fundamental QCD spectrum within the LQG approach to quantum gravity. More precisely, we have computed the matrix elements of the Yang-Mills contribution to the Hamiltonian analytically in closed form as far as the gluon field is concerned, while for the gravitational degrees of freedom a fully analytical analysis is not possible due to the necessity of computing the spectrum of the volume operator, which is known to be possible only numerically. Obviously, more analytical and numerical work is necessary to determine the spectrum with sufficient precision. However the focus of this paper was not so much on the actual computation of the spectrum, but rather to prepare the necessary analytical tools. The other message that we wanted to communicate is that the Hamiltonian that we considered in this paper needs to be improved by methods comming from renormalisation theory. For this reason, we refrain from investigating more closely the spectrum of the Hamiltonian considered here from \cite{GT06_1}, but one should rather analyse the improved Hamiltonian. We hope that, once one has found a Hamiltonian description of renormalisation, its fixed point Hamiltonian can be used, as this Hamiltonian has minimal if not vanishing discretisation errors. Once this point has been understood, we can address the important question of how the picture of the running of the Yang-Mills coupling on a gravitational background is changed in the context of the quantum gravity coupled system. Namely it transpires that the background dependent Hamiltonian depends on a cut-off while the background independent one does not. Thus, the mechanism for the running of the coupling is very different for these two theories. We reserve this analysis for future research.
\section*{Acknowledgements}
This project was supported in part by an Emerging Fields Initiative granted by the Friedrich-Alexander University. KL thanks the German National Merit Foundation for financial support.
\appendix
\section{Brief review on the $3j$'s and $6j$'s for $SU(2)$}
\label{app:recoupling}
For self-containedness some important properties of $nj$-Symbols for the group $SU(2)$ are listed here. Introductions to recoupling theory can be found in various textbooks on quantum mechanics and quantum angular momentum, e.g. \cite{BS68}. For an extensive list of properties of $nj$-symbols see e.g. \cite{wolfram}

\begin{itemize}
\item[] {\bf 3j-Symbols}
	\begin{itemize}
		\item[] \emph{Relation to Clebsh-Gordan coefficients:}
			\[
			\scal{a,\alpha; b,\beta| c,\gamma}=(-)^{b-a+\gamma}\sqrt{2c+1}
			\threej{a}{\alpha}{b}{\beta}{c}{-\gamma}
			\]
			where $\ket{b,\beta;a,\alpha}= \ket{b,\beta}\otimes\ket{a,\alpha}$
			
		\item[]  \emph{Compatibility criteria}\\
			If one (or several) of the following rules is violated, 
			then $\threej{a}{\alpha}{b}{\beta}{c}{\gamma}$ is vanishing:
			\begin{itemize}
				\item $a,b,c\in\frac{1}{2}\N$, $a\pm\alpha\in\N$,  $-a\leq\alpha\leq a$, $\cdots$
				\item $\alpha+\beta+\gamma=0$
				\item $a+b+c\in\N$, $|a-b|\leq c\leq a+b$ (triangle inequality)
			\end{itemize}
		
		\item[]  \emph{Symmetries}
			\[
			\threej{a}{\alpha}{b}{\beta}{c}{\gamma}=(-)^{a+b+c}\threej{a}{-\alpha}{b}{-\beta}{c}{-\gamma}
			=(-)^{a+b+c}\threej{b}{\beta}{a}{\alpha}{c}{\gamma}=\threej{b}{\beta}{c}{\gamma}{a}{\alpha}
			\]
	\end{itemize}
\item[] {\bf 6j-Symbols}
	\begin{itemize}
		\item[]  \emph{Definition in terms of $3j$'s}
			\begin{align*}
			&\sixj{j_1}{j_4}{j_2}{j_5}{j_3}{j_6}=\sum_{\mu_1,\cdots,\mu_6} (-)^{\sum\limits_{i=1}^6 (j_i-\mu_i)}\\
			&
			\threej{j_1}{\mu_1}{j_2}{\mu_2}{j_3}{-\mu_3}\threej{j_1}{-\mu_1}{j_5}{\mu_5}{j_6}{\mu_6}
			\threej{j_4}{\mu_4}{j_5}{-\mu_5}{j_3}{\mu_3}\threej{j_4}{-\mu_4}{j_2}{-\mu_2}{j_6}{-\mu_6}
			\end{align*}
		\item[]  \emph{Symmetries}
			\begin{align*}
			\sixj{a}{d}{b}{e}{c}{f}=\sixj{b}{e}{a}{d}{c}{f}=\sixj{b}{e}{c}{f}{a}{d}=
			\end{align*}
			\begin{align*}
			\sixj{d}{a}{e}{b}{c}{f}=\sixj{d}{a}{b}{e}{f}{c}=\sixj{a}{d}{e}{b}{f}{c}
			\end{align*}
		\item[]  \emph{Compatibility}
			\\[3pt]
			\begin{align*}
			\sixj{a}{d}{b}{e}{c}{f}=0
			\end{align*}
			unless the triangle inequalities hold for $\{a,b,c\}, \{a,e,f\},\{d,b,f\}$ 
			and $\{d,e,c\}$\\[3pt]
		\item[]  \emph{Orthogonality}
			\[
			\sum_x d_x \sixj{a}{d}{b}{e}{x}{c}\sixj{a}{d}{b}{e}{x}{c'}=\delta_{c,c'}\frac{1}{d_c}
			\]
			if the compatibility requirements are fulfilled.
	\end{itemize}

\item[] {\bf Graphical Calculus of $SU(2)$}
\begin{itemize}
	\item[]  The \emph{definitions of the basic objects} in this graphical calculus are the same as in \cite{ATZ11, ALZ13} and thus reduce to the same labeling as has been done for the $SU(3)$ case.
	\item[] Some of the \emph{rules for changing the graphs} however have altered, e.g. since the magnetic numbers are now in $\frac{1}{2}\mathbb{N}$ an arrow may change its direction by adding a sign factor of $(-)^{2a}$
	\begin{align*}
	\begin{array}{c}\includegraphics{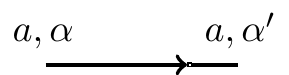}\end{array}
	=\left(-1\right)^{2a}
	\begin{array}{c}\includegraphics{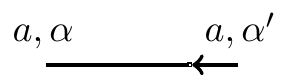}\end{array}
	\end{align*}
	\item[] This changes some of the more compley recoupling schemes (for a full list see \cite{BS68}). E.g. the \emph{extraction of a 6j-symbol}
	
\[
\begin{array}{c}\includegraphics{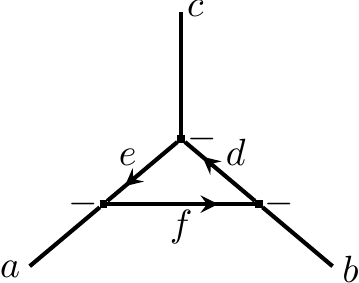}\end{array}
=
\begin{array}{c}\includegraphics{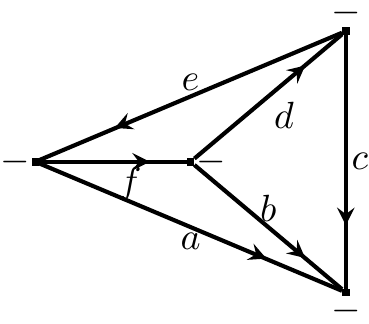}\end{array}
\cdot
\begin{array}{c}\includegraphics{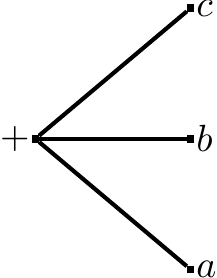}\end{array}
=
\]

\begin{equation} \label{extraction}
\left\{ \begin{array}{ccc}
a & f & e\\
d & c & b
\end{array}\right\} \cdot
\begin{array}{c}\includegraphics{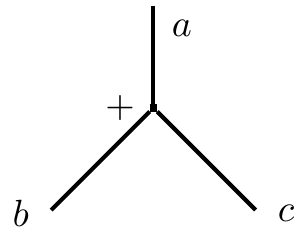}\end{array}
\end{equation}	
\end{itemize}
\end{itemize}


\begin{thebibliography}{99}
  \bibitem{KS75}
 J.~Kogut and L.~Susskind,
  ``Hamiltonian formulation of Wilson's lattice gauge theories'',
  Phys.\ Rev.\ D {\bf 11, 2} (1975) 395
  \bibitem{OS72}
 K.~Osterwalder and R.~Schrader,
  ``Axioms for Euclidean Green's Functions'',
  Commun.\ math.\ Phys.\ 31 (1973) 83-112
    
  \bibitem{GJ87}
 J.~Glimm and A.~Jaffe,
  ``Quantum Physics - A functional integral Point of View'',
  Springer-Verlag, New York, (1987)
  
  \bibitem{Bal89a}
 Tadeusz~Balaban,
  ``Large field renormalization I. The Basic Step of the R Operation'',
  Commun.\ math.\ Phys.\ {\bf 122} (1989) 207-263

  \bibitem{Bal89b}
 Tadeusz~Balaban,
  ``Large field renormalization II. Localization, Exponentation, and Bounds for the R. Operation'',
  Commun.\ math.\ Phys.\ {\bf 122} (1989) 355-392
 
   \bibitem{Fr78}
 Juerg~Froehlich,
  ``An introduction to some topics in Constructive QFT'',
  Springer-Verlag, New York, (1978)
 
   \bibitem{Riv00}
 Vincent~Rivasseau,
  ``Constructive Field Theory and Applications: Perspectives and Open Problems'',
  J.\ Math.\ Phys.\ {\bf 41} (2000) 3764-3775
\bibitem{GL10}
 C.~Gattringer and C.B.~Lang,
  ``Quantum Chromodynamics on the Lattice'',
  Springer-Verlag, New York, (2010)
\bibitem{Cr83}
 Micheal~Creutz,
  ``Quarks, Gluons and Lattices'',
 Cambridge University Press (1983)  
\bibitem{BFV03}
	R.Brunetti, K. Fredenhagen and R. Verch.
	``The Generally Covaraiant Locality Principle - A New Paradigm for Local Quantum Field Theory''
	{\it Commun.Math.Phys.}
	{\bf 237}
	(2003)
	31-68   
\bibitem{Buch00}
	Detlev Buchholz.
	``Algebraic Quantum Field Theory: A Status Report''
	(2000),	
	[arXiv:math-ph/0011044v1]
\bibitem{FRS07}
	K. Fredenhagen, K.-H. Rehren and E. Seiler.
	``Quantum Field Theory: Where we are''
	{\it Lect.NotesPhys.}
	{\bf 721}
	(2007)	
	61-87,
	[arXiv:hep-th/0603155v1]
\bibitem{JR06}
	A. Jaffe and G. Ritter.
	``Quantum Field Theory on Curved Backgrounds. II. Spacetime Symmetries''
	(2007)
	[arXiv:0704.0052v1]  
\bibitem{Thi07}
	Thomas Thiemann.
	Modern Canonical Quantum General Relativity.
	{\it Cambridge University Press}
	(2007)	
\bibitem{Rov04}
	Carlo Rovelli.
	Quantum Gravity.
	{\it Cambridge University Press}
	(2004)
\bibitem{GS13}
	K. Giesel and H. Sahlmann.
	From Classical To Quantum Gravity: Introduction to Loop Quantum Gravity.
	(2013),
	[arXiv:1203.2733v2]    
\bibitem{Thi98}
	Thomas Thiemann.
	\emph{Quantum Spin Dynamics V: Quantum Gravity as the Natural Regulator of Matter Quantum Field Theories}.
	{\it Class. Quant. Grav.}
	{\bf 15}
	(1998)
	arXiv:gr-qc/9705019
\bibitem{Thi96_1}
	Thomas Thiemann,
	\emph{Quantum Spin Dynamics (QSD)},
	arXiv:gr-qc/9606089v1,
	1996.	
\bibitem{Thi96_2}
	Thomas Thiemann,
	\emph{Quantum Spin Dynamics (QSD) II},
	arXiv:gr-qc/9606090v1,
	1996.      
\bibitem{GT12}
	K. Giesel and T Thiemann.
	Scalar Material Reference Systems and Loop Quantum Gravity.
	(2012),
	[arXiv:1206.3807v2]
\bibitem{DGKL10}
	M. Domagala, K. Giesel, W. Kaminski and J. Lewandowski.
	Gravity quantized.
	{\it Phys.Rev.D}
	{\bf 82}
	(2010)
\bibitem{HP11}
	V. Husain and T. Pawlowski.
	Time and a physical Hamiltonian for a quantum gravity.
	(2011)
	[arXiv:1108.1145v2]
	
\bibitem{HP13}
	V. Husain and T. Pawloski.
	A computable framework for Loop Quantum Gravity.
	(2013)
	[arXiv:1305.5203v1]  
\bibitem{KT89}
	 K.V. Kuchar and C.G. Torre.
	 World sheet diffeomorphisms and the canonical string.
	 {\it Journal of Mathematical Physics}
	 {\bf 30}
	 (1989)
	 1769
\bibitem{GT06_1}
	K. Giesel and T. Thiemann.
	Algebraic Quantum Gravity (AQG) I. Conceptual Setup.
	{\it Class.Quant.Grav.}
	{\bf 24}
	2465-2498,	
\bibitem{GT06_2}
	K. Giesel and T. Thiemann.
	Algebraic Quantum Gravity (AQG) II. Semiclassical Analysis.
	{\it Class.Quant.Grav.}
	{\bf 24}
	(2006)	
	2499-2564	
\bibitem{GT06_3}
	K. Giesel and T. Thiemann.
	Algebraic Quantum Gravity (AQG) III. Semiclassical Perturbation Theory.
	{\it Class.Quant.Grav.}
	{\bf 24}
	(2006)
	2499-2564	
\bibitem{GT07_4}
	K. Giesel and T. Thiemann.
	Algebraic Quantum Gravity (AQG) IV. Reduced Phase Space Quantisation of Loop Quantum Gravity.
	(2007)
	[arXiv:0711.0119v1]  
\bibitem{LOST06}
	J. Lewandowski, A. Okolow, H. Sahlmann and T. Thiemann.
	Uniqueness of diffeomorphism invariant states on holonomy-flux algebras.
	{\it Commun.Math.Phys.}
	{\bf 267}
	(2006)	
	703-733
	[arXiv:gr-qc/0504147v2]		
\bibitem{Flei06}
	Christian Fleischhack.
	Quantization Restrictions for Diffeomorphism Invariant Gauge Theories.
	{\it Quantization and Analysis on Symmetric Spaces}
	(Proceedings, Wien, 2006)
	55-63
\bibitem{STW01}
	H. Sahlmann, T. Thiemann and O. Winkler.
	Coherent States for Canonical Quantum General Relativity and the Infinte tensor Product Extension.
	{\it Nucl.Phys. B}
	{\bf 606}
	(2001)
	401-440
\bibitem{Wal84}
 	Robert Wald.
 	General Relativity.
 	{\it University of Chicago Press}
 	(1984)
\bibitem{Ash91}
	A. Ashtekar.
	Lectures on non perturbative canonical gravity.
	{\it Word Scientifitc}
	(1991)	
\bibitem{Bar94}
	J. Fernando Barbero.
	A real polynomial formulationof general relativity in terms of connection.
	{\it Phys. Rev. D}
	{\bf 49}
	(1994)
	6935-6938
\bibitem{Bar95}
	J. Fernando Barbero.
	Real ashtekar variables for lorentzian signature space times.
	{\it Phys. Rev. D}
	{\bf 51}
	(1995)
	5507-5510
\bibitem{Bar96}
	J. Fernando Barbero.
	From euclidean to lorentzian general reativity: The real way.
	{\it Phys. Rev. D}
	{\bf 54}
	(1996)
	1492-1499
\bibitem{GHTW10}
	K. Giesel, S. Hofmann, T. Thiemann and O. Winkler.
	Manifestly Gauge-Invariant General Relativistic Perturbation Theory : I. Foundations.
	{\it Class.Quant.Grav.}
	{\bf 27}
	(2010)
\bibitem{AL97}
	A. Ashtekar	and J. Lewandowski.
	``Quantum theory of geometry II: volume operators'',
	{\it Adv. Theo. Math Phys.}
	{\bf 1}
	(1997),
	388-4297
\bibitem{Brunnemann:2004xi}
  J.~Brunnemann, T.~Thiemann,
  ``Simplification of the spectral analysis of the volume operator in loop
  quantum gravity,''
  Class.\ Quant.\ Grav.\  {\bf 23}, 1289 (2006).  
\bibitem{BR06}
	J. Brunnemann, D. Rideout,
	``Spectral analysis of the Volume Operator in Loop Quantum Gravity'',
	(2006),	
	[arXiv:gr-qc/0612147]  
\bibitem{ATZ11}
	E. Alesci, T. Thiemann and A. Zipfel.
	Linking covariant and canonical LQG: new solutions to the Euclidean scalar Constraint.
	{\it Phys. Rev. D} 
	{\bf 86}
	(2011),
	[arXiv:1109.1290]

\bibitem{ALZ13}
	E. Alesci, K. Liegener and A. Zipfel.
	Matrix Elements of Lorentzian Hamiltonian Constraint in LQG.
	{\it Phys. Rev. D}
	{\bf 88}
	(2013),
	[arXiv:1306.0861v2]  
   
\bibitem{BS68}
  D.M. Brink and G.R. Satchler.
  Angular Momentum.
  {\it Clarendon Press Oxford}
  (1968)
  
\bibitem{Cvi08}
	Predrag Cvitanovi\'{c}.
	Group Theory: Birdtracks, Lie's, and Exceptional Groups.
	{\it Princeton University Press}
	2008

\bibitem{Gri84}
	Marius Grigorescu.
	SU(3) Clebsch-Gordan Coefficients.
	{\it Stud.CercetariFiz} {\bf 36}
	(1984)
	[arXiv:math-ph/0007033v1]

\bibitem{Hall03}
	Brian C. Hall.
	Lie Groups, Lie Algebras, and Representations.
	{\it Springer-Verlag New York}
	(2003)

\bibitem{Hum72}
	James E. Humphreys.
	Introduction to Lie algebras and respresentation theory.
	{\it Springer Verlag}
	(1972)

\bibitem{VK91}
	N. Ja. Vilenkin and A. U. Klimyk.
	Representations of Lie Groups and Special Functions,
	Vol. 1: Simplest Lie Groups, Special Functions and Integral Transforms.
	\emph{Kluwer Academic Publishers}
	(1991)
	
\bibitem{VK95}
	N. Ja. Vilenkin and A. U. Klimyk.
	Representations of Lie Groups and Special Functions,
	Recent Advances.
	{\it Kluwer Academic Publishers}
	(1995)

\bibitem{PST86}
	Z Pluhar, Yu F Smirnov and V N Tolstoy.
	Clebsch-Gordan coefficients of SU(3) with simple symmetry properties.
	{\it J. Phys. A: Math. Gen.}
	{\bf 19}
	(1986)
	29-34
	
\bibitem{PWH86}
	Z Pluhar, L J Weigert and P Holan.
	Symmetry properties of s-classified SU(3) 3j-, 6j- and 9j-symbols.
	{\it J. Phys A: Math Gen}
	{\bf 19}
	(1986)	
	
\bibitem{DS64}
	J.-R. Derome and W. T. Sharp.
	Racah Algebra for an Arbitrary Group.
	{\it Journal of Mathematical Physics}
	{\bf 6}
	(1965)
	10
	
\bibitem{YLV62}
	A.Yutsis, I. Levinson, V. Vanagas.
	The Theory of Angular Momentum.
	Isreal Program for Scientific Translations.
	(1962)
	
\bibitem{MS54}
	J. Meixner and F. W. Schaefke.
	Mathieusche Funktionen und Sphaeroidenfunktionen.
	{\it Springer-Verlag}
	(1954)
	
\bibitem{RW80}
	D. Robson and D. M. Webber.
	Gauge Theories on a Small Lattice.
	{\it Z. Physik C, Particles and Fields}
	{\bf 7}
	(1980)
	
\bibitem{RW81}
	D. Robson and D. M. Webber.
	The Gauge Boson Sector of U(1) Lattice Theories.
	{\it Z. Phys. C, Particles and Fields}
	{\bf 10}
	(1981)

\bibitem{RW82}
	D. Robson and D. M. Webber.
	Gauge Covariance in Lattice Field Theories.
	{\it Z. Phys. C, Particles and Fields}
	{\bf 15}
	(1982)	
\bibitem{BPMUV99}
	G. Burgio, R. De Pietri, H. A. Morales-Tecotl, L. F. Urrutia and J. D. Vergara.
	Matrix elements of the plaquette operator of Lattice Gauge Theory.
	(1999)
	[arXiv:hep-lat/9911-10v1]
\bibitem{BPMUV00}
	G. Burgio, R. De Pietri, H. A. Morales-Tecotl, L. F. Urrutia and J. D. Vergara.
	The basis of the physical Hilbert space of Lattice Gauge Theory.
	{\it Nucl.Phys. B}
	{\bf 566}
	(2000)
	547-561,
	[arXiv:hep-lat/9906036v1]
\bibitem{Mat08}
	Manu Mathur.
	Loop States in Lattice Gauge Theories.
	{\it Phys.Lett. B}
	{\bf 640}
	(2006)
	292-296,
	[arXiv:hep-lat/0510101v2]
\bibitem{KHLK09}
	H. Kroeger, A. Hosseinizadeh, J.F. Laprise and J. Kroeger.
	Spectrum and Wave Functions of Exicted States in Lattice Gauge Theory.
	{\it PoS LATTICE2008}
	{\bf 235}
	(2008)
	[arXiv:0902.2944v1]
\bibitem{CDR88}
	J.-W. Choe, A. Duncan and R. Roskies.
	Lanczos calculation of the spectrum of Hamiltonian lattice gauge theory.
	{it Physical Review D }
	{\bf 37}
	(1988)
	2
\bibitem{DR85}
	A. Duncan and R. Roskies.
	Variational estimates for spectra in lattice Hamiltonian theories.
	{\it Physical Review D}
	{\bf 31}
	(1985)
	2	
\bibitem{AR07}
	E. Alesci and C. Rovelli.
	The complete LQG propagator I. Difficulties with the Barrett-Crane vertex.
	{\it Phys.Rev.D}
	{\bf 76}
	(2007),
	[arXiv:0708.0883]		
\bibitem{wolfram}
http://functions.wolfram.com/HypergeometricFunctions/ThreeJSymbol/
\end{thebibliography}
\end{document}